\documentstyle[sprocl,epsf]{article}

\def\Journal#1#2#3#4{{#1} {\bf #2}, #3 (#4)}
\def\add#1#2#3{{\bf #1}, #2 (#3)}

\def\NPB{{\em Nucl. Phys.} B}
\def\PLB{{\em Phys. Lett.}  B}
\def\PRL{{\em Phys. Rev. Lett.}}
\def\PRD{{\em Phys. Rev.} D}

\def\SJNP{{\em Sov. J. Nucl. Phys.}}
\def\AnnP{{\em Ann. Phys.}}
\def\JETPL{{\em JETP Lett.}}

\def\CMP{{\em Comm. Math. Phys.}}

\def\beq{\begin{equation}}
\def\eeq#1{\label{#1}\end{equation}}
\def\eeqn{\end{equation}}
\def\beqa{\begin{eqnarray}}
\def\eeqa#1{\label{#1}\end{eqnarray}}
\def\eeqan{\end{eqnarray}}
\def\CR{\nonumber \\ }
\def\leqn#1{(\ref{#1})}

\let\littlebar=\bar
\let\bar=\overbar
\def\Dslash{\not{\hbox{\kern-4pt $D$}}}
\def\half{{1\over 2}}

\def\F{{\cal F}}
\def\L{{\cal L}}

\def\W{{\cal W}}

\def\tr{{\mbox{\rm tr}}}

\def\One{{\bf 1}}
\def\VEV#1{\left\langle{ #1} \right\rangle}
\def\vev#1{\langle #1 \rangle}
\def\eff{{\mbox{\rm eff}}}
\def\ttau{\mbox{\boldmath $\tau$}}
\def\a{\mbox{\bf a}}
\def\wNc{{\widetilde N_c}}

\def\hc{{\rm h.c.}}
\def\lowsim#1{\,\mathrel{\rlap{\lower7pt\hbox{$\sim$}}
    \hskip1pt\hbox{$#1$}}\,}   
\def\longvec#1{\,\mathrel{\rlap{\raise7pt\hbox{$\longrightarrow$}}
    \hbox{$#1$}}\,}   

\def\.{\hskip -4pt plus 10000pt}
\def\vr{\vrule width .6pt height 14pt depth 5pt}
\def\srarrow{\rlap{\raise 8pt \hbox{\vr}}\.\rightarrow}

\newcommand\pubnumber{SLAC-PUB-7393}
\newcommand\pubdate{February, 1997}

\def\Title#1{\begin{center} {\Large #1 } \end{center}}
\def\Author#1{\begin{center}{ \sc #1} \end{center}}
\def\Address#1{\begin{center}{ \it #1} \end{center}}

\def\doeack{\footnote{Work supported by the Department of Energy,
                     contract DE--AC03--76SF00515.}}
\def\SLAC{Stanford Linear Accelerator Center\\
    Stanford University, Stanford, California 94309 USA}
\newcommand\pubblock{\rightline{\begin{tabular}{l} \pubnumber\\
         \pubdate \end{tabular}}}
\newenvironment{Abstract}{\begin{quotation} \begin{center}
                       ABSTRACT
     \end{center}\bigskip  }{\end{quotation}}

\begin{document}
\begin{titlepage}
\pubblock

\vfill
\Title{Duality in Supersymmetric Yang-Mills Theory}
\vfill
\Author{Michael E. Peskin\doeack}
\Address{\SLAC}

\vfill
\begin{Abstract}
These lectures provide an introduction to  the behavior of
 strong\-ly-coupled supersymmetric gauge theories.  After a discussion of 
 the effective Lagrangian in nonsupersymmetric and supersymmetric
field theories, I analyze the qualitative behavior of the simplest 
illustrative models.  These include supersymmetric QCD for $N_f < N_c$,
in which  the  superpotential is generated nonperturbatively, $N=2$
$SU(2)$ Yang-Mills theory (the Seiberg-Witten model), in which the
nonperturbative behavior of the effective coupling is described 
geometrically, and supersymmetric QCD for $N_f$ large, in which the 
theory illustrates a non-Abelian generalization of electric-magnetic 
duality.
\end{Abstract}
\vfill

\vfill

\begin{center}
 to appear in the proceedings of the\\ 
1996 Theoretical Advanced Study Institute\\
{\em Fields, Strings, and Duality}\\
Boulder, Colorado, June 2--28, 1996
\end{center}
\vfill

\end{titlepage}

\vfill
\begin{center} \mbox{ } \end{center}
\newpage

\title{DUALITY IN SUPERSYMMETRIC YANG-MILLS THEORY}

\author{MICHAEL  E. PESKIN}

\address{Stanford Linear Accelerator Center, Stanford University\\
     Stanford, CA 94309, USA}

\maketitle\abstracts{
These lectures provide an introduction to  the behavior of
 strongly-coupled supersymmetric gauge theories.  After a discussion of 
 the effective Lagrangian in nonsupersymmetric and supersymmetric
field theories, I analyze the qualitative behavior of the simplest 
illustrative models.  These include supersymmetric QCD for $N_f < N_c$,
in which  the  superpotential is generated nonperturbatively, $N=2$
$SU(2)$ Yang-Mills theory (the Seiberg-Witten model), in which the
nonperturbative behavior of the effective coupling is described 
geometrically, and supersymmetric QCD for $N_f$ large, in which the 
theory illustrates a non-Abelian generalization of electric-magnetic 
duality.}

\section{Introduction}
\label{sec:aa}

Despite the recent dramatic progress in string theory, our understanding 
of string phenomena is still grounded in our understanding of quantum 
field theory.  Though string theory has magical properties that might make 
ordinary local quantum field theory feel drab and envious, field theory often
allows a tactile understanding of issues that string theory  still leaves
mysterious.  So it is useful to look for field theory realizations of the
phenomena of string theory, in order to find a more complete understanding
of these phenomena.

In fact, much of the impetus for the recent developments in string theory
has come from new discoveries in field theory.  For the past several years,
 Seiberg has led an effort to 
exploit the special simplifications of supersymmetric field theory to 
discover the behavior of these theories in the region of strong coupling.
His investigations led to many wonderful realizations about these theories.
In particular, he discovered that  many cases have  remarkable nontrivial dual
descriptions. 

In addition, 
however one considers the relative role of field theory and string theory,
it is certainly true that physics at distances well below the Planck scale
is described by a local quantum
field theory.  If, as phenomenological studies suggest, this field theory
is approximately supersymmetric, then the basic building blocks for any
theory of elementary particle physics are supersymmetric field theories, 
and, most probably, supersymmetric Yang-Mills  theories.  Any special 
properties of these systems could well be  reflected directly in the physics
of elementary particles.

Thus, we have three reasons to explore the physics of supersymmetric Yang-Mills
theory, for its relevance to the mathematical physics of fields, for its
relevance to the mathematical physics of strings, and for its own direct
application to theories of Nature.  But the best reason to explore this 
subject is that it justifies itself through its beauty and richness.
In these lectures, I will provide an introduction to the physics of 
supersymmetric Yang-Mills theory, and I will try to capture at least a bit
of the underlying beauty.

These lectures will analyze the physics of supersymmetric Yang-Mills
theories through the analysis
of effective Lagrangians constructed to 
describe their low-energy dynamics.
 In Section 2, I will discuss the general 
idea of an effective Lagrangian description of a strongly-coupled 
quantum field theory. In Section 3,
I will discuss the 
special properties of supersymmetric effective Lagrangians and, in the 
process, introduce the most important tools that we will use in our study.
In Section 4, I will give a first illustration of these tools by describing
the Affleck-Dine-Seiberg picture\cite{ADS} of the dynamics in 
the supersymmetric generalization of QCD.

In Section 5, I will present the Seiberg-Witten solution\cite{SW,SWII}
of the $SU(2)$ 
Yang-Mills theory with $N=2$ supersymmetry.  In this solution, magnetic
monopoles which appear as solitons in the weak-coupling analysis of the 
theory play a crucial dynamical role at strong coupling.  The dynamics
of this theory illustrates a role reversal of electrically and magnetically
charged fields which illustrates electric-magnetic duality in a quite 
unusual context.  
 This analysis, which
showed how solitons could take on the dynamical properties of quantum 
particles, has become an important touchstone in many aspects of field theory
and string theory duality, and in mathematical studies which make use of 
concepts of quantum field theory.  In Section 6, I will present some 
generalizations of the Seiberg-Witten theory which illustrate additional
novel effects that may be found  in these models.

In Section 7, I will return to supersymmetric QCD and consider this theory
for the case of many quark and squark flavors.  In this case, 
Seiberg\cite{SnA} has 
given evidence for a new type of dual description, which he calls
`non-Abelian electric-magnetic duality'.  I will explain this duality
and its relation to new nontrivial renormalization group fixed points in 
four dimensions.  Finally, in Section 8,
 I will discuss some generalizations of
non-Abelian duality and the connection of this idea to the Abelian
electric-magnetic duality of the Seiberg-Witten model.

\section{General Principles}
\label{sec:a}

As I have noted in the introduction, Yang-Mills theories are the basic
 building-blocks for models of the
fundamental interactions.  Our current understanding of the strong, weak,
and electromagnetic interactions rests on our knowledge of how the specific
Yang-Mills theories which appear in Nature behave.  In fact, among the 
most difficult
steps in the creation of the present `standard model' of particle physics
was the realization that Yang-Mills theory can reproduce the observed
qualitative
features of the major forces of Nature.

In trying to create theories of Nature at shorter distances, we can try to 
use again the qualitative features that we have already
found in Yang-Mills theory
or we can discover new ways in which  these theories can behave. 
 The most basic
information we can give about the qualitative behavior of a 
quantum field theory
is the manner in which its symmetries are realized in the vacuum state.
So the general question that we will be interested in is the following:
Given a Yang-Mills theory with gauge group $G_c$ and global symmetry $G$, 
how are $G_c$ and $G$ realized in the vacuum state of the theory?

\subsection{A familiar example}

In this section, I will give a specific example of an answer to this question
 and a survey of the possible choices for this 
 qualitative behavior.  The example I 
would like to consider is an $SU(3)$ gauge theory with three flavors of 
massless fermions.  The Lagrangian of the theory is 
\beq
\L = -\frac{1}{4} \left(F^a_{\mu\nu}\right)^2+ \bar q^f_Li\Dslash q^f_L +
\bar q^f_R i\Dslash q^f_R \ ,
\eeq{eq:a}
for $f = 1,2,3$.
The gauge symmetry is $G_c = SU(3)$.  At the classical level, the 
global symmetry is $U(3)\times U(3)$, separate general unitary transformations
on $q^f_L$ and $q^f_R$.  However, the $U(1)$ transformation
\beq
q_L\rightarrow e^{i\alpha}q_L \qquad q_R\rightarrow e^{-i\alpha}q_R
\eeq{eq:b} 
is spoiled by the anomaly, so that the global symmetry of the quantum theory
is $G = SU(3)\times SU(3) \times U(1)$.  

This example is, of course, QCD, the correct theory of the strong interactions.
I have only made the idealization of ignoring the masses of the light quarks
$u$, $d$, and $s$.  For this case, there is an enormous amount of evidence
from experiment, theoretical considerations, and simulations which leads to a
definite picture of the realization of $G_c$ and $G$.  For $G_c$, we observe
experimentally the permanent confinement of quarks in to color-singlet bound
states.  This property is also seen in studies of the strong-coupling limit
of QCD on a lattice\cite{LGT}, and in lattice simulations of QCD in 
the region of 
intermediate coupling.\cite{sims}  The theory is asymptotically free at 
short distances, and the lattice results show that there is no barrier to 
the coupling becoming strong at large distances.

 For $G$, the $SU(3)\times U(1)$ subgroup
is observed as a classification symmetry of hadrons; $SU(3)$ gives  the flavor
quantum numbers and the $U(1)$ charge is baryon number.  The remaining
generators of $G$ must be broken in some way.  In fact, it makes sense 
intuitively that at strong coupling quarks and antiquarks should bind into 
pairs, and that the vacuum would be filled by a condensate of these pairs. 
This intuition can be supported by explicit calculations in various
approximation schemes.\cite{chiralsymm}  To connect this intuition with
experimental observations, we have to take a few further steps.

A quark-antiquark pair condensate is characterized by a vacuum expectation
value of a scalar color singlet quark bilinear $\bar q^f_L q^{f^\prime}_R$.
The simplest form that this expectation value could take is
\beq
\VEV{\bar q^f_L q^{f^\prime}_R} = \Delta \delta^{ff^\prime} \ .
\eeq{eq:d}
Since separate $SU(3)$ rotations of $q_L^f$ and $q^f_R$ do not leave this
form invariant, \leqn{eq:d}
signals the spontaneous breaking of $G$, in the pattern
\beq
SU(3)\times SU(3)\times U(1)\ \to \ 
SU(3)\times U(1) \ .
\eeq{eq:e}
Eight global symmetries are broken, and so eight Goldstone bosons must appear.
These belong to the adjoint representation of the unbroken $SU(3)$.   
Phenomenologically, these bosons can be identified with the eight 
exceptionally light pseudoscalar mesons $\pi$, $K$, $\bar K$, $\eta$.

In fact, we now have enough information to build a quantitative theory
of the couplings of the pseudoscalar mesons.  An $SU(3)$ rotation of the 
$q_L^f$ or $q^f_R$ separately converts the vacuum expectation value
\leqn{eq:d} into 
\beq
\VEV{\bar q^f_L q^{f^\prime}_R} = \Delta U^{ff^\prime} \ ,
\eeq{eq:dplus}
where $U$ is an $SU(3)$ matrix.  Thus, the model has a manifold of vacuum
states which is  isomorphic to the group $SU(3)$. The low energy degrees 
of freedom of the theory should correspond to slow point-to-point changes
in the vacuum orientation.  We can parametrize these by a field $U(x)$ which
gives the local vacuum orientation at each point.
The $U(1)$ symmetry corresponding to baryon number leaves $U(x)$ invariant.
An $SU(3)\times SU(3)$ transformation acts on $U(x)$ by
\beq
U(x) \rightarrow \Lambda^\dagger_L U(x)\Lambda_R \ ,
\eeq{eq:h}
where $\Lambda_L$, $\Lambda_R$ are independent $3\times 3$ unitary matrices.

The dynamics of the model should be described by a Lagrangian written 
in terms of the variables $U(x)$ which is invariant to the full $G$ symmetry.
To construct the possible terms in this Lagrangian, we can consider the 
terms with each possible number of derivatives.  There are no terms without
derivatives, since any $G$-invariant can contain $U(x)$ only in the combination
$U^\dagger U = 1$.  There is a unique term with two derivatives, and additional
possible terms with higher derivatives:
\beq
\L = f^2_\pi\, \tr \left[\partial_\mu U^+\partial^\mu U\right] +
 \kappa \,\tr \left[\partial_\mu U^+\partial^\mu U \partial_\nu U^+
\partial^\nu U\right] + \cdots \ .
\eeq{eq:i}
Since there are no nonderivative terms, the eight degrees of freedom in $U(x)$
are massless, as required by Goldstone's theorem.  At sufficiently low 
energies, the interactions of these eight fields should be well described
by the term with two derivatives.  The corrections due to four- and 
higher-derivative terms are proportional to powers of $k^2/M^2$, where 
$M$ is an intrinsic mass scale of the theory.  Thus, we find definite 
predictions for the low-energy scattering amplitudes of the mesons, in 
terms of a single parameter $f_\pi$.

In principle, the Lagrangian \leqn{eq:i} could be derived starting from the
QCD Lagrangian \leqn{eq:a} by integrating out the high-momentum degrees of 
freedom.  However, this would be a very difficult analysis that would need
essential information about the strong-coupling region of the theory.  On the
other hand, we know in advance that 
 the final answer must have the form \leqn{eq:i}, since this 
is the most general Lagrangian depending on $U(x)$ which has the symmetries
of the original problem.  When combined with terms representing the 
weak $G$ symmetry breaking due to nonzero quark masses, the 
 Lagrangian \leqn{eq:i} in fact does a good job of representing the 
low-energy interactions of the pseudoscalar mesons.\cite{Georgi,GL,DGH}

For QCD, then, all of the pieces of the story fit together neatly.  Basic
theoretical considerations, the results of numerical simulations, and 
experimental observations all reinforce this qualitative picture of the 
physics of the QCD Lagrangian.  But what is the situation for other possible
Yang-Mills theories?  Need there be confinement of the gauge charges?  Could
we find another pattern of global symmetry breaking?  Does the low-energy
spectrum consist only of Goldstone bosons, or can it contain additional
bosons and fermions?

The example of QCD demonstrates that, once one has a definite qualitative
picture of the dynamics in a Yang-Mills theory, much more can be learned by 
writing the {\it effective Lagrangian} which contains the degrees of freedom
relevant at low energies and gives the most general form of  their interaction
consistent with the symmetries of the problem.  But, in nonsupersymmetric
gauge theories, there are very few methods known to constrain the qualitative
pattern of symmetry-breaking.

This is a place that supersymmetry can add powerfully to our technology.
We will see that, in the case of supersymmetric Yang-Mills theory, the 
effective Lagrangian obeys strong constraints which can test the consistency
of different schemes of global symmetry breaking.  In these lectures,
the construction of effective Lagrangians will be one of our major tools in 
working out the qualitative behavior of a variety of supersymmetric 
theories.

\subsection{Phases of gauge theories}

Before going on to supersymmetric theories, I must review one more set
of insights gained from nonsupersymmetric gauge theories, which gives
the possible 
patterns in which the gauge symmetry can be realized.

  The
original gauge symmetry $G_c$ could be completely spontaneously broken.
  Alternatively, the vector bosons could
 mediate long-ranged interactions.  These might give rise either to 
 potentials associated with 
 vector boson exchange or to confinement of the gauge
charge.  It is common to characterize these various types of behavior as
possible {\it phases} in which the gauge symmetry can be realized:
\begin{itemize}
\item Higgs phase: spontaneous breaking of $G_c$, all vector bosons obtain
 mass.
\item Coulomb phase: $G_c$ vector bosons remain massless and mediate $1/r$
 interactions.
\item Wilson phase: $G_c$ color sources are permanently bound into $G_s$
         singlets.
\end{itemize}
It is possible to have intermediate situations, for example, a gauge theory
spontaneously broken from $G_c$ to a subgroup $H_c$ which is then confined.
In such situations, I will describe the phase by the behavior of the subgroup
that survives to the lowest energy.  I should also note that the presence of
a Coulomb phase is not unique to electrodynamics.  A Yang-Mills theory with
sufficiently many fermions that it is no longer asymptotically free gives a
long-ranged potential between color charges of the form $1/r$ times a 
coefficient which decreases slowly as the logarithm of the separation.

The relation of these phases is especially well understood for the 
Abelian case
$G_c=U(1)$. There, the Coulomb phase can 
contain both electric and magnetic charges, with dual coupling strengths.
 A vacuum expectation value for an {\it electrically}
charged field takes us to the Higgs phase.  This phase has solitons
which have the form of  magnetic flux tubes.  Dually, the
appearance of a vacuum expectation value for a {\it magnetically}
charged field gives a phase with electric flux tubes which permanently
confine electric charge.\cite{NO,Tassie} 
  This is a
Wilson  phase.  In Abelian lattice gauge theories, one can 
make this duality manifest.\cite{MeMTD} 
Certain of these theories show all three
phases, with two second-order phase transitions as a function of the
coupling strength.\cite{ALGT,ALGTx}  Such distinct phases can also arise in 
non-Abelian gauge theories.

On the other hand, the relation of the Higgs and confinement phases in
the non-Abelian case is often more subtle.  In many examples, there is
no invariant distinction between the Higgs and confinement phases and
one can, as a matter of principle, move continuously from one to the
other.   Fradkin and Shenker made this possibility concrete by exhibiting
lattice
gauge theory models in which it was possible to prove that these 
phases were continuous connected.\cite{FandS}

Here is an interesting illustrative example:\cite{DRSc} 
 Consider an $SU(2)$ gauge
theory like the standard electroweak theory,
 with a Higgs scalar doublet $\phi$, an $SU(2)$
singlet right-handed fermion $e_R$, and a left-handed fermion doublet
$L = (\nu_L, e_L)$.  In the  realization of the $SU(2)$ gauge symmetry
which is standard in electroweak theory, the electron obtains a mass
through the interaction
\beq
\L_m = \lambda\, \bar L\cdot\phi e_R + \hc 
\eeq{eq:j}
The scalar $\phi$ receives the vacuum expectation value
\beq
       \VEV\phi = \frac{1}{\sqrt 2} {0\choose v} \ , 
\eeq{eq:jplus}
which breaks the $SU(2)$ gauge symmetry completely, giving mass to all
three vector bosons.  Inserting \leqn{eq:jplus} into \leqn{eq:j}, we find
a mass for the electron
\beq
m_e = \frac{\lambda v}{\sqrt 2} \ ,
\eeq{eq:k}
while $\nu_L$ remains massless.

 Now consider what would happen if the theory were realized with $SU(2)$
color confinement.  Again, there are no massless gauge bosons.  The 
fermions and the Higgs bosons would bind into the $SU(2)$ singlet combinations
\beq
E_L = \phi^\dagger\cdot L \qquad N_L=\epsilon_{ab}\phi_aL_b \qquad e_R \ .
\eeq{eq:l}
The coupling \leqn{eq:j} then takes the form of a  mass term
for the color-singlet combinations $E_L$ and $e_R$,
\beq
\L_m= \ m\, \bar E_L e_R + \hc
\eeq{eq:m}
In this way, $E_L$ and $e_R$ pair and become massive, while $N_L$ remains
massless.  

In this example, the qualitative form of the spectrum is the same in the two
cases, and in fact there is no gauge-invariant expectation value that
distinguishes them.  Of course, the two situations are quantitatively
distinguishable.  For example, because the $E_L$ is composite, its
pair-production would be suppressed by a form factor which is not
observed in high-energy experiments.  Thus, we know experimentally that the
electroweak interactions are realized in a Higgs phase and not a Wilson phase.
  However, this example indicates the
possibility that, by adjusting some parameters of a gauge theory, we can 
move continuously from one type of phase to the other.  Such transitions
will occur frequently in the examples that I will discuss later. By trying
to visualize how these transformations occur, one can acquire the 
flexibility of intuition needed to understand the global features of these 
models.

\section{Supersymmetric Effective Lagrangians}
\label{sec:b}

In the previous section, I introduced the general question of the 
realization of symmetry in a Yang-Mills theory.
 I discussed the utility of constructing an 
effective Lagrangian as a way of analyzing the qualitative features of the
model in the regime of strong coupling.  So far, all of my remarks apply
equally well to conventional and supersymmetric quantum field theory.
In this section, I would like to discuss the additional restrictions and
tools for analysis that appear in the supersymmetric case.

\subsection{The general supersymmetric Lagrangian}

In these lectures, I will only discuss models with global supersymmetry.
I will be concerned with models that, at the fundamental level, are 
renormalizable gauge theories.  However, when we describe these models by
writing effective Lagrangians, we will often be interested in models which
are not renormalizable and may contain no gauge interactions.
  For the nonsupersymmetric case, \leqn{eq:i}
provides an example.
 Thus, it is best to begin by writing down the
most general form of a supersymmetric Lagrangian and understanding what
additional restrictions supersymmetry implies.

The basic ingredients for the construction of supersymmetric field theories
are chiral superfields $\Phi^i$, antichiral superfields $\Phi^{\dagger i}$,
and vector superfields $V^a$.  For simplicity, I will represent with vector
fields only the subgroup of the gauge group $H_c$ which is realized manifestly.
Then the $V^a$ belong to the adjoint representation of $H_c$.  The $\Phi^i$ 
belong to some representation $r$ of $H_c$; call the generators of $H_c$ 
in this representation $t^a$.  Methods for the construction of Lagrangians
for these fields are described in the lectures of Lykken.\cite{Lykken}
The most general Lagrangian for the
$\Phi^i$ and $V^a$ with at most two derivatives takes the form
\beqa
\L &=& \int d^4\theta \,K(\Phi^\dagger,e^{V\cdot t}\Phi) +
 \left(\frac{-i}{16\pi}\right)\int
d^2\theta\,\tau (\Phi)\, \W^{\alpha a}\W_{\alpha a}+ \hc \CR
& & \hskip 0.8in + \int
d^2\theta\, W(\Phi) + \hc
\eeqa{eq:n}
The first term of \leqn{eq:n} is a nonlinear sigma model for the fields
$\Phi^i$, that is, a nonlinear model in which the bosonic components of 
$\Phi^i$ may be thought of as coordinates on a manifold.  This is a complex
 manifold with metric derived from the K\"ahler potential 
$K(\Phi^\dagger,\Phi)$.  Thus, the terms involving the bosonic components
of $\Phi$ only, with two derivatives, are
\beq
\L = g_{ij} \partial_\mu \Phi^{\dagger i}\partial^\mu\Phi^j + \cdots
\eeq{eq:o}
where
\beq
g_{ij} = \frac{\partial^2 K}{\partial \Phi^{\dagger i}\partial\Phi^j} \ .
\eeq{eq:p}

The second term in \leqn{eq:n} is the kinetic term for the gauge fields.
The superfield $\W^{\alpha a}$ contains as its components the gaugino 
fields and the gauge field strengths.  The coefficient of this term should
be proportional to $(1/g^2)$.  More generally, $\tau$ is the natural 
combination of the gauge coupling and the $\theta$ parameter,
\beq
\tau = \left(\frac{\theta}{2\pi} + i\, \frac{4\pi}{g^2}\right) \ .
\eeq{eq:q}
In an effective Lagrangian, $\tau$ represents a large-distance coupling,
which differs from the short-distance coupling by some renormalization effects.
Since these renormalizations can depend on the nature of the vacuum state,
$\tau$ can depend on the values of chiral superfields that indicate which
vacuum has been chosen.  If the gauge group $H_c$ is not simple, $\tau$
should  be generalized to a matrix  $\tau^{ab}$.

The last terms in \leqn{eq:n} contain the superpotential $W(\Phi)$.  This term
leads to the nonderivative interactions of the chiral superfields.  An 
important property of the superpotential is its nonrenormalization:  In any
order of perturbation theory, the superpotential can be  modified only by 
field rescalings.  In particular, if the superpotential is zero in the 
underlying theory at short distances, a superpotential can be generated in the
effective Lagrangian only by nonperturbative effects.\cite{nonren}

In the Lagrangian \leqn{eq:n}, the K\"ahler potential $K$ can be a general
real-valued function of $\Phi$ and $\Phi^\dagger$.  However, the coefficient
functions which appear under chiral fermion integrals, $\tau(\Phi)$ and 
$W(\Phi)$, must be {\em holomorphic} functions of the chiral fields. If
these functions carry any dependence on $\Phi^\dagger$, the Lagrangian will
not be supersymmetric.  This restriction is apparently straightforward, but it
will turn out to be a very powerful constraint on the effective Lagrangian.

I should note an important subtlety which is contained in this statement.
The transition from a fundamental Lagrangian to an effective Lagrangian
involves integrating out high-momentum degrees of freedom.  Alternatively, we
might just integrate out all of the degrees of freedom and calculate the 
Green's functions of the original theory.  The generating functional of 
Green's functions is the {\it effective action} $\Gamma$, which is often 
interpreted as a sort of effective Lagrangian.  However, 
$\Gamma$ typically does not have the form of a supersymmetric Lagrangian
with holomorphic coefficients.  

An important example arises in the renormalization of gauge couplings.
Let $g^2$ be the short-distance coupling defined at a large scale $M$.
Integrating out a charged field with vacuum expectation value $\Phi$ will 
produce a renormalized gauge coupling.  If we compute this coupling using
the one-loop $\beta$ function only,
\beq
           \beta(g) = -  \frac{b_0}{(4\pi)^2} g^3 \ ,
\eeq{eq:qplus}
 we find a holomorphic result of the form
\beq
       \tau_\eff = \frac{4\pi i}{g^2}  - \frac{ib_0}{2\pi}\log\frac{M}{\Phi}\ .
\eeq{eq:qpplus}
However, the result of integrating the two-loop renormalization group equation
involves $\log\log(|\Phi|^2)$ and is not properly holomorphic.  Shifman and 
Vainshtein have explained how to reconcile this result with the 
supersymmetry of the effective action.\cite{SV,DineSh} 
 Integrating out only high-momentum degrees
of freedom leads to the result \leqn{eq:qpplus}.  This gives the coefficient
of the gauge kinetic term in the effective Lagrangian.  To emphasize that
only high-momentum degrees of freedom are considered, they call this result
the `Wilsonian effective Lagrangian'.  If one continues to integrate out 
degrees of freedom down to zero momentum, one finds the additional terms
in the effective action which convert the coefficient of $(F_{\mu\nu}^a)^2$ to
the solution of the two-loop 
renormalization group equation. In these lectures, I will typically be 
carrying out manipulations at the level of the Wilsonian effective action, and
so the one-loop $\beta$ functions will be not only sufficient but exact.

\subsection{Conditions for the vacuum state}

Once we have written a supersymmetric effective Lagrangian in the form 
\leqn{eq:n}, we can try to find the vacuum state of the theory.  In 
supersymmetric theories, the  energy of any state
satisfies $\VEV{H} \geq 0$, where 
equality holds if the state is annihilated by the supersymmetry generators.
 Thus, if a supersymmetric state exists, it will be a vacuum state of 
zero energy.

To find the vacuum state of from the effective Lagrangian, we minimize
the potential energy.  The Lagrangian \leqn{eq:n} leads to 
a potential energy of the form
\beq
V =  F^\dagger_i\, g^{ij}F_j + \half\, g^2(D^a)^2
\eeq{eq:r}
where $g$ is the coupling constant defined by \leqn{eq:q}, 
 $g^{ij}$ is the inverse of the metric \leqn{eq:p}, and
the Lagrange multiplier fields $F_j$ and $D^a$ are given by
\beqa
F_j &=& \frac{\partial}{\partial \Phi^j}\, W \nonumber \\[1ex]
D^a &=& \sum_i \Phi^{\dagger i} t^a\Phi^i \ ,
\eeqa{eq:s}
where $t^a$ represents the gauge group generators on $\Phi$.
 $F_j$ and
$D^a$ transform nontrivially under supersymmetry, in such a way that
the conditions
\beq
 \VEV{F_j}\ne 0 \qquad \hbox{\rm or}\qquad \VEV{D^a} \ne 0
\eeq{eq:t}
signal the breaking of supersymmetry.  On the other hand, if
supersymmetry is exact, the formula \leqn{eq:r} gives $V=0$.

The conditions $F_j=0$ and $D^a = 0$ are called, respectively, 
`$F$-flatness' and `$D$-flatness'.   Typically, these conditions can be
satisfied simultaneously, leading to a supersymmetric vacuum state.
For example, if $W$ is a polynomial in unconstrained fields $\Phi^i$, the
conditions
\beq
\frac{\partial W}{\partial \Phi_i} = 0 \qquad i=1, \ldots, n
\eeq{eq:u}
are $n$ polynomial equations in $n$ unknowns, to be solved over the
complex domain.  A solution will exist unless we are in an exceptional
case.  One way to arrange such an exceptional case is to choose $W$ in such
a way that, for some particular value of $i$,
$\Phi^i$ does not appear in \leqn{eq:u}. Then  some remaining
$\Phi^i$ is doubly constrained.  This is how the O'Raifeartaigh model 
of supersymmetry breaking works.\cite{ORaif}

The conditions $F_j=0$ are holomorphic in fields.  The $D$-flatness conditions
are not holomorphic, but the solutions to 
 $D^a=0$ can be parametrized
holomorphically.  The reason for this is that the fundamental gauge
symmetry of a supersymmetric gauge theory is
\beq
\Phi \rightarrow e^{i\alpha\cdot t}\Phi
\eeq{eq:v}
where $\alpha$ is a chiral superfield.  The bosonic part of $\alpha$ is
thus a complex parameter.  The $F$-flatness conditions are invariant
under this complex extension of the gauge group.
The $D$-flatness condition may be thought of as a gauge-fixing term which
breaks this complex
gauge symmetry down to the actual gauge group $G_c$.\cite{LT}
That is, fixing the gauge symmetry
$G_c$ and  imposing of the conditions $D^a=0$ is equivalent to 
fixing the complex extension of $G_c$.
Then the solution of the $D^a=0$ conditions are described by
gauge-invariant combinations of holomorphic
fields.  Luty and Taylor\cite{LT} have shown, further,
that 
it is possible to parametrize the
space of solutions of the $D$-flatness conditions simply 
by gauge-invariant
polynomials.  We will see examples in Section 4 in which both
descriptions of the $D$-flat configurations, that in terms of expection
values of the fundamental fields, and that in terms of the gauge-invariant
polynomials, are useful.

\subsection{Consequences of holomorphicity}

The holomorphic structures involving the coupling constant and the 
superpotential have some additional consequences that I will make use of
in my analysis.  Let me discuss three of these points here.
 
First, the description of supersymmetric Lagrangians in superspace 
naturally suggests that the complex rotation of the fermionic 
coordinate $\theta^\alpha$ should be a symmetry,
\beq
\theta \rightarrow e^{-i\alpha}\theta \ .
\eeq{eq:z}
This transformation, called `$R$ symmetry', is realized on the component fields
as chiral rotations of the  fermionic fields of the model.  If we denote
the fermionic components of chiral superfields by $\psi^i$ and the gaugino
fields by $\lambda^a$, then the transformation \leqn{eq:z} can be written
alternatively as 
\beq
\psi^i \rightarrow e^{-i\alpha}\psi^i \qquad \lambda^a \rightarrow
e^{i\alpha}\lambda^a \ .
\eeq{eq:y}

$R$-symmetry may be broken if the superpotential does not transform
correctly.  Since the term in the Lagrangian following from the
superpotential is
\beq
\L_W = \int d^2\theta\, W = (\mbox{coefficient of $\theta^2$ in $W$)} \ ,
\eeq{eq:a1}
the superpotential should have charge 2 under $R$,
\beq
W\rightarrow e^{+2i\alpha}W \ .
\eeq{eq:a1plus}
If the fundamental theory has $R$ symmetry, the effective Lagrangian should
respect this, and so 
the
superpotential of the effective Lagrangian should have $R$-charge equal
to 2.

It often happens that the `canonical'
$R$ symmetry just described is anomalous.  In that case, it is often 
possible to form a non-anomalous $U(1)$ symmetry by combining the  canonical
$R$ symmetry with some global $U(1)$ transformation that acts on chiral 
multiplets.  If the anomaly-free $R$ transformation is to be a symmetry,
the superpotential must have charge 2 under this modified transformation.
In the following sections, when I apply $R$ symmetry, I will state
explicitly whether I am discussing
the canonical or the anomaly-free $R$ transformation.

The second of these consequences concerns the symmetry-breaking
dynamics of gauginos.
Because the gauginos of 
supersymmmetric Yang-Mills theories are massless, strongly interacting
fermions, it will be interesting to ask whether these particles undergo pair
condensation like the quarks in QCD.  Holomorphicity gives us a useful
tool to examine this question. 

 The gaugino condensate analogous to 
\leqn{eq:d} is
\beq
\VEV{\lambda^{\alpha a}\lambda^a_\alpha} \ .
\eeq{eq:b1}
The fermion bilinear in \leqn{eq:b1} is also the scalar component of the 
superfield $\W^{\alpha a} \W_\alpha^a$.  This means that we can extract
the expectation value of this operator by differentiating the Lagrangian
\leqn{eq:n} with respect to $F_\tau$, the 
 $F$ component of $\tau$.  The fundamental definition of the operator is
given by differentiating with respect the $F$ term of the
short-distance coupling 
constant $\tau_0 = 4\pi i/g^2$.  However, according to \leqn{eq:qpplus},
 the effective  
gauge coupling $\tau_\eff$ is related to the short-distance coupling
$\tau$ by an additive term, so we could equally well simply
differentiate with respect to the $F$ terms of 
$\tau_\eff$.  In any event, we have
\beq
\VEV{\lambda^{\alpha a}\lambda^a_\alpha} = 16\pi\,
\frac{\partial}{\partial F_\tau} \log Z, \qquad \mbox{\rm where} \quad
Z = \int e^{i\int\L} \ .
\eeq{eq:c1}
If we integrate out the gauge fields and describe the theory using an 
effective Lagrangian with chiral fields only, we can still recover the 
value of the gaugino condensate through the dependence of the effective 
superpotential $W_\eff$ on $\tau$,
\beq
\VEV{\lambda\lambda} = 16\pi i\, \frac{\partial}{\partial F_\tau} \int
d^2\theta\, W_\eff(\tau,\phi) = 16\pi i\,
\frac{\partial}{\partial\tau}\, W_\eff(\tau,\phi) \ .
\eeq{eq:d1}

Finally, it is interesting to think about the 
relation between the 
effective Lagrangians of related supersymmetric models.  An example we
will often encounter is the relation between a Yang-Mills theory with
with $(n+1)$ matter flavors
to that with $n$ flavors.  We can obtain the second of these theories
from the first by adding a mass term for the $(n+1)$st flavor
and then taking this  mass to be large.

The result of this  procedure on the effective superpotential is very 
simple to analyze.  Typically, if the chiral
field $\Phi_{n+1}$ is not yet integrated out, the mass operator for this 
field in the effective Lagrangian will be simply the
original mass term for this field.  Then, if
the theory with $(n+1)$ massless flavors has superpotential $W_\eff$,
the  superpotential with the mass perturbation will be 
\beq
W_\eff(\Phi) + m \Phi^2_{n+1} \ .
\eeq{eq:e1}
We can then solve the $F$-flatness conditions which involve $m$ and use
these to eliminate the field $\Phi_{n+1}$ from the effective Lagrangian.
This procedure generates a new holomorphic effective superpotential from the 
original one. 

I will refer to the relation of these two superpotentials as `holomorphic 
decoupling'.  If we have the exact form of the 
effective superpotential for some number of 
flavors $n$, decoupling
allows us to compute the effective superpotentials explicitly for any smaller
number of flavors.
Even more remarkably, holomorphic decoupling also turns out to be a powerful
tool for determining the effective superpotentials in model with a larger
number of flavors, since it provides a stringent
consistency condition on any proposed superpotential for these models.

\section{Supersymmetric QCD}
\label{sec:bb}

As a first example for the application of these methods, I would like
to consider the supersymmetric generalization of QCD, $SU(N_c)$ gauge
theory with $N_f$ flavors of quark superfields 
in the fundamental representation 
of the gauge group.  At least for the case of a
small number of flavors, this theory was
analyzed many years ago by Veneziano, Taylor, and Yankielowicz\cite{VTY}
and by Affleck, Dine, and Seiberg.\cite{ADS}  Naively, one might expect
the same behavior found in ordinary QCD---chiral symmetry breaking caused
by pair condensation of the quarks.  Instead, we will find
many surprises.

\subsection{Lagrangian and symmetries}

Let me first set up some basic notation for this theory.  The $N_f$ flavors
of quarks can be described as $N_f$ left-handed fermions in the 
  $(N_c+\bar N_c)$
representation of the gauge group.
These belong to chiral supermultiplets that I will call $Q_i$ and $\bar Q_i$,
$i = 1, \ldots, N_f$.  Note that the bar refers to a chiral superfield in 
the $\bar N_c$ representation, while an antichiral superfield will be 
denoted by a dagger.  I will use the symbol $Q_i$ to denote both the superfield
and its scalar component (with the precise meaning hopefully evident from 
context) and denote the fermionic components of $Q_i$, $\bar Q_i$ by 
$\psi_{Qi}$, ${\psi}_{\littlebar{Q}i}$.   

The Lagrangian of supersymmetric QCD is
\beq
\L = \int d^4\theta\left(Q^\dagger_i e^VQ_i+\bar Q_ie^V\bar Q^\dagger_i\right)
 -
\frac{i}{16\pi}\int d^2\theta\, \tau\, W^{\alpha a} W^a_\alpha + \hc  \ , 
\eeq{eq:f1}
$i = 1, \ldots, N_f$, with no superpotential.  At the classical level, this 
theory has the  $R$ symmetry \leqn{eq:z}.  In the   quantum theory this
symmetry is anomalous, though it will still be useful to us, as we will see
in a moment. On the other hand, the $R$ symmetry can be combined with the 
anomalous $U(1)$ flavor symmetry to form an anomaly free $R$ symmetry.  The 
full global symmetry of the model is then
\beq
    G = SU(N_f) \times SU(N_f) \times U_B(1) \times U_R(1) \ ,
\eeq{eq:f1plus}
where the first $U(1)$ factor is proportional to 
baryon number and the second is the 
anomaly-free $R$ symmetry.  I will define the chiral multiplets $Q_i$ 
to have $U_B(1)$ charge $B = +1$; the chiral multiplets $\bar Q_i$, whose
fermionic components are left-handed antiquarks, will have $B = -1$.

If we wish to work with the anomalous $R$ symmetry, we must take into account
the effect of the anomaly.  To do this, note that the chiral rotation of a
left-handed fermion field
\beq
 \psi \to e^{i\alpha}\psi
\eeq{eq:f1pplus}
changes the measure of integration over $\psi$ in such a way as to shift the 
$\theta$ parameter of the Yang-Mills theory by 
\beq
     \theta \to \theta - n \alpha
\eeq{eq:f1xplus}
where $n$ is the coefficient of the anomaly term in the conservation law
for the corresponding chiral current.  (Equivalently, $n$ is the number of 
zero modes of $\psi$ in a one-instanton solution of the Yang-Mills equations.)
Thus, an anomalous chiral symmetry can be combined with a transformation of 
$\theta$ or $\tau$ to give a symmetry of the theory.

A  supersymmetric Yang-Mills theory with gauge group $G_c$ and chiral
 superfields in the representations $r_i$ has a one-loop $\beta$ function of 
the form \leqn{eq:qplus}, with 
\beq
b_0 = 3C_2(G_c) - \sum_i C(r_i)\ , 
\eeq{eq:h1}
where $C_2(r) \One = (t^at^a)_r$ is the quadratic Casimir operator and 
$C(r)\delta^{ab}=\tr_r[t^at^b]$, and $G_c$ denotes the adjoint representation.
In the same notation, the anomaly coefficient $n$ for fermions in the 
representation $r$ is given by
\beq
   n = 2C(r) = \cases{1 & $r = N_c$ or  $\bar N_c$ of $SU(N_c)$ \cr
                      2N_c & $r =$ adjoint of $SU(N_c)$\cr}
\eeq{eq:h1plus}

 In the case of supersymmetric QCD, the
formula for the $\beta$ function becomes
\beq
b_0 = 3N_c-N_f \ .
\eeq{eq:g1}
If the fundamental coupling constant $g^2$ is defined at the large mass
scale $M$, the 
effective running coupling constant of the theory is given by 
\beq
  \frac{4\pi}{g^2}(Q)
 = \frac{4\pi i}{g^2}  -  \frac{3N_c - N_f}{2\pi}\log\frac{M}{Q}\ .
\eeq{eq:g1plus}
It is convenient to define $\Lambda$ to be the scale at which this expression
formally diverges,
\beq
  \Lambda^{b_0} =  M^{b_0} e^{-8\pi^2/g^2} = M^{b_0} e^{2\pi i \tau} \ .
\eeq{eq:g1pplus}

Note that, in any particular perturbative
scheme for defining $g^2$,  such as $\bar{MS}$
or $\bar{DR}$, there may be a scheme-dependent constant added to the right-hand
side of \leqn{eq:g1plus}, which generates an overall constant rescaling of
$\Lambda$.   I will ignore these constants, since they can be 
absorbed by a redefinition of $M$.  However, to 
 compare  exact results for 
supersymmetric Yang-Mills theory to explicit perturbative or instanton
calculations, it is necessary to keep track of these terms.  A careful 
treatment is given in \cite{FP}.

\subsection{$N_f = 0$}

Let us begin by considering the pure supersymmetric Yang-Mills theory, 
the case $N_f = 0$.  The Lagrangian of this theory is written in 
terms of component fields as
\beq
\L = - \frac{1}{4g^2}\, (F^a_{\mu\nu})^2 + \frac{1}{g^2}\, \bar
\lambda^a i \Dslash \lambda^a + \frac{i\theta}{32\pi^2}
 F^a_{\mu\nu}{\tilde F}^{a\mu\nu} \ .
\eeq{eq:i1}
This Lagrangian looks just like that of ordinary QCD, but with the massless
quarks replaced by one flavor in the adjoint representation of the gauge
group.  In \leqn{eq:i1}, the gaugino $\lambda^a$ is a left-handed field, but
this is not an essential difference because $\lambda^a$ belongs to a real
representation of $G_c$ and thus can have a gauge-invariant mass term.
It is very tempting to conjecture that this theory behaves exactly like 
QCD: The gauge coupling becomes strong and confines color, the gauginos
condense into the vacuum in pairs and break the chiral symmetry.\cite{VY}

To analyze this theory, we should first understand its global symmetries.
Because there are no quark flavors, it is not possible to build an 
anomaly-free $R$ symmetry in this case.  However, a discrete subgroup of the
canonical $R$ symmetry is left unbroken.  One way to see this is to note 
that, according to \leqn{eq:f1xplus},
 a chiral rotation of the gaugino field becomes a symmetry if we 
combine it with a shift of the $\theta$ parameter
\beq
\theta \rightarrow \theta + 2N_c\alpha\ ,\quad \mbox{\rm or} \quad
\tau \rightarrow \tau + \frac{2N_c}{2\pi}\, \alpha \ .
\eeq{eq:m1}
Since the physics of Yang-Mills theory is periodic in $\theta$ with period
$2\pi$, no compensation is necessary if $\alpha$ is a multiple of $2\pi/2N_c$.
Thus, a $Z_{2N_c}$ subgroup of the original $R$ symmetry survives as a 
symmetry of the quantum theory. 

 On the other hand, it is often appropriate
to think of $\tau$ as an adjustable background superfield.  In string theory,
$\tau$ is proportional to the dilaton superfield $S$.  If the pure Yang-Mills
theory is derived by integrating out fields which are massive due to the 
vacuum expectation value of some chiral superfield $\Phi$, the effective
coupling $\tau$ will be a function of $\Phi$.   If we
take this point of view that $\tau$ may be 
treated as  a background superfield, then the supersymmetric Yang-Mills
theory should be 
invariant under the full continuous $R$-symmetry combined with the 
 shift of this 
superfield given in \leqn{eq:m1}.

This statement has an interesting consequence.  Under our hypothesis, the 
pure supersymmetric Yang-Mills theory has no massless particles.  The gluons
and gluinos combine into massive color-singlet bound states $gg$, 
$\lambda\lambda$, and $g \lambda$.  Thus, the low-energy 
effective Lagrangian of the theory contains  only the background 
superfield $\tau$.  In principle, this Lagrangian should have a superpotential
which is a function of $\tau$.  The requirement that the superpotential
should have $R$ charge 2 specifies its form uniquely:
\beq
W_\eff = c\, M^3\, e^{2\pi i\tau/N_c} \ ,
\eeq{eq:n1}
where $c$ is a constant and I have supplied factors of the large scale $M$
to give $W_\eff$ the correct mass dimension.  Given \leqn{eq:n1}, we can 
use \leqn{eq:d1} to compute the gaugino condensate, 
\beq
\VEV{\lambda\lambda} = 16\pi i\, \frac{\partial}{\partial\tau}\, W_\eff
=   - \frac{32\pi^2}{N_c} \cdot c\, M^3\, e^{2\pi i\tau/N_c} \ , 
\eeq{eq:o1}
or, at $\theta=0$
\beq
\VEV{\lambda\lambda} = - \frac{32\pi^2}{N_c}\cdot c\, M^3\, 
e^{-8\pi^2/N_cg^2} \ .
\eeq{eq:p1} 

 This formula accords with our physical intuition in two ways.  First, 
if a gaugino condensate is generated nonperturbatively, the renormalization
group requires that the size of this condensate should be set by the 
 nonperturbative QCD scale
$\Lambda(g^2,M)$ given in \leqn{eq:g1pplus}.  Specifically,
we must have
\beq
\VEV{\lambda\lambda} =  A \Lambda^3 \ ,
\eeq{eq:p1plus} 
where $A$ is a pure number.  Evaluating \leqn{eq:g1pplus}
with $b_0 = 3N_c$, we can see that \leqn{eq:p1} has precisely this form.

Second, as I explained below \leqn{eq:m1}, the pure supersymmetric
 Yang-Mills theory
should have a $Z_{2N_c}$ global symmetry, $\lambda \to e^{i\alpha} \lambda$
with $\alpha = 2\pi m/2N_c$.   Under this symmetry, the gaugino bilinear
is invariant to this transformation for $\alpha = \pi$ or $m = N_c$; thus, 
an expectation value of this bilinear should break the $Z_{2N_c}$ symmetry
spontaneously to $Z_2$.  This symmetry-breaking would result in $N_c$ 
inequivalent vacuum states. These states appear explicitly 
in the formula \leqn{eq:o1}, since
the transformations $\theta \to \theta + 2\pi$ or $\tau\to \tau+1$ 
which are invariances of the Yang-Mills theory sweep out $N_c$ 
distinct values of the gaugino condensate.  

Thus, it is reasonable that supersymmetric Yang-Mills theory should acquire
a superpotential of the form \leqn{eq:n1}.  Still, this line of reasoning
is not quite satisfactory.
Though we have shown that the appearance of the superpotential \leqn{eq:n1}
is consistent, we still had to assume that this nonperturbative effect was
nonvanishing.
To justify this assumption, we must examine some further 
examples of supersymmetric gauge theories. 

\subsection{The Affleck-Dine-Seiberg superpotential}

Consider next supersymmetric QCD with $N_f$ flavors, for $N_f < N_c$.
I have presented the non-anomalous global symmetry of this theory in 
\leqn{eq:f1plus}. 

We have seen in the previous section that we
can also consider the transformation of the theory under anomalous global
symmetries as long as we compensate the anomalous transformation laws
by an appropriate shift of $\theta$ or $\tau$.  Thus, I will
analyze this theory by making use of the larger symmetry group
\beq
SU(N_f)\times SU(N_f)\times U_B(1)\times U_A(1)\times U_R(1)
\eeq{eq:u1a}
which includes the following two anomalous transformations:
\beqa
A&:& \psi_Q \rightarrow e^{i\alpha}\psi_Q\ ,  \quad \psi_{\bar Q}
\rightarrow e^{i\alpha}\psi_{\littlebar Q}\ , \quad  \mbox{and}\quad 
 \theta \rightarrow \theta + 2N_f\alpha\nonumber\CR
R&:& \psi_Q \rightarrow e^{-i\alpha}\psi_R\ , \quad \psi_{\bar
Q}\rightarrow e^{-i \alpha}\psi_{\littlebar Q}\ , \quad \lambda \rightarrow
e^{-i\alpha}\lambda\ ,\CR
& & \hskip 0.9in 
\mbox{and}\quad \theta \rightarrow \theta +
(2N_c-2N_f)\alpha \ .
\eeqa{eq:v1}

The anomaly-free $R$ symmetry is the combination of these two operations
which does not require a transformation of $\theta$.  Its $U(1)$ charge is
\beq
R_{AF} = R + \frac{N_f-N_c}{N_f} A \ .
\eeq{eq:w1}
This symmetry will be important to us at a later stage of our analysis.

It is useful to tabulate the transformation properties of the various
fields under the four $U(1)$ symmetries that we have defined:
\beq
\begin{array}{c@{\hspace{.5in}\extracolsep{.2in}}ccccl}
               &  B  & A & R    & R_{AF} & \\[2ex]
Q_i        &  +1     & +1  & 0           & (N_f-N_c)/N_f & \\[1ex]
\psi_{Qi}  &  +1     & +1  & 0           &  -N_c/N_f & \\[1ex]
\bar Q_i        &  -1     & +1  & -1           & (N_f-N_c)/N_f & \\[1ex]
\psi_{\littlebar Q i}  &  -1     & +1  & -1           &  -N_c/N_f & \\[1ex]
\lambda  &  0     & 0  &  +1           &  +1 & \\
\end{array}
\eeq{eq:uaa}

There are two additional quantities whose quantum numbers will also be
 important to us. The first is the unique gauge-invariant chiral superfield 
that we can build from $Q_i$ and $\bar Q_i$, 
\beq
T_{ij} = Q_i\cdot\bar Q_j  \ .
\eeq{eq:x1} 
We might think of $T_{ij}$ as a meson superfield; its scalar component is a 
color-singlet combination of scalar quarks.  Since $T_{ij}$ transforms as
a $(N_f, \bar N_f)$ under the $SU(N_f)\times SU(N_f)$ global symmetries, it
is especially useful to consider the determinant of this $N_f\times N_f$ 
matrix, which is invariant under the non-Abelian part of $G$.
 The second important quantity is the nonperturbative
scale $\Lambda$, which transforms under the anomalous $U(1)$ symmetries by
virtue of the transformation of $\theta$. The quantum numbers of these 
objects are:
\beq
\begin{array}{c@{\hspace{.5in}\extracolsep{.2in}}ccccl}
               & B & A & R     & R_{AF} & \\[2ex]
\det T         & 0      & 2N_f      & 0          & 2(N_f-N_c) & \\[1ex]
\Lambda^{b_0}  & 0      & 2N_f   & 2(N_c-N_f) & 0          & \\
\end{array}
\eeq{eq:ubb}

It is natural to represent the low-energy dynamics of  supersymmetric
QCD by an effective Lagrangian which is built out of 
gauge-invariant chiral superfields.  This Lagrangian would generalize 
the structure \leqn{eq:i} that we wrote for non-supersymmetric QCD.
If we build this Lagrangian out of gauge-invariant combinations of 
$Q_i$ and $\bar Q_i$, it must be a function of components of $T_{ij}$.

The superpotential of this effective Lagrangian must be a 
holomorphic function of $T_{ij}$ and $\tau$ which is invariant to the
global symmetry group except that it transforms with charge 2 under the 
$R$ symmetry. There is only one possible function that satisfies these
requirements,  
\beq
W_\eff = c \cdot \left(\frac{\Lambda^{b_0}}{\det T}\right)
^{1/(N_c-N_f)} \ .
\eeq{eq:y1}
This is the Affleck-Dine-Seiberg superpotential.\cite{ADS,VTY}  
An alternative method for constructing possible superpotentials would be
to construct a function of $T$ that is invariant to the non-anomalous
global symmetry group  except that it transforms with charge 2 under 
$R_{AF}$.  Factors of $\Lambda$ can then be supplied to give the 
effective superpotential the correct mass dimension 3.  This argument
also gives \leqn{eq:y1} as the unique superpotential for this theory.

\begin{figure}
\begin{center}
\leavevmode
\epsfbox{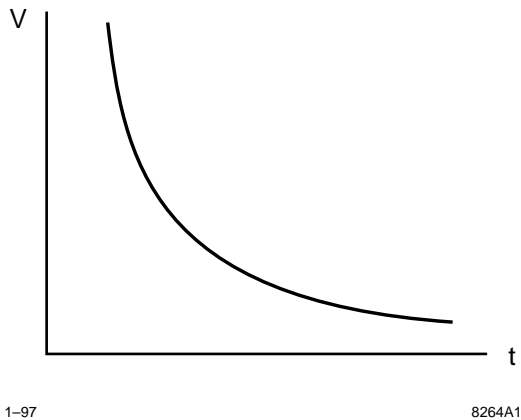}
\end{center}
 \caption{Form of the potential for supersymmetric Yang-Mills theory
 with $N_f< N_c$.}
\label{fig:one}
\end{figure}

At first sight, the result \leqn{eq:y1} 
looks very bizarre.  Differentiating the formula
to construct the $F$ term of $Q_i$, we find
\beq
 F_i = \frac{\partial W_\eff}{\partial Q_i} \sim  \frac{1}{Q} \cdot
         \left(\frac{1}{\det Q}\right)^{1/(N_c-N_f)} \ .
\eeq{eq:z1}
This expression decreases as the expectation value of $Q_i$ or $T_{ij}$ becomes
large.  But  this is the weak-coupling region where the K\"ahler 
potential for $Q$ should take the simple canonical form.  Thus, the 
potential \leqn{eq:r} derived from the effective Lagrangian tends to zero
as $\VEV{T}$ tends to infinity as shown in Figure \ref{fig:one}.  In fact,
this potential  pushes the theory to a 
vacuum state at infinity.

However, some careful thinking shows that this is in fact the correct behavior
of the model.  Consider the alternative possibility that supersymmetric
QCD leads to confinement and chiral symmetry breaking, just as in ordinary
QCD.  In that case, we would expect the quarks to condense in pairs 
as in \leqn{eq:d} or, in our new notation, 
\beq
\VEV{\psi_{Qi}^\alpha \cdot \psi_{\littlebar Q j\alpha}} \ne 0 \ .
\eeq{eq:a2}
The gaugino bilinear could acquire an expectation value consistently
with supersymmetry.  But for the quark bilinear, this is not true; 
$\psi_{Qi} \cdot \psi_{\bar Q j}$ is the $F$ term of $T_{ij}$, and so an 
expectation value for this expression signals supersymmetry breaking.
Since the vacuum energy of a supersymmetric theory is zero only when 
supersymmetry is unbroken, a QCD-like vacuum with a quark pair condensate
is thus unstable with respect to any field configuration---no matter
how bizarre---which can allow supersymmetry to remain manifest.
   
Here is a way to find such a configuration:  The $D$-flatness condition 
of supersymmetric QCD is
\beq
D^a = Q^\dagger t^aQ - \bar Q t^a\bar Q^\dagger = 0 \ .
\eeq{eq:d2}
This condition is satisfied by any set of expectation values with 
$\VEV{Q_i} = \langle\littlebar Q_i^\dagger\rangle$. 
 By choosing gauge and flavor 
rotations to diagonalize the $\VEV{Q_{ik}}$, where $k$ is the gauge group
index, we can write this expectation value in the form
\beq
\VEV Q_{ik} = \left(
\begin{array}{lcccccr}
a_1 &   &        & 0 &  & 0 & \ddots \\
  & a_2 &        &   &  & 0 &   \\
  &   & \ddots &   & \\
  &   &        & a_{N_f-1} &  & 0 & \\
  & 0 &        &   & a_{N_f}  & 0 & \ddots
  \end{array}\right) \qquad \VEV{\bar Q^\dagger_{ik}} = 0 \ .
\eeq{eq:d2plus} 
Note that this solution has the form of a diagonal matrix acted on by 
two $U(N_f)$ flavor matrices, just the amount of information encoded in 
$\VEV{T_{ij}}$.  For $a_1, \ldots, a_{N_f}$ large, the gauge symmetry is 
spontaneously broken
\beq
SU(N_c) \rightarrow SU(N_c-N_f) \ .
\eeq{eq:f2}
In the process, all fermions and bosons which transform under
the residual gauge group
$SU(N_c-N_f)$ obtain mass.  If we send the parameters $a_i$ to infinity,
the situation reverts to that of the pure gauge theory, for which we have
argued that there is a supersymmetric vacuum with gaugino condensation.
This is a vacuum state with zero energy and is therefore the preferred
configuration for this theory.

It is interesting to work out some further details of this picture.  For
simplicity, I will consider the symmetrical configuration $a_1= \ldots = 
a_{N_f} = v$.  Then, for momenta scales $Q > v$, the theory looks like
supersymmetric QCD with $N_f$ flavors, while for $Q< v$ it look like a 
pure supersymmetry gauge theory with gauge group $SU(N_c-N_f)$.

  We can compute
the effective coupling constant in the pure gauge theory at low energies by
matching to the high-energy coupling constant at the scale $v$. The high-energy
behavior, valid for $Q> v$, is
\beq
\frac{4\pi}{g^2}\, (Q) = \frac{3N_c-N_f}{2\pi}\, \log \frac{Q}{\Lambda}
\ .
\eeq{eq:g2}
For $Q<v$, the $\beta$ function of the theory is given by $b_0 = 3(N_c-N_f)$.
We parametrize the value of the running coupling constant by a scale 
$\Lambda_\eff$,
\beq
\frac{4\pi}{g^2}\, (Q) = \frac{3(N_c-N_f)}{2\pi}\, 
\log\frac{Q}{\Lambda_\eff} \ .
\eeq{eq:h2}
We may  obtain an expression for $\Lambda_\eff$ by insisting that the 
running coupling constant should change continuously.  Thus, we should set
the expressions \leqn{eq:g2} and \leqn{eq:h2} equal at the scale $v$.
This gives
\beq
\left(\frac{\Lambda_\eff}{v}\right)^{3(N_c-N_f)} =
\left(\frac{\Lambda}{v}\right)^{3N_c-N_f}
\eeq{eq:i2}
or
\beq
\Lambda^3_\eff =
\left(\frac{\Lambda^{3N_c-N_f)}}{v^{2N_f}}\right)
^{\frac{1}{N_c-N_f}} \ .
\eeq{eq:j2}
Finally, we can make use of the result of the previous section that the
pure supersymmetric Yang-Mills theory has gaugino pair condensation 
 $\VEV{\lambda\lambda} \sim
(\Lambda_\eff)^3$, which is the consequence of an effective superpotential
\beq
W_\eff = c \cdot \Lambda^3_\eff \ .
\eeq{eq:k2}
Substituting \leqn{eq:j2} into \leqn{eq:k2}, we find precisely the 
Affleck-Dine-Seiberg effective superpotential \leqn{eq:y1}.

There is a second way to check the validity of the superpotential \leqn{eq:y1},
by holomorphic decoupling.  Start with the effective superpotential for 
$N_f$ flavors, and add a mass term for the $N_f$th flavor.  The supersymmetric
mass term is 
\beq
\Delta W = m\, Q_{N_f}\cdot\bar Q_{N_f} = m\, T_{N_fN_f} \ .
\eeq{eq:l2}
Then the superpotential of the massive theory is
\beq
W = c\left(\frac{\Lambda^{b_0}}{\det T}\right)^{\frac{1}{N_c-N_f}}
+ m\, T_{N_fN_f} \ .
\eeq{eq:m2}

Now work out the $F$-flatness conditions for this superpotential.  The 
vanishing of the $F$-flatness conditions for $T_{N_f i}$ (or for 
$Q_i$) imply that $T_{N_f i} = 0$ for $i \neq N_f$.  Similarly,
 $T_{i N_f} = 0$ for $i \neq N_f$.  Then $T$ takes the block form
\beq
T =  
\left( \begin{array}{ccc}\widetilde T & 0 \\ 
 0 & t
\end{array}\right) \ .
\eeq{eq:n2}
The $F$-flatness condition for $T_{N_f N_f} = t$ is 
\beq
    - \frac{c}{N_c - N_f} 
\left(\frac{\Lambda^{b_0}}{\det\widetilde T}\right)^{1/(N_c -N_f)}
\left(\frac{1}{t}\right)^{1 + 1/(N_c-N_f)}  + m = 0 \ ,
\eeq{eq:o2}
which implies
\beq
t = 
\left(\frac{N_c-N_f}{c}\,m\,
\left(\frac{\Lambda^{b_0}}{\det\widetilde T}\right)^{1/(N_c-N_f)}\right)
^{(N_c-N_f)/(N_c-N_f+1)} \ .
\eeq{eq:p2}
Putting this back into the superpotential, we obtain
\beq
W = c^\prime \left(\frac{(m\Lambda^{b_0})}
{\det{\widetilde T}}\right)^{1/(N_c-N_f+1)}\ ;
\eeq{eq:q2}
if $c = (N_c-N_f)$, $c^\prime = (N_c-N_f+1)$. 
This is precisely the form of the Affleck-Dine-Seiberg superpotential for
$(N_f-1)$ flavors.  Thus, the various effective superpotentials of the 
family \leqn{eq:y1} are consistent with one another by decoupling.
In these decoupling relations, the various nonperturbative scales $\Lambda$
are related by the formula
\beq
\left(\Lambda^{b_0}\right)_{\eff,N_f-1} = m\,
\left(\Lambda^{b_0}\right)_{N_f}
\eeq{eq:r2}
It is not difficult to check that this is precisely the relation that 
is required by a renormalization group analysis similar to the derivation
of \leqn{eq:j2}, in which we match  running coupling constants
above and below the scale $Q=m$.

\subsection{$N_f = N_c -1$}

We have now shown that the Affleck-Dine-Seiberg superpotentials are linked
to one another by holomorphic decoupling.  Then, if this superpotential is
known explicitly for any particular value of $N_f$, we can compute its 
coefficient for all values of $N_f < N_c$.  Thus, to complete the derivation
of these superpotentials, we need only find one value of $N_f$ at which we 
can derive them directly.

Affleck, Dine, and Seiberg showed that there is a direct derivation of the 
superpotential for the case $N_f = N_c -1$.  In this case, the expectation
values for $Q_i$ and $\bar Q_i$ given in \leqn{eq:d2plus} break the 
$SU(N_c)$ gauge symmetry completely. For large values of the $a_i$, the
gauge theory never reaches strong coupling and so any terms that appear in the
effective Lagrangian must be visible in a weak-coupling analysis.  On the other
hand, because of the nonrenormalization theorem, a superpotential cannot be
generated in any order of perturbation theory.  The only 
gap between these two
requirements is the possibility
 that a superpotential may be generated through a systematic
instanton calculation.

The instanton is the leading  nonperturbative contribution to gauge theory
amplitudes which appears in a weak-coupling expansion.
Methods for performing instanton calculations are reviewed in \cite{inst}.
In this article, I will not attempt to obtain the correct coefficient of 
the instanton amplitude but only to show that it is nonzero.  For this purpose,
one can view an instanton as a source of chiral fermions. More precisely,
if $\psi$ is a fermion matter field in the representation $r$, the instanton
creates $n = 2C(r)$ units of $\psi$ charge.  In the model at hand, an 
instanton creates one each of the $\psi_{Qi}$ and $\psi_{\littlebar Qi}$ and 
$2N_c$ of the $\lambda$.  On the other hand, the supersymmetric gauge
theory contains a vertex proportional to 
$Q^\dagger \lambda^\alpha \psi_{Q\alpha}$, which can annihilate a $\lambda$
and a $\psi_Q$ (or $\psi_{\bar Q}$) in the presence of a vacuum expectation
value of $Q$ (or $\bar Q$).  Annihilating all of the $\lambda$'s, we are
left with an operator of the form
\beq 
   \Delta \L =   F(Q^\dagger,\bar Q^\dagger)
 \psi^{\dagger\dot\alpha}\psi^\dagger_{\dot\alpha}
\eeq{eq:r2plus}
which can be rewritten as the Hermitian conjugate of the  superpotential term
\beq
    \Delta \L = \int d^2\theta  W(Q,\bar Q)
\eeq{eq:r2pplus}
The amplitude is proportional to one power of 
\beq
      M^{b_0} e^{-8\pi^2/g^2 + i\theta} =  \Lambda^{b_0} \ ,
\eeq{eq:r2xplus}
where the dependence on $g^2$ follows from the instanton action and the $M$
dependence appears because this factor must be a renormalization group 
invariant.  The dependence of $W$ on $Q$ and $\bar Q$ then follows from 
the fact that $W$ must be an $SU(N_f)\times SU(N_f)$ invariant of mass
dimension 3.  From these considerations, we obtain
\beq
W = c\ \frac{\Lambda^{b_0}}{\det T} \ .
\eeq{eq:s2}
with a nonzero value of $c$.  The exact value of $c$ has been 
obtained by carrying out the instanton calculation explicitly, which has been
done in a series of papers by  
Cordes,\cite{Cordes}  Shifman and Vainshtein,\cite{SVpot} 
and Finnell and Pouliot.\cite{FP}  Thus, the Affleck-Dine-Seiberg 
superpotential  can be carefully justified for this case and, by extension,
for all cases $N_f < N_c$.  

As a final comment on these models, I would like to note that the 
potential we have found, which pushes the vacuum state to infinity, is 
actually not so inconsistent with the familiar symmetry-breaking pattern 
of nonsupersymmetric QCD.  Given the potential in the supersymmetric case,
we can break supersymmetry explicitly by adding a positive mass term for the 
scalar quarks only,
\beq
\Delta\L = - m^2\left( |Q|^2+|\bar Q|^2\right) \ .
\eeq{eq:t2}
This term pulls the minimum of the potential back from infinity to some large
but finite value of $\VEV{T_{ij}}$.  The minimum occurs for an expectation
value
\beq
            \VEV{T_{ij}} =  A \delta_{ij} \ ,
\eeq{eq:t2plus}
and so the vacuum of the modified theory spontaneously breaks
 $SU(N_f)\times SU(N_f)$ to the diagonal  $SU(N_f)$. This is just the 
symmetry-breaking pattern of nonsupersymmetric QCD.  As $m^2$ increases,
the expectation value $\VEV{T}$ decreases while $\VEV{F_T}
 = \langle\psi_Q \cdot \psi_{\littlebar Q}\rangle$ increases.
  Thus, it is reasonable that,
as $m^2$ is sent to infinity, the vacuum state we have found goes over 
smoothly to the QCD vacuum with a nonzero quark pair
 condensate.\cite{Aetal}
  The only
thing that is still peculiar about this transition is that its starting
point, for small $m^2$, is a Higgs phase and its endpoint, at large 
$m^2$, is a confining or Wilson phase.  But we have already seen that 
these two situations are not distinguished by any gauge-invariant 
expectation values and that 
it is possible to make a smooth transition between them.  We will see 
additional examples of such transitions as we proceed.

\subsection{$N_f = N_c$}

Now that we have understood the behavior of supersymmetric QCD
for $N_f < N_c$, it is natural to ask what happens for larger values of 
$N_f$.  I will discuss this question in full detail in Section \ref{sec:e}.
But I would like to give a preview of that discussion now by considering
the case $N_f = N_c$.

It is tempting to think of this next case as a smooth extrapolation of the
cases discussed in this section.  However, it cannot be.  Most clearly, 
the formula \leqn{eq:y1}
for the Affleck-Dine-Seiberg superpotential is singular
or meaningless at $N_f = N_c$.  To see the origin of this difficulty, notice
from \leqn{eq:w1} that the canonical $R$ symmetry is  has no anomaly 
and from \leqn{eq:uaa} that the elementary fields $Q$ and $\bar Q$ have 
$R$ charge zero.  Thus, it is impossible to build a superpotential
with $R$ charge 2 out of these ingredients.

There is another new feature in the case $N_f = N_c$.  This is the first
case in which it is possible to build gauge-invariant chiral fields with 
the quantum numbers of baryons.  We have two such terms here,
\beqa
B &=& \epsilon_{a_1\cdots a_{N_c}}\ Q_1^{a_1}\cdots Q_{N_c}^{a_{N_c}}
\nonumber\\[1ex]
\bar B &=& \epsilon_{a_1\cdots a_{N_c}}\ \bar Q_1^a \cdots \bar
Q_{N_c}^{a_{N_c}} \ ;
\eeqa{eq:v2}
the lowered indices denote the flavor, as before, and the raised indices
denote the color. 

 I pointed out earlier that the solutions of the 
$D$-flatness equations are parametrized by gauge-invariant polynomials.
Thus, the appearance of new gauge-invariants should be accompanied by 
the appearance of new families of the solutions to the $D$-flatness
conditions.  In this case, there is a new solution of the form
\beq
\VEV Q = \left(
\begin{array}{lcccr}
a &   &        & 0 & \\
  & a &        &   & \\
  &   & \ddots &   & \\
  &   &        & a & \\
  & 0 &        &   & a 
  \end{array}\right) \qquad \langle\bar Q^\dagger\rangle = 0
\eeq{eq:w2}
A second solution is obtained by reversing the roles of $Q$ and
 $\bar Q^\dagger$
in \leqn{eq:w2}.  However, this should not be counted as a new solution, 
since it is a combination of the above and a solution with
$\VEV Q = \langle\bar Q^\dagger\rangle$.  Through the correspondence between 
solutions and gauge-invariant polynomials, this implies that the 
three polynomials $T$, $B$, and $\bar B$ should not be independent.
Indeed, classically, they obey the relation
\beq
\det T = B \bar B \ . 
\eeq{eq:x2}

It is very tempting to think of the low-energy dynamics of this theory as 
being described by the fields $T$, $B$, and $\bar B$ fluctuating subject 
to the constraint \leqn{eq:x2}.
  There can be  no superpotential generated, and 
so the composite
chiral fields sweep out a manifold of supersymmetric vacuum states. 

However, Seiberg has argued that this manifold of vacua is distorted
by nonperturbative effects.\cite{Sqmod} 
 In fact, there is no symmetry which prohibits
the modification of the constraint \leqn{eq:x2} to 
\beq
\det T - B\bar B = \Lambda^{2N_c} \ .
\eeq{eq:y2}
All of the terms in this equation have $R=0$ and mass dimension $2N_c$.
In addition, \leqn{eq:y2} gives a different result from that of \leqn{eq:x2}
under holomorphic decoupling.  To see this, add a mass term
for the last flavor by 
adding to the theory the superpotential
\beq
W = m\, T_{N_fN_f} \ .
\eeq{eq:z2}
Let $t = T_{N_fN_f}$, and consider this field to be determined in terms of
the other fields by the constraint.  The $F$-flatness conditions for 
$B$, $\bar B$, and $T_{N_f j}$ for $j < N_f$ are then solved by setting these
components equal to zero.  This leaves $T$ in the form that we have seen 
in \leqn{eq:n2},
\beq
T =  
\left( \begin{array}{ccc}\widetilde T & 0 \\ 
 0 & t
\end{array}\right) \ ,
\eeq{eq:n2twice}
and the constraint \leqn{eq:y2} now implies
 $\det \widetilde T \cdot t = \Lambda^{2N_c}$.  Inserting the 
constrained value of $t$ into \leqn{eq:z2}, we find
\beq
W = \frac{m\, \Lambda^{2N_c}}{\det\widetilde T} \ .
\eeq{eq:b3}
The renormalization group relation for the effective 
$\Lambda$ parameter \leqn{eq:r2} implies that the numerator of \leqn{eq:b3}
can be replaced by $\Lambda^{b_0}$ for the effective theory with $(N_c-1)$
flavors.  This is precisely the Affleck-Dine-Seiberg superpotential.

Among  plausible forms for the constraint among the gauge-invariant
effective fields, only the version \leqn{eq:y2} with Seiberg's 
quantum modification is consistent through decoupling with our results for 
$N_f < N_c$.  Thus, we find a space of supersymmetric vacuum states
parametrized by $T$, $B$, and $\bar B$ obeying this constraint.  The space
of vacuum states resembles the space of solutions to the classical 
$D$-flatness conditions when the vacuum expectation values of these fields
are large.  However, when the vacuum expectation values become small, 
the space becomes distorted in such a way that it no longer contains the
point  $T = B = \bar B = 0$.  I will have more to say about this case, and
about the cases for $N_f > N_c$, in Section \ref{sec:e}.

The case of supersymmetric QCD with $N_f=N_c$ provides a first example
of a theory with a manifold of vacuum states.  Actually, this is a common
situation in supersymmetric Yang-Mills theories.  In any situation for which
there is a continuous family of solutions to the conditions for unbroken
supersymmetry, we will find a manifold of degenerate vacuum states with 
zero energy.  This manifold will typically
be parametrized by the expectation values of
chiral superfields; thus, it will be a complex K\"ahler manifold.  It is 
common to call this space the `moduli space' of the theory, and I will use that
terminology from here on.

\section{The Seiberg--Witten Model}
\label{sec:c}

In the previous examples, the low-energy dyanamics of the gauge theory
contained only chiral multiplets, while all of the gauge charges were either
confined or spontaneously broken.  So it would be good to  illustrate that
it is also possible for the low-energy gauge symmetry to be realized in the
Coulomb phase.  The simplest illustrative model of this type is the 
celebrated model of Seiberg and Witten.\cite{SW}

Consider $SU(2)$ Yang-Mills theory with an extra chiral superfield
$\phi$ in the {\it adjoint} representation of the gauge group.  With the 
superpotential set to zero, the
Lagrangian of the theory as
\beq
\L = \int d^4\theta\, \frac{1}{g^2}\, \phi^\dagger e^V\phi -
\frac{i}{16\pi}\int d^2\theta\, \tau\, W^{\alpha a}W^a_\alpha + \hc
\eeq{eq:c3}
 This model is in fact the pure Yang-Mills theory
with $N=2$ supersymmetry.  The two gauginos of the theory are $\lambda$
and $\psi_\phi$; I have put a factor $1/g^2$ in front of the $\phi$ kinetic
energy term to make the symmetry relation of these two fields more clear.
  My discussion of this model will be given 
mainly in $N=1$ notation, and the conclusions will apply to similar 
models which 
are only $N=1$ supersymmetric.  Nevertheless, as I will discuss later, 
the $N=2$ supersymmetry has interesting consequences that will help us in 
our analysis.

\subsection{Parametrization of the vacuum states}

The classical potential of the model comes only from the $D$-term 
contribution
\beq
V = \frac{g^2}{2}\, (D^a)^2 \ ,  \quad  \mbox{where}\quad
D^a = \frac{1}{g^2}\, \phi^+t^a\phi \ .
\eeq{eq:d3}
The $D$ term is most clearly written by expressing $\phi$ and $D$ as 
matrices: $\phi =\phi^a t^a$, $D = D^a t^a$.  Then
\beq
D = \frac{i}{g^2}\ [\phi^\dagger,\phi] \ .
\eeq{eq:e3}
Thus $D=0$ if $\phi$ and $\phi^\dagger$ can be simultaneously diagonalized.
Since these are $SU(2)$ matrices, this condition implies that there is a 
gauge rotation such that
\beq
          \VEV{\phi^b} = a \delta^{b3} \ ,
\eeq{eq:f3}
with $a$ a complex number.
This expectation value spontaneously breaks $SU(2)$ to $U(1)$.  Notice that 
the classical potential equals zero for any value of $a$.

  Expanding the 
classical Lagrangian about any of the points \leqn{eq:f3},
 one finds that all of the 
the fields
charged under the $U(1)$ receive mass from the $\phi$ vacuum expectation value.
The fields that remain massless are the $U(1)$ gauge boson $A^3_\mu$, the
fermions $\lambda^3$ and $\psi_\phi^3$, and the complex scalar $\phi^3$.   It 
is clear that the vector and the scalar must remain massless: The vector
field is the gauge field of an unbroken gauge symmetry, and the scalar field
is the fluctuation along a manifold of degenerate vacuum states.  Together
with their superpartners, these states fit together into an $N=2$ supersymmetry
multiplet.

  Since the massless fields of the model are noninteracting at large 
distances, all of the vacuum states \leqn{eq:f3} belong to the Coulomb phase
of the  $U(1)$ gauge symmetry.   And some additional structure is present:
Because a non-Abelian gauge
group is spontaneously broken to $U(1)$, this theory has 't Hooft-Polyakov
magnetic monopoles.\cite{tHm,P}
 The $N=2$ supersymmetry of the model and the flatness
of the potential for $a$ implies that these monopoles are regulated by a 
Bogomolny-Prasad-Sommerfield (BPS) inequality.\cite{B,PS}  The general 
properties of these magnetic monopole solutions are described in Harvey's
lectures at this school.\cite{Harvey}

 Classically, the vacuum states of the theory are related by a $U(1)$
symmetry 
\beq 
\phi \rightarrow e^{i\alpha}\phi\ , \qquad \psi_\phi \rightarrow
e^{i\alpha}\psi_\phi \ .
\eeq{eq:g3} 
However, as in supersymmetric QCD, this symmetry is broken by a gauge
anomaly.  Equivalently, the transformation \leqn{eq:g3} is equivalent to 
a shift
 of the $\theta$ parameter,
\beq
 \theta \rightarrow \theta - 4\alpha\ , \quad \mbox{\rm or} \quad \tau
\rightarrow \tau -\frac{4}{2\pi}\, \alpha \ .
\eeq{eq:g3plus}
Since a shift of $\theta$ by $2\pi$ is a symmetry of the theory, we could 
also say that the original model is invariant under the discrete symmetry
\beq 
\phi \rightarrow e^{i\frac{\pi}{2}}\phi \ .
\eeq{eq:h3}

Though we can reasonably parametrize the vacuum states at weak coupling by
the expectation value of $\phi$, this is not a useful way to describe these
vacua in the strong-coupling region, because $\phi$ is not a gauge-invariant
quantity.
  I will now propose two different
ways to generalize the definition of the parameter $a$ introduced above
so that it makes sense in all regions in which we would like to analyze the 
theory. First of all, we could characterize the vacuum by the 
vacuum expectation value of the gauge-invariant operator
\beq
u = \VEV{(\phi^a)^2} \ .
\eeq{eq:i3}
In the weak-coupling region, $u \approx a^2$.
The chiral symmetry \leqn{eq:h3} is realized on $u$ as a $Z_2$ symmetry
\beq
 u \rightarrow -u \  .
\eeq{eq:i3plus}

Another generalization of $a$ involves the particle mass spectrum.  At
weak coupling, in the normalization introduced above, the $W$ bosons
acquire mass $m_W = \sqrt{2} a$ from the Higgs mechanism, where $a = 
\langle\phi^3\rangle$. The
magnetic monopoles have mass $m_M = 4\pi \sqrt{2}a/g^2$.  The BPS inequality
implies that, at all values of the coupling, these
 particle masses satisfy a relation of the form
\beq 
m  =  \sqrt 2 \ \left| a\, Q_e + a_DQ_M\right|
\eeq{eq:j3}
where $Q_e\, Q_M$ are the electric and magnetic charges and $a$ and
$a_D$ are some constants.  Thus, we can consider the coefficient $a$ in 
this formula to be the gauge-invariant generalization of the vacuum 
expectation value of $\phi$.   The coefficient $a_D$ should be determined
uniquely by the value of $a$ and the effective large-distance coupling
constant $g^2$ or $\tau$. 

 At weak coupling, $a_D$ obeys the relation
\beq
a_D  \approx \frac{4\pi i}{g^2}\, a = \tau\cdot a \ .
\eeq{eq:l3}
However, this cannot be an exact relation in the theory.  The effective 
coupling $\tau$ is determined, at least in part, by the renormalization
group running of the coupling constant in the $SU(2)$ gauge theory from the
fundamental short-distance scale down to the scale $a$.  But the formula
$a_D = \tau(a) a$ is not renormalization-group invariant.  Seiberg and Witten
proposed the formula
\beq 
 \tau = \frac{da_D}{da}  \ . 
\eeq{eq:n3}
This relation is consistent with a nontrivial dependence of $\tau$ on $a$.
It also suggests a duality symmetry
\beq
a \leftrightarrow a_D\ , \qquad  \tau \leftrightarrow -1/\tau = \tau_D\ .
\eeq{eq:n3plus}

Some motivation for the formula \leqn{eq:n3} is given by computing the 
magnetic monopole mass in the weak-coupling limit of the effective $U(1)$
gauge theory.  In that limit, the monopole mass is given by 
\beq
      m =  \int d^3x \left(\frac{1}{g^2}\, |\vec\nabla a|^2 +
\frac{1}{2g^2}\, (\vec B)^2\right) \ .
\eeq{eq:o3}
Using \leqn{eq:n3}, we can transform the first term using the relation
$\tau \vec\nabla  a = \vec \nabla  a_D$, to give
\beqa
      m &=&  \int d^3x \left({g^2\over (4\pi)^2} |\vec\nabla a_D|^2 + 
\frac{1}{2g^2}\, (\vec B)^2\right) \CR
    &=&  \int d^3x \left| {g\over 4\pi} \vec\nabla a_D \pm 
\frac{1}{\sqrt{2}g}\, \vec B\right|^2 \ 
   \mp \ {\sqrt{2}\over 4\pi}\int d^3x \vec\nabla (a_D \vec B) \ .
\eeqa{eq:p3}
Then, finally
\beq 
m \geq  {\sqrt{2}\over 4\pi} \int d^2s\, \hat  n\cdot a_D \vec B \ ,
\eeq{eq:q3}
consistent with \leqn{eq:j3}.  However, we will obtain much stronger tests 
of the relation \leqn{eq:n3}, which also can be made in the strong-coupling 
region, by examining the properties this relation predicts for the
 $\theta$-dependence of the properties of magnetic monopoles.

I have already noted that the vacuum parameter, or `modulus', $a$, the $U(1)$
gauge boson, and the fermionic partners of these fields fit together into an
$N=2$ supermultiplet.  Now that we have seen that these fields can be 
characterized in a gauge-invariant way, it makes sense to write an effective
Lagrangian which could describe their dynamics.  The most general possible
such Lagrangian is
\beq
\L = \int d^4\theta\, K(a,\bar a) - \frac{i}{16\pi}\int d^2\theta\,
\tau(a)\, W^\alpha W_\alpha + \hc 
\eeq{eq:s3}
The $N=2$ supersymmetry forbids a superpotential.  It also relates $\tau$
and $K$ through a `prepotential' $\F(a)$: 
\beq
\tau = \frac{\partial^2\F}{\partial a^2} \qquad K = \frac{1}{4\pi}\,
{\rm Im}\ \frac{\partial\F}{\partial a}\, \bar a \ .
\eeq{eq:t3}
Using \leqn{eq:n3}, we can evaluate
\beq
 K = \frac{1}{4\pi}\,
\mbox{Im}\ a_D\bar a \ ,
\eeq{eq:u3}
which is also symmetric under electric-magnetic duality.

In \leqn{eq:t3}, I have written $\tau$ as a function of $a$.  In fact, 
the three complex variables $\tau$, $a_D$, and $a$ are tied together
by the relation $\leqn{eq:n3}$.  All three of these variables 
can be thought of as functions of $u$ defined in 
 \leqn{eq:i3}.  To understand the qualitative 
beahavior of the model, we should try to determine the explicit 
dependence of these
quantities on $u$.

To determine $\tau(u)$, we will make essential use of the fact that the 
relations among $\tau$, $a$, $a_D$, and $u$ are holomorphic.  In particular,
the holomorphic function $\tau(u)$ can be reconstructed from the knowledge
of its singularities and its behavior at infinity.  However, a singularity
in the effective coupling constant must be associated with divergent
coupling constant renormalization, and this is possible only if very light
states appear in the physical spectrum.  The strategy of Seiberg and Witten
is then to determine the singularities of $\tau(u)$ from physical arguments
and then to construct the global function from the properties of these 
singularities.

\subsection{Weak-coupling behavior of $\tau(u)$}

To begin this program, we might first analyze the behavior of $\tau(u)$
in the limit $u \to \infty$. 
 This corresponds to the weak-coupling region of the theory,
and so we can obtain the relation between $\tau$ and $u$ from a weak-coupling
renormalization group analysis.

\begin{figure}
\begin{center}
\leavevmode
\epsfbox{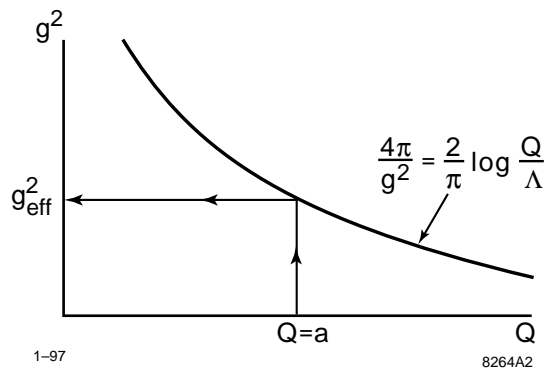}
\end{center}
 \caption{Determination of the effective coupling constant in the 
         weak-coupling limit of the Seiberg-Witten model.}
\label{fig:two}
\end{figure}

 In pure $N=2$ Yang-Mills theory, the 
formula \leqn{eq:h1} gives 
\beq
b_0 = 2N_c \ , 
\eeq{eq:v3}
or $b_0 =  4$ in the $SU(2)$ theory.  The running coupling constant is then 
given by \leqn{eq:qpplus}:
\beq
\frac{4\pi}{g^2}\, (Q) = \frac{4\pi i}{g^2} + \frac{4}{2\pi}\,\log 
\frac{Q}{M}
\eeq{eq:w3}
  In the theory with spontaneously broken symmetry, 
the coupling constant will run from the short-distance scale $M$ to the 
scale $a$, and then stop at the mass scale of  the particles with nonzero
$U(1)$ charge, as shown in Figure \ref{fig:two}.
 Thus, the effective coupling at $\theta = 0$ should be given
 by  $\tau(a) = 4\pi/g^2(a)$, plus a  possible constant shift from one-loop
corrections at the scale $a$.  We can absorb this shift into the $\Lambda$
parameter.  Using also the relation \leqn{eq:i3}, we can write the holomorphic
relation between $\tau$ and $u$ as  
\beq
\tau(u) = \frac{i}{\pi}\, \log \frac{u}{\Lambda^2}
\ .
\eeq{eq:x3}

We can check this formula in a nontrivial way by thinking about the 
implications of this formula for the dependence of the effective Lagrangian
and the monopole masses on the phase of $u$.  First of all, the phase
rotation $u \to e^{2i\alpha} u $ should be equivalent to the shift of 
$\theta$ by \leqn{eq:g3}, and this relation is nicely realized in \leqn{eq:x3}.
The weak-coupling relation $u = a^2$ can be combined with  \leqn{eq:n3} to 
evalute $a_D$ as
\beq
a_D = \frac{2i}{\pi}\, \left( a\log \frac{a}{\Lambda}-a\right) \ .
\eeq{eq:a4}
Then under the rotation $a \rightarrow e^{i\alpha/4}a$, which corresponds
to  $\theta \to  \theta - \alpha$,
\beq
a_D \to  e^{i\alpha/4} \left(a_D-\frac{\alpha}{2\pi}\, a
\right) \ .
\eeq{eq:b4}
Then the BPS bound  becomes
\beq
m \to \sqrt{2} \left|
 a\left(Q_e  - \frac{\alpha}{2\pi}\,Q_M\right)
+ a_DQ_M\right| \ .
\eeq{eq:c4}
This is just right.  In the presence of a nonzero $\theta$ parameter, 
a magnetic monopole acquires an additional
electric charge $\theta Q_M/2\pi$.\cite{Wtheta}
 The effect of the nonzero $\theta$ generated by the rotation of $a$
thus shifts the monopole electric charges in precisely the manner 
indicated in \leqn{eq:c4}.

\begin{figure}[t]
\begin{center}
\leavevmode
\epsfbox{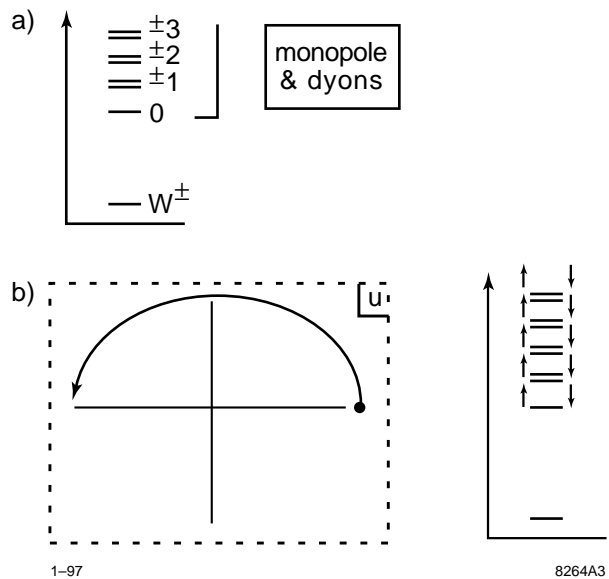}
\end{center}
 \caption{(a) Spectrum of $W$ bosons, monopoles, and dyons in the 
weak-coupling limit of the Seiberg-Witten model with $\theta = 0$.
(b) Transformation of the spectrum of monopole and dyon states
    as we turn on $\theta$ or rotate $u$ in the weak-coupling region.}
\label{fig:three}
\end{figure}

To push this picture a little further, recall that the 
classical solutions of the theory we are considering include not only 
magnetic monopoles but also dyons.\cite{Harvey}  The magnetic
monopole solution can be deformed to a solution rotating in the $U(1)$ 
direction, and each such solution with quantized angular momentum gives a 
new, electrically charged, solution.  Thus, in weak coupling at $\theta= 0$,
 the 
spectrum of the model includes a tower of states with $Q_M = 1$ and all 
integer values of the electric charge, and a similar tower for $Q_M = -1$. 
All of these particles obey the BPS mass formula \leqn{eq:j3}. The particle
spectrum of the theory at weak coupling is shown in Figure \ref{fig:three}(a).

All of these states are affected by the shift of $\theta$ induced by 
a phase rotation of $u$ or $a$.  Under the transformation \leqn{eq:b4}, 
the positively charged dyons become lighter while the negatively
charged dyons become heavier.  When we have gone half-way around the 
$u$ plane, $u\to e^{i\pi} u$ or $a \to e^{i\pi/2} a$, the spectrum 
goes back to its original form, but with 
the dyon which 
originally had charge $Q_E = 1$ becoming the lightest particle with 
magnetic charge. This transformation is shown in Figure \ref{fig:three}(b).
If we had rotated around the $u$ plane in the other direction, we would have
found as the lightest monopole the dyon which had $Q_E = -1$ at $\theta=0$.

The fact that the model has the same spectrum when we carry $u \to -u$ should
be no surprise, because these points are related by the $Z_2$ symmetry 
\leqn{eq:i3plus}.  It is a surprise, though, that this identity of the 
spectra is realized thorough a rearrangement of the states. Similarly, when
we come back to the original value of $u$ after a $2\pi$ circuit of the $u$
plane, we find the same spectrum shifted by 2 units.  This behavior is
suggested by the form of \leqn{eq:x3}, which is a branched function of $u$.
In fact, we now see that the whole theory has a branched structure in $u$. 
rather than being single-valued as a function of this variable.

\subsection{Strong-coupling singularities of $\tau(u)$}

If the function $\tau(u)$ has a branch cut singularity
at large values of $u$, this branch cut must originate at some point or 
points in the interior of the $u$ plane.  We will now try to find and 
characterize these points.

At first sight, it seems possible that \leqn{eq:x3} could be exact.  It is 
true that there are no perturbative corrections to \leqn{eq:x2}, since any
modification of this equation by logarithms of $u$ would destroy the 
transformation properties of $a$ and $a_D$ under a shift of $\theta$ that
we have just discussed.  However, \leqn{eq:x3} can be corrected by 
nonperturbative effects
\beq
\tau(u) = \frac{i}{\pi}\, \log \frac{u}{\Lambda^2} + a u^{-2}
    + b u^{-4} + \ldots
\eeq{eq:x3x}
Because of the $Z_2$ symmetry \leqn{eq:i3plus}, only even powers of $u$
can appear. Solving for $u$ perturbatively, one can see that $u^{-2} \sim
e^{-8\pi^2/g^2}$, characteristic of a one-instanton correction.  In fact, 
these leading
instanton corrections have been evaluated and are nonzero.\cite{FP,Mattis}
So we will need a more sophisticated hypothesis.

\begin{figure}[t]
\begin{center}
\leavevmode
\epsfbox{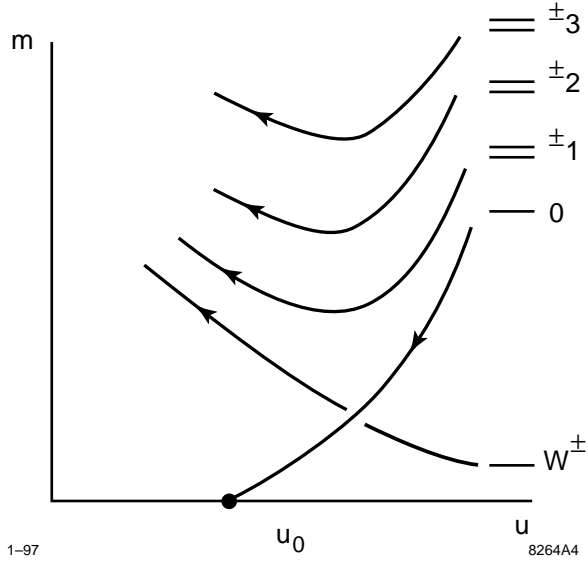}
\end{center}
 \caption{Dependence of monopole and dyon masses on $u$ along the 
   positive real axis of the $u$ plane.}
\label{fig:four}
\end{figure}

The next simplest idea is that nonperturbative effects in the theory generate
a scale $u_0$ proportional to $\Lambda$, and that $\tau(u)$ has a pair of 
singularities at $u = \pm u_0$.  The presence of a 
pair of singularities is required by the $Z_2$ symmetry.  In fact, there is a
pleasing physical picture of the origin of these singularities.  As $u$
decreases, the coupling constant should increase.  This will cause the 
parameter $a_D$ to decrease and thus should lower the masses of the 
monopoles.  As long as $a$ is nonzero, the dyons with nonzero $Q_E$ must
remain massive, but the lightest monopole with $Q_E = 0$ could come down
to zero mass.  This evolution is shown in Figure \ref{fig:four}.  We will
then assume that $a_D$ has a zero at a point $u_0$ on the real axis.
  At the reflected point
$u = -u_0$, the dyon which has charge 1 for real positive $u$, which becomes
the lightest dyon on the negative real axis, comes down to zero mass in 
the same way.

This picture leads to an explicit expression for the singularity of 
$\tau(u)$ at $u = 0$.  Near this point, the only light states in the theory
are magnetic monopoles with zero electric charge.  These monopoles 
renormalize the effective coupling in such a way as to screen 
the dual coupling constant $\tau_D = -1/\tau$.  The $\beta$ function of this
dual theory is the same as that in supersymmetric quantum electrodynamics
with one charged species, $b_0 = -2$.  (This is \leqn{eq:h1} with 
$C_2(G_c) = 0$  and $C(r) = 1$ for each chiral multiplet.)  Then 
\beq
\tau_D = - \frac{2i}{2\pi}\, \log m_M
\eeq{eq:d4}
where $m_M$ is the monopole mass.  I will assume that $a_D$ is nonsingular
at $u_0$ with a simple zero,
\beq
a_D \approx  b(u-u_0)\ .
\eeq{eq:e4}
Then, since $m_M = \sqrt{2} a_D$, 
\beq
\tau_D = - \frac{1}{\tau(u)} =  \frac{-i}{\pi}\, \log(u-u_0)\ .
\eeq{eq:f4}

From the expressions for $\tau$ and $a_D$, we can reconstruct the formula 
for $a$ near $u = u_0$.  Since 
\beq
 \frac{da}{da_D} = -\tau_D =  \frac{i}{\pi}\, \log a_D \ , 
\eeq{eq:g4}
we find
\beq
a = \frac{i}{\pi}\, (a_D \log a_D-a_D) .
\eeq{eq:h4}

The singularity of $\tau(u)$ at $u = -u_0$ must be the mirror image of the 
singularity at $u = u_0$.  To compute the behavior of $\tau$, $a$, and $a_D$
at this point, start at large real positive $u$, and go around the outside
of the complex $u$ plane from to the point $ue^{i\pi}$ on the negative 
real axis.  The new values of $a$ and $a_D$ are
\beq
   a \to \tilde a = ia    \ , \qquad   a_D \to \tilde a_D = i(a_D - a)
\eeq{eq:h4plus}
The transformed $a_D$ must have a simple zero which is the image of 
\leqn{eq:e4}, 
\beq
      \tilde a_D \sim (u + u_0)\ . 
\eeq{eq:h4pplus}
Then, in the vicinity of $u = -u_0$, $\tau(u)$ has a singularity given 
by 
\beq
\tau_D = - \frac{1}{\tau(u)} =  \frac{-i}{\pi}\, \log(u+u_0)\ .
\eeq{eq:h4xplus}
and $a$ has the singular behavior
\beq
\tilde a = \frac{i}{\pi}\, (\tilde a_D \log \tilde a_D-\tilde a_D) .
\eeq{eq:h4yplus}

The branched behavior of $\tau$ around each of these singularities is most
clearly demonstrated by the transformation of $a$ and $a_D$ around each 
of the singularities.  If we make a $2\pi$ circuit of the $u$ plane for 
large $u$, \leqn{eq:b4} implies that $a$ and $a_D$ return to the values
\beq
    a \to -a \ , \qquad  a_D \to - (a_D -a) \ .
\eeq{eq:h4zplus} 
Around the singularity at $u= u_0$, \leqn{eq:h4} implies the transformation
\beq
     a \to a - 2a_D \ , \qquad a_D \to a_D \ .
\eeq{eq:h4oplus}
Around the singularity at $u=-u_0$, $\tilde a$ and $\tilde a_D$ go through
the same transformation.  Replacing these by $a$ and $a_D$ using 
\leqn{eq:h4plus},
we find
\beq
a \to 3a - 2a_D \qquad a_D \to 2a - a_D \ .
\eeq{eq:l4}

Such transformations of functions around a complex singularity are called
`monodromies'. In this case, we can characterize each singularity by a 
$2\times 2$ monodromy matrix $M$,  by writing
\beq
 {a_Dchoose a} \rightarrow M \, {a_D\choose a}  \ .
\eeq{eq:l4plus}
For the three singularities,
\beq
   M_\infty = \pmatrix{-1 & 2 \cr 0 & -1} \ , \qquad
   M_{u_0} = \pmatrix{1 & 0 \cr -2 & 1} \ , \qquad
   M_{-u_0} = \pmatrix{-1 & 2 \cr -2  & 3} \ .
\eeq{eq:l4pplus}
Since $u$ is by definition nonsingular on the $u$ plane, the doublet
\beq
{ da_D/du \choose da/du}
\eeq{eq:o4}
has the same monodromies.  From the components of this vector, we can 
reconstruct $\tau(u)$ as
\beq
\tau(u) = \frac{(da_D/du)}{(da/du)} \ .
\eeq{eq:p4}

\begin{figure}
\begin{center}
\leavevmode
\epsfbox{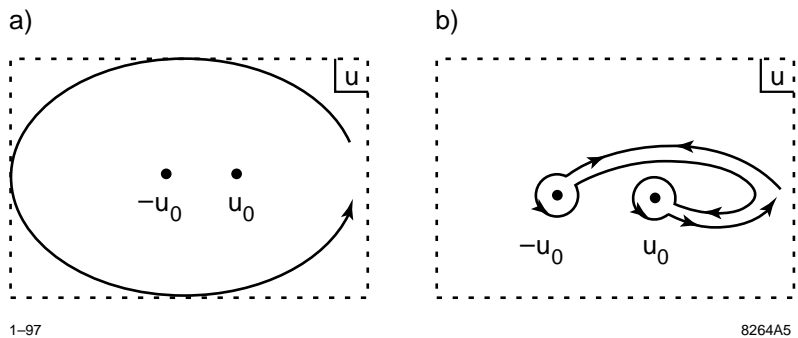}
\end{center}
 \caption{Relation of the monodromies about the three singularities
      of $\tau(u)$.}
\label{fig:five}
\end{figure}

At this moment, there is no guarantee that we have discovered all of the
singularities of $\tau(u)$.  However, it is possible to check that the 
branch cuts which originate at $u_0$ and $-u_0$ are sufficient to account
for the branched behavior of $\tau(u)$ at infinity. Around the path shown 
in Figure \ref{fig:five}(a), $a$ and $a_D$ should have the monodromy 
$M_\infty$.  If $\tau(u)$ has no sigularities other than those we have
already identified, this path can be 
 deformed continuously to that shown in 
Figure  \ref{fig:five}(b).  Notice that the path to and from $u = -u_0$ 
passes above the singularity at $u=u_0$ and thus belongs to  the branch for 
which we have derived  \leqn{eq:l4}.  The test that no other singularities
are needed is the equality of these transformations, that is, 
\beq
      M_{u_0} M_{-u_0} = M_\infty \ .
\eeq{eq:q4minus}
And, indeed, this follows from \leqn{eq:l4pplus}.

\subsection{Geometry of the moduli space}

Now that we have determined the singularity structure of $\tau(u)$, we should
be able to reconstruct the function explicitly.  In principle, this could be
done completely algebraically.  However, there are two clues in the 
information we have uncovered which suggested to Seiberg and Witten a 
geometrical solution to this problem.  The first is the general property
that the effective coupling $g^2(u)$ must be positive.  This restricts
$\tau(u)$ by   
 \beq
\mbox{Im}\ \tau = \frac{4\pi}{g^2} > 0 \ .
\eeq{eq:q4}
The explicit formula for $\tau$ must naturally respect this relation.
The second is the set of monodromy relations, which induce specific quantized
shifts of $\tau$ as we move around each singularity.  Both of these properties
suggest that $\tau$ is the modulus of a torus.

\begin{figure}
\begin{center}
\leavevmode
\epsfbox{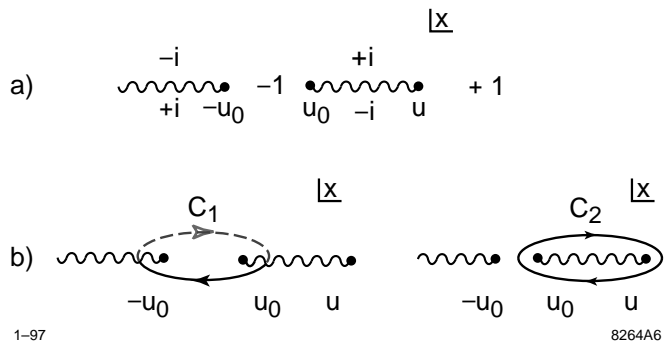}
\end{center}
 \caption{(a) The singularities of $y(x)$, and a branch choice for the 
square root.  (b) The contours $C_1$ and $C_2$ used to define the 
translations $z_1$ and $z_2$.}
\label{fig:six}
\end{figure}

A convenient way to construct these tori is to use the following 
representation:  Let     
\beq
y^2 = (x-u_0)(x+u_0)(x-u)
\eeq{eq:r4}
and consider the integral
\beq
z = \int^x_{u_0}\
\frac{dx}{\left[(x-u_0)(x+u_0)(x-u)\right]^{1/2}} = 
\int^x_{u_0} \frac{dx}{y}
\eeq{eq:s4}
For definiteness, choose the branch of the square root such that, when
$u$ is real, positive, and greater than $u_0$, the square root has
the phases shown in Figure \ref{fig:six}(a).

The integral $z(x)$ is a mapping from the two-sheeted $x$ plane to a torus.
When $u$ is real and positive, 
as $x$ moves from $u_0$ along the positive real axis, $z$ moves in the
imaginary direction  in
the complex plane.  As $x$ is moved to the left of $u_0$, $z$ moves along
the positive real axis. Complete circuits along the contours $C_1$
and $C_2$ shown in Figure \ref{fig:six}(b) carry $z$ into the values
 \beq
z_1 = \oint_{c_1} \frac{dx}{y}\ , \qquad z_2 \oint_{c_2}
\frac{dx}{y} \ .
\eeq{eq:t4}
These are the fundamental translations on a torus of modulus
\beq
         \tau(u) =  z_2/z_1 \ .
\eeq{eq:t4plus}
It is not difficult to see that the double-sheeted $x$ plane is mapped
1-to-1 into this torus.  For example, the upper half plane on the 
first sheet in $x$ is mapped into the rectangle
whose corners are $z = 0,
z_1/2, z_2/2, (z_1 + z_2)/2$.
Then
\beq
     \frac{d a}{du} = A z_1 \ , \frac{da_D}{d u} = A z_2 \ ,
\eeq{eq:t4pplus}
where the common constant $A= 1/4\pi$ can be determined from the relation
$a \to \sqrt{u}$ as $u \to \infty$.

It is straightforward to see that \leqn{eq:t4plus} and \leqn{eq:t4} do 
indeed construct a function with the properties of $\tau(u)$. The function
that we have defined has singularities at only at $u = \pm u_0$.
By
taking $u \gg u_0$, one can verify the form \leqn{eq:x3}. By carrying $u$
around $u_0$, and paying close attention to the branches of the square root
in the various segments of the integral, one can verify the monodromy 
relation \leqn{eq:h4oplus}.  

Once $\tau(u)$ has been determined in this way, it is possible to make a 
quite nontrivial check on the solution by comparing the coefficients $a$
and $b$ in \leqn{eq:x3x} with the explicit results of one- and two-instanton
calculations.  The check confirms the Seiberg-Witten solution.\cite{FP,Mattis}

\begin{figure}
\begin{center}
\leavevmode
\epsfbox{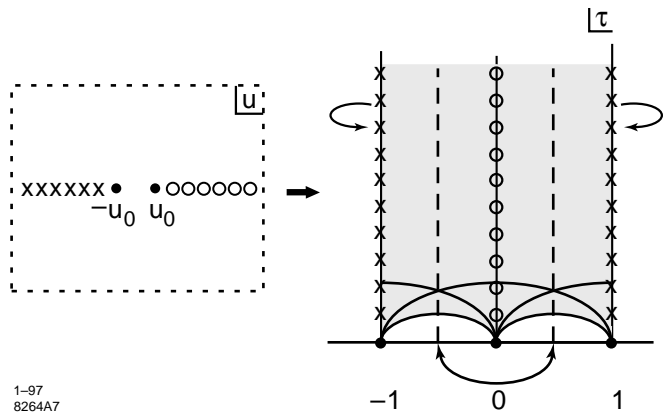}
\end{center}
 \caption{The mapping from the $u$ plane to the space of torus moduli $\tau$.}
\label{fig:seven}
\end{figure}

  The mapping from the $u$ plane to the space of moduli $\tau$ is quite
interesting.  As $u\to \infty$, $z_2 \to i\infty$ with $z_1$ fixed, so 
we find a tall, thin torus (the `Witten torus').  As $u \to u_0$, $z_2\to 0$,
and so we find a short, fat torus (the `Peskin torus'). As $u\to -u_0$ from
above or below the real axis, $z_2 \to \pm 1$ and we find a short 
torus twisted through $2\pi$.
  The full mapping of the $u$ plane is shown in Figure \ref{fig:seven}.
The image of the $u$ plane covers four copies of the fundamental region
of the modular group.  It is, in fact, the fundamental region of the 
subgroup $\Gamma(2)$ of $SL(2,Z)$. 

\subsection{Relation to $N=1$ Yang-Mills theory}

Another way to confirm our understanding of the $N=2$ $SU(2)$ Yang-Mills
theory is to explicitly break the $N=2$ supersymmetry to $N=1$.  This is
easily done by adding a mass term for the $\phi$ supermultiplet to the 
Lagrangian \leqn{eq:c3}.  When the field $\phi$ and its fermionic partner
are decoupled, the theory should revert to the $N=1$ pure Yang-Mills theory
that we discussed in Section 4.2.  It is not at all obvious that this 
correspondence can be made.  In our earlier discussion, we analyzed the
$N=1$ Yang-Mills theory as a theory of confinement; our analysis of the 
$N=2$ theory was based on the realization of this model in the Coulomb
phase.  How could these descriptions be connected?

In any case, we can carry out the analysis.  To add a mass term for $\phi$,
add the superpotential
\beq
\Delta W = \half\, m\,\phi^2 = \half\, m\, u \ .
\eeq{eq:v4}
At a typical point in the strong-coupling region, $u$ is the only light 
chiral superfield, so \leqn{eq:v4} is the full effective superpotential.
Then the $F$-flatness condition   $\partial W/\partial u = 0$ 
cannot be satisfied.

Near $u=u_0$, there is a better situation.  A set of  magnetic monopoles
become light and so we should include magnetic monopole fields in the 
effective Lagrangian.  To write a mass term, we need a pair of chiral 
superfields $M$ and $\bar M$, which create the $Q_E=0$ monopole and 
antimonopole.  These two fields form an $N=2$ hypermultiplet.  The effective
superpotential then takes the form
\beq
W_\eff = \sqrt 2\, b(u-u_0)\ M\bar M + \half\, m\, u \ ,
\eeq{eq:w4}
where I have used the expression \leqn{eq:e4} to write
 the monopole mass as a 
function of $u$.
 
The $F$-flatness conditions following from \leqn{eq:w4} are
\beq
(u-u_0)M=(u-u_0)\bar M = 0 \qquad
\sqrt 2\, b\, M\bar M + \half\, m = 0 \ . 
\eeq{eq:x4}
The solution of these conditions is
\beq
u=u_0 \qquad   M = \bar M = \left(\frac{-m}{2\sqrt 2\,b}\right)^{1/2} \ ,
\eeq{eq:y4}
up to a $U(1)$ gauge transformation on $M$, $\bar M$. 

Thus, in the presence of the superpotential \leqn{eq:v4}, the manifold of 
vacuum states characteristic of the Coulomb phase is lifted away from 
zero energy.  Only two discrete supersymmetric configurations remain,
the vacuum we have found at $u = u_0$ and a mirror-image vacuum at 
$u = -u_0$.   In fact, we showed in Section 4.2 that the $N=1$ supersymmetric
gauge theory should have precisely two vacuum states, reflecting the 
spontaneous global symmetry breaking from $G = Z_4$ to $Z_2$.

 A remarkable property of the vacuum states at $u = \pm u_0$ is that the 
magnetic monopole fields acquire vacuum expectation values.   
These vacuum states are realized in the Higgs phase of the magnetic
$U(1)$ theory.  Dually, they belong to the confining phase of the 
original Yang-Mills theory, according to the criteria for confinement that
we discussed in Section 2.2.

\section{More Phenomena of the Coulomb Phase}
\label{sec:d}

There is much more to say about properties of the Coulomb phase of
supersymmetric gauge theories.  In this section, I would like to highlight
two particularly interesting physical 
phenomena which appear already in the 
simplest extension of the Seiberg-Witten model.  Then I will discuss
some models which generalize the geometrical  structure of the space of
vacua which we found in Section 5.4.

\subsection{$N=2$ $SU(2)$ Yang-Mills theory with matter}

  It is a natural generalization of the Seiberg-Witten model discussed in the
previous section to add some number of matter fields which couple to the 
gauge symmetry.  In an $N=2$ supersymmetric theory, matter fields belong to 
$N=2$ hypermultiplets, which are pairs of $N=1$ chiral supermultiplets 
 $(Q_i,\bar Q_i)$ in 
conjugate representations of the gauge group. In $N=1$ language, the 
Lagrangian consists of the standard coupling of the gauge multiplet to these
fields, plus the superpotential
\beq
W =  2 \sum_i \bar Q_i  \phi^a t^a Q_i \ , 
\eeq{eq:z4}
which couples the fields $\phi$, $\psi$ of the $N=2$ gauge multiplet to the 
hypermultiplets.  

For $SU(N_c)$ gauge theories with matter in the fundamental representation,
the $\beta$ function is given by
\beq 
        b_0 = 2N_c -N_f \ ;
\eeq{eq:z4plus}
thus, the theories are asymptotically free with any number of matter multiplets
up to $2N_c$.  For $SU(N_c)$ gauge theories with matter in the adjoint 
representation, adding one hypermultiplet already gives $b_0= 0$.  This latter
theory, which has a total of
four chiral fermions and six real scalars in the adjoint
representation, is precisely the $N=4$ supersymmetric Yang-Mills theory.

  Since the $N=2$ theories with matter have a superpotential, the classical 
vacuum states are determined both by $D$-flatness and $F$-flatness conditions.
There are two classes of solutions to these conditions.  The first gives a 
Coulomb phase similar to that of the previous section, with 
\beq
\VEV\phi \ne 0 \qquad \VEV{Q} = \VEV{\bar  Q} = 0 \ .
\eeq{eq:e5}
The second gives a Higgs phase with
\beq
\VEV\phi = 0 \qquad \VEV{Q} = \VEV{\bar Q^\dagger} \neq 0 \ .
\eeq{eq:f5}
In these lectures, I will only discuss the properties of the Coulomb phase. 
The behavior of the Coulomb phase in all four possible cases,
$N_f = 1, 2, 3, 4$, was worked out by Seiberg and Witten.\cite{SWII}  

To analyze the Coulomb phase, we must work out the global symmetries of the 
theory.
For a general $SU(N_c)$ gauge group, the theory has the continuous
global symmetry
of supersymmetric QCD, $SU(N_f) \times SU(N_f) \times U_B(1) \times U_R(1)$
(where the last factor is the anomaly-free $R$), broken by the superpotential
coupling \leqn{eq:z4} to $SU(N_f)\times U_B(1)$. If the gauge group is $SU(2)$,
however, there is additional symmetry because the spinor  of $SU(2)$ is a real
representation, equivalent to its conjugate.  For this case, the continuous
global symmetry of supersymmetric QCD is $SU(2N_f) \times U_R(1)$.  Since 
an $SU(2)$ vector couples to two spinors in the symmetric combination, the 
coupling \leqn{eq:z4} preserves an $SO(2N_f)$ subgroup of this group.

If we wish to generalize the analysis of the previous section, it 
will also be interesting to understand the anomalous global symmetry
corresponding to the 
phase rotation
\beq
      \phi \to e^{i\alpha} \phi \ , \mbox{\rm or}\  u \to e^{2i\alpha} u \ .
\eeq{eq:b5}
To preserve the superpotential \leqn{eq:z4}, this rotation must be 
carried out together with 
\beq
Q \to e^{-i\alpha/2} Q  \qquad
\bar Q \to e^{-i\alpha/2}\bar Q \ .
\eeq{eq:c5}
The rotations \leqn{eq:b5} and \leqn{eq:c5} together give a symmetry of 
classical $N=2$ Yang-Mills theory. In the quantum theory, the anomaly
generates a shift of $\theta$ or $\tau$.  When we include the 
effect of \leqn{eq:c5}, our previous relation \leqn{eq:g3plus} is shifted
to 
\beq
 \theta \rightarrow \theta - (4-N_f)\alpha\ , \quad \mbox{\rm or} \quad \tau
\rightarrow \tau -\frac{4-N_f}{2\pi}\, \alpha \ .
\eeq{eq:g3xplus}
This transformation is an exact discrete symmetry of the theory when 
$\tau$ is shifted by an integer.  Thus, the action of this symmetry on $u$
gives a $Z_2$ symmetry for $N_f = 0$, a $Z_3$ symmetry for $N_f = 1$, and a 
$Z_2$ symmetry for $N_f = 2$.  For $N_f=4$, the full $U(1)$ symmetry is
present, as should be expected for a theory with $\beta$ function equal to 
zero.

  From the $\beta$ function \leqn{eq:z4plus}, we can deduce the behavior of 
$\tau(u)$ in the weak-coupling region at large $u$.  Analogously to 
\leqn{eq:x3}, we find
\beq
\tau(u) = \frac{i}{4\pi}\, (4-N_f)\log \frac{u}{\Lambda^2}
\ .
\eeq{eq:x3again}
Following the logic of Section 5.2, we find for this case the monodromy 
matrix at infinity
\beq
   M_\infty = \pmatrix{-1 & \frac{1}{2}(4-N_f) \cr 0 & -1} \ .
\eeq{eq:g3pplus}
This formula is somewhat awkward to use
for the cases in which the matrix elements 
of $M$ are not integers.  For this reason, Seiberg and Witten change their
conventions for this case and define a rescaled $\tau$ and $\a_D$, 
\beq
{\ttau}  = 2\tau \ , \qquad \a_D = 2a_D \ . 
\eeq{eq:i5}
Then
\beq
{\ttau} = \frac{i}{2\pi}\, (4-N_f)\, \log \frac{u}{\Lambda^2}
\eeq{eq:j5}
and the new doublet $(a, \a_D)$ has monodromy
\beq
M_\infty = \pmatrix{-1 & (4-N_f) \cr 0 & -1} \ .
\eeq{eq:k5}

\subsection{More about $N_f = 0$}

For nonzero values of $N_f$, we might expect to be able to construct the 
effective coupling $\ttau$ using the method described for $N_f=0$ in 
Section 5.  That is, we consider $\ttau(u)$ to be the modulus of a torus.
We consider the points in the $u$ plane where this torus degenerates 
to be points where some particles of the theory become massless.
The we determine the geometry of the tori as a function of $u$ by 
finding the analytic function $\ttau(u)$ consistent with these 
singularities.  This program is carried out in detail in \cite{SWII}.
In these notes, I would like to focus on the cases $N_f = 1, 2$ to 
call attention to some interesting physical features of the solution.

As a point of reference for these cases, however, we should first rewrite
the solution for $N_f = 0$ in the new notation.  For $N_f= 0$, 
$\ttau$ goes through
\beq
         \ttau \to \ttau - 4
\eeq{eq:l5}
as $u \to e^{2\pi i} u$.  Thus, the $\ttau$ plane is a double cover of
the $u$ plane and also of the shaded region in Figure \ref{fig:seven}.
Seiberg and Witten suggest that we can parametrize these tori by writing,
instead of \leqn{eq:r4}, the family of cubic polynomials
\beq
y^2 = x^3 - u x^2  + \frac{1}{4}\, \Lambda^4x \ ;
\eeq{eq:n5}
where  I have $u_0 = \Lambda^2$.  Since the magnitude of $u_0$ is given
by the nonperturbative scale of the theory, the quantity $\Lambda^2$
defined in this way is equal to that in \leqn{eq:x3again} up to an overall
constant.  This notation will be useful to us when we study the decoupling
relation of the solutions for different values of $N_f$. 

To see that this new family of cubics gives the same physics as \leqn{eq:r4},
we should study its singularities.
The cubic \leqn{eq:n5} has its zeros at
\beq
x  = 0\ ,  \qquad x = x_\pm =
 \frac{1}{2} \left( u -  \sqrt{u^2-\Lambda^4}\right) \ . 
\eeq{eq:o5}
Define $z_1$ and $z_2$ as in \leqn{eq:t4} where the contours $C_1$ and 
$C_2$ wrap around $(0,x_-)$ and $(x_-, x_+)$, respectively, in the manner
indicated in Figure \ref{fig:six}. The singular tori occur where pairs of
zeros \leqn{eq:o5} coincide. This happens at $u = \pm \Lambda^2$ (that is,
at $u = \pm u_0$), and at $u = \infty$.  As $u\to \infty$, it is easy to 
directly evaluate the integrals and see that the formula 
$\ttau = z_2/z_1$ reproduces \leqn{eq:j5} with $N_f = 0$. In particular,
\beq
 z_2 \sim 2i \int^u_{\Lambda^4/4u} {dx\over x} {1\over \sqrt{u}} \sim
           {2i\over \sqrt{u}} \cdot 2\log{u\over \Lambda^2} \ ;
\eeq{eq:o5plus}
this accounts for the extra factor of 2 in $\ttau$.
  As $u$ makes a 
complete circle around the point $\Lambda^2$, one can observe that 
the two zeros $x_\pm$ exchange places.  By playing with the contours, it is
not hard to see that this leads to the monodromy
\beq
     a \to a - \a_D \ , \qquad \a_D \to \a_D \ .
\eeq{eq:h4oagain}
which is the correct transcription of  \leqn{eq:h4oplus} for 
$(\a_D,a)$.

For the cubic \leqn{eq:n5}, and for other cubics that we will encounter in
this section, it is not immediately obvious which values of $u$ correspond to
singular tori.  For this case, we could find these values by solving a 
quadratic equation.  A more generally applicable procedure is to compute
the {\em discriminant} $\Delta$.  If $e_1, e_2,  e_3$ are the three roots 
of the cubic, $\Delta$ is defined by
\beq
          \Delta  = \prod_{i<j} (e_i-e_j)^2 \ .
\eeq{eq:o5xplus}
On the other hand, for a cubic polynomial
\beq
       x^3 + B x^2 + C x + D
\eeq{eq:o5yplus}
it is straightforward to show that
\beq
        \Delta = B^2 C^2 - 4 C^3 - 4 B^3 D + 18 BCD - 27 D^2 \ .
\eeq{eq:o5zplus}
For \leqn{eq:n5}, we find $\Delta = (u^2 - \Lambda^4)\Lambda^8/16$. The 
singular tori occur where two zeros of the cubic collide, that is, at 
the zeros of $\Delta(u)$.  Thus, the discriminant easily picks out the 
singularities at $u = \pm \Lambda^2$ found above.

\subsection{The tori for $N_f = 1, 2$}

In this section, I will review the generalization of the structure
just described for $N_f = 0$ to nonzero $N_f$.  I will work through
explicitly the two simplest cases, $N_f =1$ and $2$.

As a step toward generalizing to nonzero $N_f$, consider
first the consequence of adding massive matter fields to the theory.
Hypermultiplets of $N=2$ Yang-Mills theory can recieve mass from a 
superpotential term 
\beq
           \Delta W = m_i \bar Q_i Q_i
\eeq{eq:p5plus}
 which preserves the full supersymmetry.  When all of the mass parameters
$m_i$ are large, we must recover the $N_f= 0$ solution for $\ttau(u)$
just discussed.  On the other hand, $\ttau(u)$ must depend holomorphically
on the $m_i$.  So it is natural that, for $N_f$ nonzero, the effective 
coupling $\ttau(u)$ is still the modulus of a torus whose geometry is a
holomorphic function of $u$ and the $m_i$.

We can identify these tori by constructing the associated cubic 
polynomials $y(x)$.   The $U(1)$ symmetry \leqn{eq:b5} provides a 
useful tool in constructing these polynomials.
So far in this discussion, we have been thinking of this transformation
as an anomalous global symmetry.
However, as in Section 4, we can supplement  this transformation
by a shift of the theta parameter, $\tau\to \tau + (4-N_f)\alpha/2\pi$
and consider it as an exact global $U(1)$ symmetry.  Under this 
transformation, $u$ has charge 2 and $\Lambda^{b_0}$ has charge
$(4-N-f)$, giving $\Lambda$ charge 1 for any $N_f$.  The cubic $y(x)$
should have a definite transformation property under this symmetry.
In fact, the following set of charge assignments make the $N_f = 0$ cubic 
\leqn{eq:n5} covariant
under this $U(1)$:
\beq
  u\ : \ 2 \ , \quad \Lambda\ : \ 1 \ , \quad
  x\ : \ 2 \ , \quad y\ : \ 3 \ .
\eeq{eq:o5aplus}
If we obtain the $N_f=0$ torus from a torus with nonzero $N_f$ by 
holomorphic decoupling, and we are careful to give  masses to
 the matter fields in a way
that preserves the $U(1)$ symmetry, we should expect these charge assignments
to hold for the tori we will find for nonzero $N_f$. The $U(1)$ symmetry
will be respected by the mass terms if the masses  $m_i$ are assigned a charge
which compensates the rotation \leqn{eq:c5}, that is
\beq
  m_i\ : \ 1 \ .
\eeq{eq:o5bplus}.

For very large values of $u$, $\ttau$ must have the asymptotic behavior
\leqn{eq:x3again}.  It is interesting to ask how the $N_f=0$ solution
joins on to this behavior.  Consider the situation in which all of the
$m_i$ are much greater than the effective $\Lambda$ of the theory with
the matter fields decoupled.  Then for $\Lambda^2 \ll |u| \ll m_i^2$, 
$\ttau$  will have the singularity \leqn{eq:l5}.  However, at large values of
$u$ we encounter a new singularity.  The full superpotential for the $i$th
flavor is
\beq
           \Delta W =  m_i \bar Q_i Q_i + 2 \bar Q_i\phi^a t^a Q_i \ , 
\eeq{eq:p5pplus}
so that when $\VEV{\phi^3} = \mp m_i$ or $u = m_i^2$,
 a pair of matter fields has zero mass.  Since these massless fields are 
charged under the unbroken $U(1)$ gauge symmetry, they renormalize the 
effective coupling toward zero.   In fact, we find that, near this point,
\beq 
          \ttau \sim   \frac{i}{2\pi} \log \frac{u - 2m_i^2}{\Lambda^2} \ . 
\eeq{eq:p5xplus}
For larger values of $|u|$, $\ttau$ shifts by one fewer unit as the phase 
of $u$ goes from 0 to $2\pi$.  In a theory with $N_f$ flavors of massive
matter fields, we will eventually pass $N_f$ of these singularities and 
recover the asymptotic behavior \leqn{eq:l5}.

\begin{figure}
\begin{center}
\leavevmode
\epsfbox{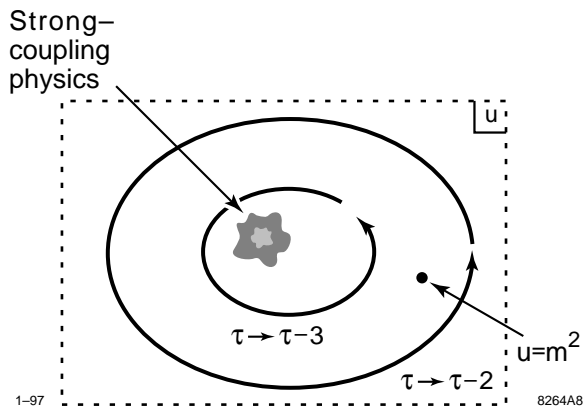}
\end{center}
 \caption{Monodromy of \boldmath$\tau$\unboldmath
 in the weak-coupling region for a theory 
      with a large mass $m$ for one flavor.}
\label{fig:eight}
\end{figure}

A similar effect occurs when we consider the decoupling of a single flavor
from a theory with nonzero $N_f$.  Consider, for definiteness, the theory 
with $N_f = 2$.  Asymptotically in $u$, $\ttau$ shifts by 2 units as the 
phase of $u$ is increased from 0 to $2\pi$.  However, if one flavor is light
and one is heavy, we find the situation shown in Figure \ref{fig:eight}.
At small values of $u$, there is a region which exhibits strong-coupling 
dynamics.  When $u$ is carried around this region, $\ttau$ shifts by 3 units.
At large $u$, there is an additional singularity which changes the shift
in $\ttau$ to the step of 2 units required by \leqn{eq:l5}.

With this orientation, we can try to obtain the family of tori which 
describe the theory for $N_f =2$ massless flavors.  By decoupling the two
flavors one at a time, we should obtain the $N_f =1$ and $N_f = 0$ theories.

The problem of finding the tori of $N_f=2$ has several features in common
with the problem we solved in the previous section for $N_f = 0$.  The 
theory has a $Z_2$ symmetry acting in $u$.  The behavior of $\ttau$ at 
infinity is just that which we required for $\tau$ in \leqn{eq:x3}.

We can try to find strong-coupling singularities of $\tau(u)$ associated
with the magnetic monopoles of the theory coming down to zero mass.
The global symmetry of the theory is $SO(4) = SU(2)\times SU(2)$. The
monopoles have zero modes for the fermionic partners of $Q_i$ and $\bar Q_i$;
when we consider the multiplet of states in which these zero modes are
filled or empty, the monopoles form spinor representations of the global
symmetry group.  The monopoles with even electric charge become  $(2,1)$ 
multiplets of
of $SU(2)\times SU(2)$; the monopoles with odd electric charge become
$(1,2)$ multiplets.  The simplest $Z_2$-invariant set of singularities is one
in which a $(2,1)$ multiplet of monopoles becomes massless at $u= \Lambda^2$
and a $(1,2)$ multiplet becomes massless at  $u= -\Lambda^2$.  Since two
pairs of monopoles are becoming massless at each of these points, we find a 
singularity in $\tau$ twice as strong as that in \leqn{eq:f4}, 
 \beq
 - \frac{1}{\ttau(u)} = - \frac{1}{2\tau(u)} =  \frac{-i}{\pi}\, \log(u-u_0)\ .
\eeq{eq:f4again}

From this data, we see that the requirements on the function $\ttau(u)$ for 
$N_f =2$ are precisely those which we found in the previous section for 
$\tau(u)$ in the case $N_f = 0$.  Thus, the effective coupling constant
$\ttau$ in this case  is
given by the family of tori associated with 
\beq
y^2 =  (x-\Lambda^2)(x+\Lambda^2)(x-u) \ ,
\eeq{eq:r4again}
just as in Section 5.4.  Notice that, with the $U(1)$ charge assignments
given in \leqn{eq:o5aplus}, this polynomial transforms covariantly with 
charge 6, as we would expect.

From this solution for $N_f =2$, we can decouple one flavor to find the 
solution for $N_f =1$.  First of all, we must determine how the mass 
perturbation affects the polynomial \leqn{eq:r4again}.  For small $m_2$
(and so, formally, for all $m_2$), the mass of a matter field
does not affect the coupling constant renormalization in perturbation theory.
This mass can enter, however, through nonperturbative corrections.
For small $m_i$, these are given by instanton effects. According
 to \leqn{eq:r2xplus}, each instanton brings with it a power of
 $\Lambda^{b_0}$. For $N_f =2$, the one-instanton amplitude is zero unless
we saturate the zero modes by supplying masses for both flavors.  Thus, the
leading effect comes from a 2-instanton contribution.  This term is 
proportional to $m_2^2 \Lambda^4$, and this term saturates the allowed 
$U(1)$ charge.  Thus, the most  general cubic possible for the $N_f =2$
theory with one massive flavor is
\beq
y^2 = (x-\Lambda^2)(x+\Lambda^2)(x-u)- c m_2^2\Lambda^4 \ .
\eeq{eq:x5}
where $c$ is a constant to be determined.  This constant can be fixed
in the following way.  We have
argued that, when $m_2 \gg \Lambda$,  we must find a singular torus when 
$u = m_2^2$. The discriminant of \leqn{eq:x5} is given by
\beq
   \Delta = (u - c m_2^2) (4 u^3 \Lambda^4 - 27(u-cm_2^2)\Lambda^8) + \cdots\ ,
\eeq{eq:x5plus}
where the omitted terms are negligible for $m_2 \gg \Lambda$.  Thus, we find
a singular torus for $u = cm_2^2$ and no other singularities except in the 
region $u \sim \Lambda^2$. This implies that  $c = 1$.

Having now determined the polynomial for $N_f = 2$ and one flavor massive,
we can find the polynomial for $N_f =1$ by holomorphic decoupling.
Take $m_2 \to \infty$, while keeping the $\Lambda$ parameter of the effective
1-flavor theory fixed.  According to \leqn{eq:r2}, this is given by
\beq
\left(\Lambda^{3}\right)_{\eff,N_f-1} = m_2\,
\left(\Lambda^{2}\right)_{N_f} \ ,
\eeq{eq:r2x5}
so we must take $\Lambda\to 0$ as $m_2\to \infty$ in such a way that the 
right-hand side of \leqn{eq:r2x5} is fixed.
Then, the family of tori for $N_f = 1$ are given by 
\beq
y^2 = x^2(x-u)- \Lambda^6 \ ,
\eeq{eq:y5}
where I have written the new
effective QCD scale simply as $\Lambda$.

As a first check, the formula \leqn{eq:y5} has the correct $U(1)$ charge.
To understand this polynomial more fully, we might compute its 
discriminant:
\beq
\Delta = - 4u^3\Lambda^6 -27\, \Lambda^{12} \ .
\eeq{eq:d6}
Pairs of zeros collide when
\beq
u = \left(- \frac{27}{4}\, \Lambda^6\right)^{1/3} \ .
\eeq{eq:e6}
There are three cube roots, and so we find three singularities in a
$Z_3$-symmetric pattern.   This realizes the $Z_3$ symmetry that we
predicted below \leqn{eq:g3xplus}.  

Another check of \leqn{eq:y5} 
is given by decoupling the remaining flavor. If we wish to add to 
\leqn{eq:y5} a term proportional to one power of $m_1$ and one
instanton factor $\Lambda^3$ in a way consistent with the $U(1)$ symmetry,
the only possibility is
\beq
y^2 = x^2(x-u)- \Lambda^6  - m\Lambda^3(a x + b u) \ ,
\eeq{eq:y5plus}
where $a$ and $b$ are to be determined.  Note that higher powers of 
$(m\Lambda^3)$ have a $U(1)$ charge higher than 6.  Computing the discriminant,
we can see that there is a singular point at $u = m_1^2$ only if $a=2$ 
and $b=0$.  Then the polynomial corresponding to $N_f=1$ with a  nonzero
mass is
\beq
y^2 = x^2(x-u)- \Lambda^6  - 2 m\Lambda^3 x\ .
\eeq{eq:y5pplus}
If we let $m_1\to \infty$, we find a theory with zero flavors and the effective
QCD parameter
\beq
\left(\Lambda^{4}\right)_{\eff,0} = m_2\,
\left(\Lambda^{3}\right)_{1} \ .
\eeq{eq:r2x5again}
The polynomial which characterizes this situation is 
\beq
y^2 = x^2(x-u) - 2 \Lambda^4 x\ .
\eeq{eq:y5xplus}
which agrees with \leqn{eq:n5} after a permitted
 constant rescaling of $\Lambda$.

Now that we understand the transition from the $N_f=2$ theory to the 
$N_f = 1$ theory at a technical level, it is worth thinking a bit more
about the physics of this transition.  In each of these problems, the 
effective coupling has singularities at specific points in the 
moduli space of $u$ where magnetic monopoles become massless.  In the 
$N_f =2$ theory, there were two such points, at each of which two 
monopole-antimonopole pairs become massless.
 In the $N_f =1$ theory, there were three points, at each of
which one monopole-antimonopole pair becomes massless.  As in the 
$N_f=0$ cases analyzed in the previous section, the monopoles which 
become massless at each point differ in their electric charges.  In the 
$N_f = 1$ case, where $\ttau$ goes through 3 units as $u$ increases its 
phase by $2\pi$, the monopoles which become massless are those which 
begin at $u$ real with the electric charges 0, 1, 2.

\begin{figure}
\begin{center}
\leavevmode
{\epsfxsize=4.5in
\epsfbox{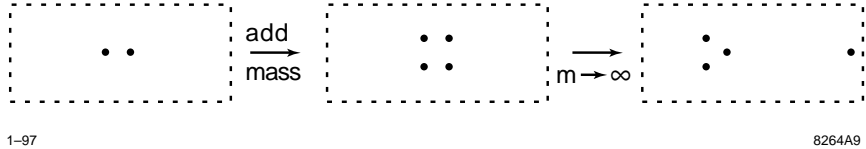}}
\end{center}
 \caption{Motion of the singularities of \boldmath$\tau$\unboldmath$(u)$
 for the Seiberg-Witten model
      with $N_f =2$ as a mass $m$ is turned on for one flavor.}
\label{fig:nine}
\end{figure}

The transition from the set of $u$-plane singularities for $N_f=2$ to that
for $N_f=1$ is shown in Figure \ref{fig:nine}.  Something strange is 
happening here.  At zero mass, we have two singularities in $Z_2$-symmetric
locations.  Under a small mass perturbation, these break up into four 
singularities, each of which corresponds to a point where one 
monopole-antimonopole pair becomes massless.  As the mass is increased, 
one of these points runs out to infinity, while the other three organize
themselves into the $Z_3$-symmetric structure required for $N_f =1$.
But when the fourth singularity comes out into the weak-coupling region, it
has the interpretation of a point at which an elementary matter field
becomes massless.  So apparently, we can pass continuously, in the 
Seiberg-Witten solution, between solitons of the theory and 
elementary particles.  This is an extreme, but perfectly permissible, 
example of the continuous connection of phases which would seem to be 
distinguished qualitatively.

The $N_f=1$ theory has one more very interesting feature. Starting from 
\leqn{eq:y5}, take the limit $\Lambda\to 0$.  The three points where 
monopole pairs become massless then approach one another and coalesce.
We obtain a theory with a singularity at $u=0$ at which 
monopoles with electric charge 0, 1, and 2 simultaneously become massless.
This is a quite unusual situation, because these three species are 
mutually nonlocal.  This general situation, in which nonlocal species are
simultaneously massless, is called an {\em Argyres-Douglas point}.\cite{AD}
The particular points
of this type in $N=2$ $SU(2)$ Yang-Mills theory have been analyzed in
 detail by 
Argyres, Plesser, Seiberg, and Witten,\cite{ADPT} who give evidence that they  
are new nontrivial scale-invariant field theories and compute some of the 
scaling dimensions of operators.

  There is one more aspect of the $SU(2)$ gauge theories which I have no 
space to discuss here.  For the case $N_f =4$, the $\beta$ function of the 
theory vanishes.  This case would then have zero coupling constant 
renormalization and might also be expected to have exact strong-weak-coupling
duality (`$S$-duality'). 
 Seiberg and Witten
argue that this case can be described by a 
family of tori described by a cubic which transforms covariantly under the 
$SL(2,Z)$ $S$-duality group.\cite{SWII} More concretely, they find
\beq
      y^2 = 4x^3 -  g_2(\ttau) x - g_2(\ttau) \ .
\eeq{eq:z5}
where $g_2$ and $g_3$ are the unique modular forms of weights 4 and 6 under
$SL(2,Z)$. I refer you to their paper for a detailed discussion of the 
$S$-duality and for a demonstration that this formula implies all of those 
given above by holomorphic decoupling.

  Unfortunately, there are still some lingering questions about the $N=2$
$SU(2)$ Yang-Mills theories.  For the case $N_f =3$, the general arguments
that I have given in this section fix the family of tori only up to one
undetermined constant, which eventually was fixed by an explicit two-instanton
computation.\cite{Mattis,Aoy}  For $N_f = 4$, it turned out that the 
explicit formula \leqn{eq:z5} was incompatible with the result of a similar
two-instanton calculation.  Presumably, this is evidence that the coupling 
constant definition  used in this calculation, the Pauli-Villars prescription,
is not invariant under $S$-duality.  The Pauli-Villars coupling would then
be related to the coupling constant definition used by Seiberg and Witten by 
an arbitrary function of $\ttau$. It would be strange and remarkable if 
$S$-duality could be exact in field theory only with the string theory 
regulator.  The precise resolution of this confusion, though, is still not
clear.

\subsection{Larger gauge groups}

To conclude this section, I would like to comment briefly on the generalization
of the Seiberg-Witten theory to larger gauge groups. 

The same analysis that predicted a Coulomb phase of the $SU(2)$ gauge theory
applies to any gauge group.  Quite generally, we find  a vacuum 
state of the classical theory by solving the $D$-flatness condition
\leqn{eq:e3}.  The matrix $\VEV\phi$ can be diagonalized; for
 example, for  $G_c = SU(N_c)$ we have
\beq
\VEV{\phi} = \left(\begin{array}{lcr} \phi_1 &&\\
& \ddots & \\ & &\phi_{N_c}\end{array}\right)\ ,
\eeq{eq:h6}
where $\phi_1, \ldots, \phi_{N_c}$ are
 complex parameters such that $\sum_i\phi_i
=0$.  At a generic point where no pair of the $\phi_i$ are equal, this
expectation value breaks the gauge group $G_c$ down to $(U(1))^r$, where $r$
is the rank of $G$.  For the case of $SU(N_c)$, we find a product of $(N_c-1)$
$U(1)$ gauge groups.  These vacua remain supersymmetric minima in the 
quantum theory.  They  are described by the effective Lagrangian
\beq
\L_\eff = \frac{-i}{16\pi}\int d^2\theta\ \tau^{ij}(\phi)\ \W^{\alpha i}
\W^i_\alpha  + h.c.\ ,
\eeq{eq:j6}
where $i,j$ are summed over $1,\ldots, r$.   The effective couplings form 
an $r\times r$ matrix, which depends on gauge-invariant functions of $\phi$.
If we gauge-fix to configurations of the form \leqn{eq:h6}, $\tau$ must 
still be invariant under all permutations of the eigenvalues $\phi_i$.
  If the gauge boson kinetic energy term in \leqn{eq:j6}
is to be positive, the matrix $\tau$ must satisfy
\beq
\mbox{Im}\, \tau > 0 
\eeq{eq:k6}
as a matrix.

\begin{figure}
\begin{center}
\leavevmode
\epsfbox{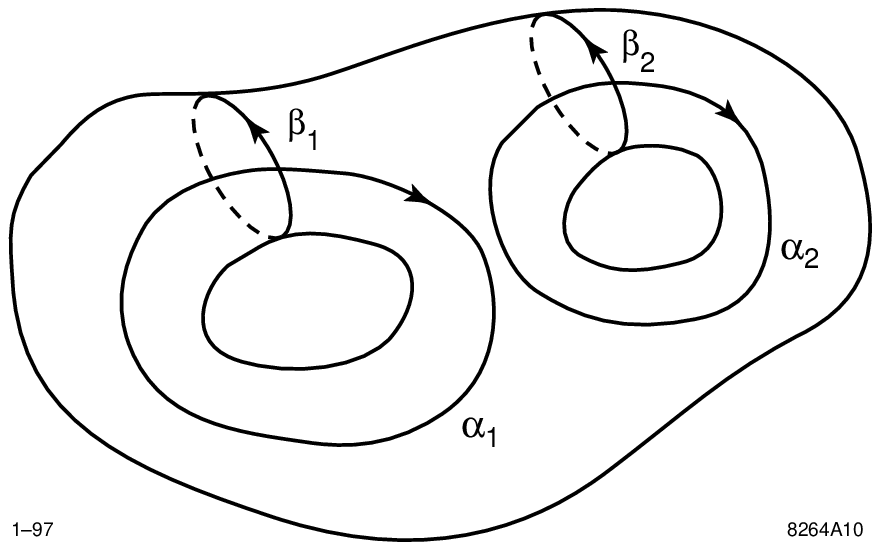}
\end{center}
 \caption{Complementary cycles on a surface of genus 2.}
\label{fig:ten}
\end{figure}

This condition is naturally satisfied if $\tau$ is the {\em period matrix}
of a 2-dimen\-sion\-al surface of genus $g = r$.  
This object is defined as follows. A surface of genus $g$ can be 
characterized by pairs of complementary cycles $\alpha_i$, $\beta_i$, 
$i=1,\ldots, g$, as shown in Figure \ref{fig:ten}.  Alternatively, such a
surface can be characterized by $g$ independent holomorphic differentials
$\lambda_\ell$.  These objects have a complementary relation; the differential
$\lambda_i$ integrated around the cycle $\alpha_i$ or $\beta_i$ gives a nonzero
result.  More generally, define
\beq
 A^i_\ell = \oint_{\alpha_i}\lambda_\ell\,  \qquad B_{j\ell} =
\oint_{\beta_j}\lambda_\ell \ .
\eeq{eq:k6plus}
 Then the period matrix of the genus $g$ surface is given by
\beq
\tau = B\, A^{-1} \ .
\eeq{eq:l6}
This generalizes the formula \leqn{eq:t4plus} for the modulus $\tau$ of 
a torus.

The most direct generalization of the construction in Section 5.4 would 
associate $\tau$ with a 2-dimensional surface defined by an integral
\beq
z = \int^x_{x_0}\frac{dx}{y}\ ,
\eeq{eq:m6}
where $y^2(x)$ is a polynomial.  If this polynomial has degree $n = 2g+2$, 
$y(x)$ is a double-sheeted surface with $(g+1)$ branch cuts; this is a 
surface of genus $g$.  (It is equivalent to write a polynomial 
of degree $n= 2g+1$; this puts 
one branch point at $\infty$.)

A surface constructed in this way is called
`hyperelliptic'.  All surfaces of genus 1 and 2 are equivalent to 
hyperelliptic surfaces by general coordinate and 
conformal transformations, but the set of 
hyperelliptic surfaces is a smaller and smaller subspace of the space of all
2-dimensional surfaces at higher genus.

Nevertheless, it was shown by Argyres and Faraggi\cite{AF} and by Klemm, 
Lerche, Thiesen, and Yankielowicz\cite{KLTY} that the Seiberg-Witten
problem for more general gauge groups is solved by a particular class of 
hyperelliptic surfaces.   For $G_c = SU(N_c)$, these surfaces can be 
constructed easily by 
generalizing the $U(1)$ symmetry described in \leqn{eq:o5aplus}.  For 
the  $N=2$ $SU(N_c)$ gauge theory with $N_f$ flavors of hypermultiplets 
in the fundamental representation, the $U(1)$ symmetry of the theory is
\beq
         \phi \to e^{i\alpha} \phi\ , \qquad 
\tau\to \tau + (2N_c-N_f)\alpha/2\pi \ .
\eeq{eq:n6}
Since the first $\beta$ function coefficient is
given by $b_0 = (2N_c - N_f)$, we find from \leqn{eq:g1pplus} that
$\Lambda$ has charge 1. The one-instanton amplitude is proportional to 
$\Lambda^{2N_c-N_f}$.

Consider first the pure $N=2$ $SU(N_c)$ gauge theory, $N_f = 0$.  Introduce
a variable $x$ with charge 1 under the $U(1)$ symmetry. (This is effectively
the square root of $x$ in Section 6.2.)  Then consider the polynomial
\beq
      y^2 = \prod_i (x-\phi_i)^2  - \Lambda^{2N_c} \ .
\eeq{eq:n6plus}
This object is covariant under the $U(1)$ and totally symmetric in the 
$\phi_i$.  The QCD scale $\Lambda$ enters as the one-instanton factor.
Thus, it is a reasonable candidate for the polynomial we are seeking.
For the case $SU(2)$, we may set $\phi_1 = - \phi_2 = \phi$, with 
$\phi^2 = u$.  Then \leqn{eq:n6plus} takes the form
\beqa
       y^2 &=& (x^2-\phi^2)^2 - \Lambda^4 \CR
           &=& (x + \sqrt{u+\Lambda^2})(x - \sqrt{u+\Lambda^2})
(x + \sqrt{u-\Lambda^2})(x - \sqrt{u-\Lambda^2}) \ .\qquad
\eeqa{eq:n6pplus}
The pairs of zeros coalesce at $u = \pm\Lambda^2, \infty$. This is in fact
another representation of the family of tori discussed in Section 5.
For more general $SU(N_c)$ groups, it is not difficult to check that
\leqn{eq:n6plus} has the correct decoupling limit near points in the 
moduli space where an $SU(2)$ subgroup of the gauge group is manifest at 
low energy.  If the vacuum expectation value of $\phi$ preserves an 
approximate $SU(2)$ symmetry, then two eigenvalues of $\phi$ are almost
equal.  Call these $i=1,2$ and write
\beq
 \phi_i = \phi + \chi\ , \quad   \phi_2 = \phi - \chi\ , \quad
   \hat x = x - \phi \ .
\eeq{eq:n6xplus}
Then \leqn{eq:n6pplus} becomes
\beq
     y^2 = (\hat x^2 - \chi^2) \prod_{i>2} (\phi - \phi_i)^2 - \Lambda^{2N_c}
                \ . 
\eeq{eq:n6yplus}
The factors $(\phi-\phi_i)^2$ are the vacuum expectation value which give
mass to  the off-diagonal vector bosons when $SU(N_c)$ is broken to $SU(2)$.
Using the analogue of the relation \leqn{eq:j2}, we can write
this equation in the form \leqn{eq:n6pplus}
in terms of the effective $\Lambda$ parameter of the $SU(2)$ theory.
Additional checks of the formula \leqn{eq:n6plus} are given in \cite{AF,KLTY}.

The analogous polynomial representing the family of surfaces for the 
Cou\-lomb phase of $SU(N_c)$ Yang-Mills theory with $N_f$ flavors, 
$N_f \leq N_c$, is\cite{HO,APS,NM}
\beq
      y^2 = \prod_i (x-\phi_i)^2  - \Lambda^{2N_c-N_f}\prod_f(x-m_j) \ ,
\eeq{eq:n6zplus}
where $i = 1, \ldots, N_c$ and $j= 1, \ldots, N_f$.  The term proportional
to one power of each mass $m_j$ is also proportional to the one-instanton
factor. The full expression \leqn{eq:n6zplus} returns to \leqn{eq:n6} 
when we decouple the massive hypermultiplets.  For $N_f > N_c$, there are
additional ambiguities of the type discussed above for $SU(2)$ gauge 
theories with $N_f = 3$ and 4.

In the moduli space of larger $SU(N_c)$ gauge groups, there are many 
families of magnetic monopoles, and thus there are many opportunities 
for Argyres-Douglas points where mutually nonlocal species becomes 
simultaneously massless.  The original example of Argyres and Douglas
was given for  the case of SU(3).\cite{AD}

Finally, I should note that the  Seiberg-Witten construction appears
in a natural way in considerations of superstring duality.  Kachru and 
Vafa\cite{KV} considered examples of heterotic string compactifications on
$K3\times T^2$.  These theories have $N=2$ space-time supersymmetry, and 
one can find examples  which give rise to effective $SU(2)$ Yang-Mills theories
at low energies. 
These theories are dual to Type IIA theories compactified on certain
Calabi-Yau manifolds.  And, indeed, the dual theory exhibits 
the moduli space of the Seiberg-Witten model. The systematics of this
phenomenon has been explored further in \cite{KKLMV,KLMVW}. 
 A review of 
this set of developments has been given in \cite{Lerche}.  More recently,
Sen\cite{Sen} and Banks, Douglas, and Seiberg\cite{probe} have shown 
that the Seiberg-Witten moduli space arises also from the 
consistency conditions for embedding 3-branes in well-chosen
Type IIB compactifications.

\section{Seiberg's Non-Abelian Duality}
\label{sec:e}

In Section 4, I discussed the behavior of strongly-coupled supersymmetric
$SU(N_c)$
Yang-Mills theory for values of the number of flavors $N_f$ from 0 to $N_c$.
It is now time that we returned to this theory and continue to explore its
properties, considering still larger numbers of flavors.  Seiberg found
a compelling picture for  the behavior of supersymmetric QCD
 in this regime.\cite{SnA} 
This picture includes a region in which 
the non-Abelian gauge symmetry is unbroken but nevertheless is realized in a 
Coulomb phase.  Seiberg argued that this region is dual to a similar 
non-Abelian Coulomb phase of a different $SU(N_c)$ gauge theory, thus 
generalizing the familiar Abelian electric-magnetic duality.

\subsection{More about  $N_f = N_c$}

To extend the picture of Section 4 to higher values of $N_f$, I would like
to begin by clarifying one aspect of the physical picture  for  $N_f = N_c$
which we discussed in Section 4.5.  I argued there that, in this case, 
supersymmetric QCD had a manifold of degenerate, supersymmetric vacuum 
states.  These vacua were parametrized by the gauge-invariant fields
$T$, $B$, and $\bar B$, subject to the Seiberg's constraint \leqn{eq:y2}.
Oscillations of the scalar components of these fields which satisfy the 
constraint correspond to local fluctuations along the manifold of vacuum
states.  Thus, they are massless composite bosons.  By supersymmetry, the 
fermionic components of these fields are then massless composite fermions.

The question of whether relativistic fermions can be tightly bound into 
massless composite states is obviously a fundamental issue in quantum field 
theory.  The question of whether massless fermionic bound states are possible
is also a matter of phenomenological relevance for people who would like to 
construct composite models of quarks and leptons.  Some time ago, `t~Hooft
proposed a general consistency condition on massless fermionic composite
states which has turned out in practice to be very stringent.\cite{tH}
  I would now like to
introduce 't Hooft's criterion and then check whether it is satisfied by 
the the physical picture we have built for supersymmetric QCD with $N_f = N_c$.

\begin{figure}
\begin{center}
\leavevmode
\epsfbox{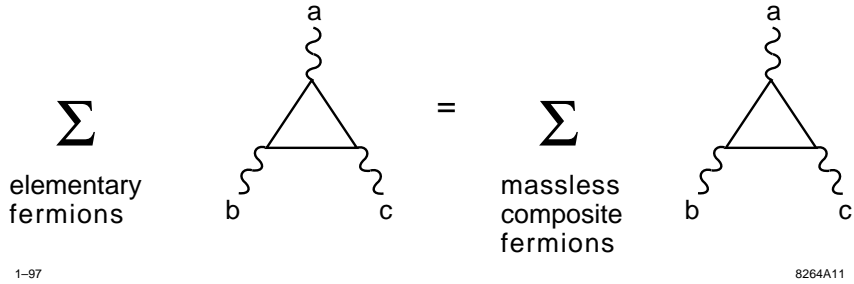}
\end{center}
 \caption{The 't Hooft anomaly matching condition.}
\label{fig:eleven}
\end{figure}

Consider, then, a Yang-Mills theory 
with gauge group $G_c$ coupled to some matter fields.  Let the 
continuous global symmetry of this theory be $G$.  Let $J_\mu^a$,  $J_\mu^b$,
 $J_\mu^c$ be three currents of the global symmetry.  Typically, the
product of these  
currents will have a nonzero axial vector anomaly, which can be evaluated
at short distances by computing the 
triangle diagram of the three currents, summed 
over all elementary matter fermions in the loop.  When we consider the theory
at low energies, the three corresponding symmetries may be spontaneously 
broken, or they may be exact symmetries of the vacuum.  If they are exact
symmetries, we can assign the massless particles in the theory definite
quantum numbers under these symmetries, and we can compute the triangle
diagram summing over the massless fermions of the effective low-energy theory.
't Hooft claimed that the anomaly computed in this way must agree with the 
anomaly obtained from the short-distance calculation using the elementary 
fields.  This is the {\em 't Hooft anomaly matching condition}.  The condition
is illustrated in 
Figure \ref{fig:eleven}.

The proof of this condition given by 't Hooft is very simple.  Add to the 
theory weakly-coupled vector bosons which gauge the global symmetry $G$, and 
add massless fermions which are neutral under $G_s$ (call them `leptons')
as necessary to 
cancel the $G$
gauge anomalies.  We have now defined a consistent gauge theory. The effective
theory at low energies should also be a consistent gauge theory of $G$.
But in this theory, the `leptons' have the same nonzero $G$ anomalies, and 
these must be cancelled by contributions of 
the physical massless fermions arising from the $G_s$ theory at low energy.
Note that {\em massive} fermions must be vectorlike under unbroken global
symmetries, so these do not contribute at all to  anomaly matching.
A more formal proof of the anomaly matching condition, which uses dispersion
relations to connect the low- and high-energy evaluations of the anomaly,
has been given in \cite{BFSY}.

Since our picture of the the behavior of supersymmetric QCD with $N_f= N_c$
contains massless composite fermions, it can only be consistent if these
satisfy the 't Hooft anomaly condition.  The original global symmetry of the
model is
\beq
G = SU(N_f)\times SU(N_f)\times U_B(1) \times U_R(1)  \ ,
\eeq{eq:r6}
In this section, the symbol $R$ will always refer to the anomaly-free
$R$ symmetry \leqn{eq:w1}; however, for $N_f = N_c$, this
 coincides with the canonical $R$ symmetry.  At a typical point in the 
moduli space, however, this symmetry is broken all the way to  $U_R(1)$ by 
vacuum expectation values of the fields $T$, $B$, and $\bar B$.  But there
are certain special points of maximal symmetry where a large part of $G$
remains unbroken.  At these points, the 't Hooft condition is especially
strong.

One set of expectation values which 
satisfies the constraint \leqn{eq:y2} and leads to a point of maximal 
symmetry is:  $T = \Lambda^2 \cdot \One$, 
$B = \bar B = 0$.  At this point, $G$ is broken to 
\beq
SU(N_f) \times U_B(1) \times U_R(1) \ .
\eeq{eq:s6}
Under this subgroup,  the elementary fermions have the quantum numbers:
\beq
\psi_Q\,:\,(N_f)_{1,-1} \qquad \psi_{\bar Q}\, :\, (\bar N_f)_{-1,-1}
\qquad \lambda \,:\, (1)_{0,+1} \ .
\eeq{eq:t6}
Among the composite fermions, we may eliminate the superpartner of $\tr\, T$
using the constraint \leqn{eq:y2}. The remaining fermionic partners $\psi_T$
form an adjoint representation of $SU(N_f)$.  The quantum numbers of the
physical composite fermions under \leqn{eq:s6} are then
\beq
\psi_T\,:\,(N^2_f-1)_{0,-1} \qquad \psi_B\,:\, (1)_{N_f,-1}\qquad
\psi_{\bar B}\,:\, (1)_{-N_f,-1} \ .
\eeq{eq:u6}

From these sets of quantum numbers, we can compute the anomaly coefficients
directly.  Recall the group theory
coefficient $C(r)$ defined below \leqn{eq:h1}
equals $\half$ in the fundamental representation of $SU(N_f)$ and equals
$N_f$ in the adjoint representation.  Similarly, let $A d^{abc}$ be the value 
of the anomaly of three $SU(N_f)$ currents due to a chiral fermion in the
fundamental representation.  Then, for example, the $(SU(N_f))^2 U_R(1)$
anomaly coefficient from the elementary fields $\psi_Q$ and 
$\psi_{\bar Q}$ in \leqn{eq:t6} equals 
$2 \cdot N_c \cdot C(N_f) \cdot (-1) = - N_f$, while the anomaly coefficient
from the composite fields comes only from $\psi_T$ and equals $C(G) \cdot 
(-1) = - N_f$.  The full set of nonvanishing
anomaly coefficients in the theory is
\beq
\begin{array}{rcc}
 & {elementary} & {composite}  \\[1.5ex]
\left(SU(N_f)\right)^2 U_R(1)\, :\ & -N_f & - N_f  
                                   \\[1ex]
\left(U_B(1)\right)^2U_R(1)\, :\   & -2N_f        & -2N_f  \\[1ex]
\left(U_R(1)\right)^3\,:\ & - (N_f^2 + 1)  & - (N_f^2 + 1) 
\end{array}
\eeq{eq:v6pre}
I have used $N_f = N_c$.  The last line is the sum of the cubes of the $U_R(1)$
charges of all of the chiral fermions.  The 't Hooft argument also applies
to the gravitational  anomaly of $U(1)$ charges,\cite{AGW}
 and therefore we should also 
check that the trace of the $U_R(1)$ charge is the same in the elementary
and composite fermion multiplets.  This is
\beq
\begin{array}{rcc}
\tr[U_R(1)]\,:\ & -(N_f^2 + 1) & - (N_f^2 + 1)
\end{array}
\eeq{eq:w6pre}
All of the anomalies match.

A similar check can be made at another point of maximal symmetry with a 
rather different unbroken gauge group.  We can satisfy the constraint
\leqn{eq:y2} without breaking the $SU(N_f)\times SU(N_f)$ global symmetry 
at the point the in the moduli space given by the vacuum expectation values
 $T = 0$, 
$B = - \bar B = \Lambda^{N_c}$.  At this point, $G$ is broken to 
\beq
SU(N_f) \times SU(N_f) \times U_R(1) \ .
\eeq{eq:v6}
Under this subgroup,  the elementary fermions have the quantum numbers:
\beq
\psi_Q\,:\,(N_f,1))_{-1} \qquad \psi_{\bar Q}\, :\, (1,\bar N_f)_{-1}
\qquad \lambda \,:\, (1,1)_{+1} \ .
\eeq{eq:w6}
For the composite fermions, we may use the constraint to eliminate the 
constraint to eliminate $\psi_{\bar B}$ or $\psi_{B}$. The quantum numbers of
the remaining composite fermions are
\beq
\psi_T\,:\,(N_f, \bar N_f)_{-1} \qquad \psi_B\,:\, (1,1)_{-1}\qquad
\psi_{\bar B}\,:\, (1)_{-N_f,-1} \ .
\eeq{eq:x6}
The various anomalies can easily be found to be
\beq
\begin{array}{rcc}
 & {elementary} & {composite}  \\[1.5ex]
\left(SU(N_f)\right)^3\, :  & A N_f & A N_f  
                                  \\[1ex]
\left(U_B(1)\right)^2U_R(1)\, :   & - \half N_f        & -\half N_f  \\[1ex]
\tr \left[ U_R(1) \right] \, :   & - (N_f^2 + 1)  & - (N_f^2 + 1)  \\[1ex]
\left(U_R(1)\right)^3\,:\ & - (N_f^2 + 1)  & - (N_f^2 + 1) 
\end{array}
\eeq{eq:y6}
Again, the anomalies match.  So Seiberg's picture of the behavior of 
supersymmetric QCD for $N_f = N_c$ passes this unexpected and quite nontrivial
consistency condition.

\subsection{$N_f = N_c + 1$}

With this insight, we can move on to discuss the case $N_f = N_c + 1$. 
For this case, the gauge-invariant chiral superfields include $T$ and also
the baryonic superfields
\beqa
 B_i &=& \epsilon_{ij_1\cdots j_{N_c}}\epsilon_{a_1\cdots
a_{N_c}} Q_{j_i}^{a_1}\cdots Q_{j_{N_c}}^{a_{N_c}} \ , \nonumber\\[1ex]
\bar B_i &=& \epsilon_{ij_1\cdots j_{N_c}} \epsilon_{a_1\cdots a_{N_c}}
\bar Q_{j_1}^{a_1}\cdots \bar Q_{j_{N_c}}^{a_{N_c}} \ .
\eeqa{eq:z6}
where the $j_i$ are flavor indices and the $a_i$ are color indices.
The fields
$B_i$ and $\bar B_i$ transform, respectively,  as a $(\bar N_f,1)$ and a 
$(1,N_f)$
 of $SU(N_f)\times SU(N_f)$. 

 Seiberg proposed that this
system is described by the superpotential\cite{Sqmod}
\beq
W = \frac{1}{\Lambda^{b_0}}\left( \det T - B_i T^{ij}\bar B_j\right) \ . 
\eeq{eq:a7}
This expression is invariant under the global symmetry of the model and 
has charge 2 under the anomaly-free $R$ symmetry \leqn{eq:w1}. 

Holomorphic decoupling provides a more stringent test.   Add a mass term
for the last flavor, to give the superpotential
\beq
W = \frac{1}{\Lambda^{b_0}} \left(\det T-B_i T^{ij}\bar B_j\right) + m\,
T_{N_fN_f} \ .
\eeq{eq:b7}
The $F$-flatness conditions for $T_{N_f i}$, $T_{i N_f}$, $B_i$, and 
$\bar B_i$ for $i < N_f$ 
reduce $T$ $B$ $\bar B$ to the form
\beq
T = \pmatrix{ \widetilde T & 0 \cr 0 & t \cr} \qquad B = \pmatrix{ 0 \cr
             B_{N_f}\cr} \qquad \bar B = \pmatrix{ 0 \cr \bar B_{N_f}\cr} \ . 
\eeq{eq:c7}
The condition $F_t=0$ is
\beq
\frac{1}{\Lambda^{b_0}} \left(\det \widetilde T - \widetilde B
\widetilde{\bar B}\right) + m = 0 \ . 
\eeq{eq:d7}
This can be rewritten as
\beq 
\det \widetilde T - \widetilde B \widetilde{\bar B} = m\, \Lambda^{b_0}
= \left(\Lambda^{b_0}\right)_{\eff, N_F-1} \ , 
\eeq{eq:e7}
where I have used the decoupling condition \leqn{eq:r2}.  Since, in the 
effective theory with $N_f = N_c$, $b_0 = 2N_c$, this is precisely the 
constraint \leqn{eq:y2}.  Thus, the effective description of this case as a 
moduli space of vacua parametrized by $T$, $B$, $\bar B$, subject to the 
equations of motion following from the superpotential \leqn{eq:a7}, does 
connect correctly to our descriptions of supersymmetric QCD for smaller 
numbers of flavors.

The precise description of the moduli space of vacuum states is given by 
solving the $F$-flatness conditions which follow from \leqn{eq:a7}. These
are
\beq
T\cdot \bar B = B \cdot T = 0 \qquad \det T(T^{-1})^{ij} = B^i\bar B^j \ .
\eeq{eq:f7}
Notice that the point
$T=B=\bar B=0$ satisfies these conditions, and so there is a point in the 
moduli space where the full global symmetry
\beq 
SU(N_f)\times SU(N_f) \times U_B(1)\times U_{R}(1)
\eeq{eq:g7}
is preserved.  At this point, the 't Hooft anomaly conditions provide an
especially stringent check of the analysis.

The quantum numbers of the elementary fermions of the theory are
\beq
\psi_Q\,:\, (N_f,1)_{1,-1+ 1/N_f} \qquad
\psi_{\bar Q}\,:\, (1,\bar N_f)_{-1,-1+ 1/N_f}
\qquad
\lambda\,:\, (1,1)_{0,+1} \ .
\eeq{eq:h7}
Note that the $U_R(1)$ quantum numbers are those of the anomaly-free
$R$ symmetry \leqn{eq:w1}.  At the point of maximal symmetry, the 
composite fermions have the quantum numbers
\beq
\psi_T\,:\, (N_f,\bar N_f)_{0,-1+ 2/N_f}\qquad
\psi_B\,:\, (\bar N_f,1)_{N_c,-1/N_f} \qquad
\psi_{\bar B}\,:\, (1,N_f)_{N_c,-1/N_f}
\eeq{eq:h7a}
You can readily check that the  anomaly coefficients due to these 
representations are the following:
\beq
\begin{array}{rcc}
 & {elementary} & {composite}  \\[1.5ex]
\left(SU(N_f)\right)^3 \, : &  A N_c & A N_c \\[1ex]
\left(SU_L(N_f)\right)^2 U_B(1)\, :   & \half N_c &  \half N_c
      \\[1ex]
\left(SU_L(N_f)\right)^2 U_R(1)\, :   & -\half N_c^2/N_f &  -\half N_c^2/N_f
      \\[1ex]
\left(U_B(1)\right)^2U_R(1)\,: & -2 N_c^2 &  -2 N_c^2 \\[1ex]
\tr \left[ U_R(1) \right] \ ,:& -N_f^2 + 2 N_f -2 & -N_f^2 + 2 N_f -2  \\[1ex]
\left(U_r(1)\right)^3 \,:& N_f(N_f-2) -2N_c^4/N_f^2 & 
 N_f(N_f-2) -2N_c^4/N_f^2  \\[1ex]
\end{array}
\eeq{eq:i7}
where, in all of the lines, it is necessary to use the relation
 $N_c = (N_f -1)$.  The last line of the table is especially tedious to 
verify, but all of the anomalies do match, providing a remarkable consistency
check on the physical picture.

The picture of the vacuum states of supersymmetric QCD that we have 
constructed for $N_f = N_c +1$ has an obvious generalization for 
higher values of $N_f$.  The gauge-invariant chiral superfields of the 
theory are
\beq
T^{ij}\qquad B_{ij\cdots k}\qquad \bar B_{ij\cdots k} \ .
\eeq{eq:k7}
Here  $B_{ij\cdots k}$ is  the baryon superfield, build as a product of 
$N_c$ quark superfields, which contains all flavors 
except $ij\cdots k$, and $\bar B$ is defined in a similar way.  An 
$SU(N_f)\times SU(N_f)$-invariant
superpotential is given by 
\beq
W \sim \left(\det T - B_{ij\cdots k}T^{i\bar i} T^{j\bar j}\cdots
T^{k\bar k}\bar B_{ij\cdots \bar k}\right) \ .
\eeq{eq:m7}
However, this superpotential does not have $R$ charge equal to 2, and the 
multiplet of 
fields $(T, B , \bar B)$  does not satisfy the 't Hooft anomaly
conditions for the unbroken gauge group.  In fact, since the anomaly of the
flavor representation with $p$ indices grows as $N_f^{p-1}$, 
the mismatch of the 't Hooft conditions grows worse with each successive 
number of flavors.   We need a better idea.

\subsection{Seiberg's dual QCD}

Seiberg addressed this challenge in the following way: The baryon superfields
in \leqn{eq:k7} have
\beq 
\wNc =N_f-N_c
\eeq{eq:m7plus}
 indices.  Thus, we can view these fields as bound states of $\wNc$ components.
Let us assume that these components are the  physical asymptotic states of
 the theory.  We can associate these components with new superfields
$q$ and $\bar q$.  To bind these constituents into the gauge-invariant 
baryon superfields, we need a Yang-Mills theory with gauge group $SU(\wNc)$,
for which the $q$ and $\bar q$ transform in the fundamental and antifundamental
representations.  Then the baryon superfields would have the dual 
description
\beq
B_{ij\cdots k} = \epsilon_{a_1\cdots a_{\widetilde N_c}}\,
q^{a_1}_i q^{a_2}_j \cdots q^{a_{\widetilde N_c}}_k \ , 
\eeq{eq:q7}
and similarly for $\bar B$. 

The complete proposal put forward by Seiberg is that supersymmetric 
QCD with $N_f$ flavors can be described, for $N_f > N_c +1$, by a 
supersymmetric Yang-Mills theory with gauge group $SU(\wNc)$ coupled to the 
chiral fields $q_i$ and $\bar q_i$, $i = 1, \ldots , N_f$, and an additional 
chiral supermultiplet $T^{ij}$, which is a gauge singlet.  The field $T$
couples to $q$ and $\bar q$ through the superpotential
\beq
W = q\, T\, \bar q \ .
\eeq{eq:w7}
Without the superpotential, the theory has an additional $U(1)$ global
symmetry which acts on $T$; however, this symmetry is broken by 
\leqn{eq:w7}.  I will check below that the superpotential preserves the 
anomaly-free $R$ symmetry.  Thus, this model has the same global 
symmetry \leqn{eq:g7}
as the original supersymmetric QCD model.  Seiberg
refers to the relation between this theory and the original $SU(N_c)$
Yang-Mills 
theory as {\em non-Abelian electric-magnetic duality}.
I will explain some aspects of the duality of these theories in a moment.

In this picture, the $SU(\wNc)$ gauge group comes out of nowhere.  Its initial
role is to parametrize the constraint  that the dual quark fields $q$, $\bar q$
should combine correctly into the baryon fields.  However, systems are
known in which a gauge field which arises in this way to parametrize a 
constraint can become dynamical.  The most famous example  is the 
$CP^N$ nonlinear sigma model in 2 dimensions.\cite{CPN,CPNW}  In any event, we 
will assume here that the $SU(\wNc)$ gauge symmetry  is realized with  a fully
dynamical Yang-Mills theory, including asymptotic gauge bosons and gauginos.
With this idea, we place the theory  in a Coulomb phase of the $SU(\wNc)$
Yang-Mills theory in which the full complement of $SU(\wNc)$ gauge bosons
are massless.  Now we would like to ask, can we find nontrivial consistency
checks of this picture?

I have emphasize that the 't Hooft anomaly condition provides a stringent
test of the low-energy particle content of a strongly-coupled gauge theory.
Let us apply the test here, at the maximally symmetric point where none of the
fields acquire vacuum expectation values and the full global symmetry 
 \leqn{eq:g7} is realized.  The original quark superfields 
 have the quantum numbers
\beq
Q\,:\, (N_f,1)_{1,1-N_c/N_f}\qquad
\bar Q\,:\, (1,\bar N_f)_{-1,1-N_c/N_f} \ ,
\eeq{eq:n7}
using \leqn{eq:w1} for the $R$ charge.
Then the quantum numbers of the elementary fermions are
\beq
\psi_Q\,:\, (N_f,1)_{1,-N_c/N_f} \qquad
\psi_{\bar Q}\,:\, (1,\bar N_f)_{1,-N_c/N_f} 
\qquad \lambda\,:\, (1,1)_{0,+1} \ .
\eeq{eq:o7}
To obtain the quantum numbers of the superfields in the dual description,
compute the quantum numbers of a baryon field from \leqn{eq:n7} and then
divide the result among its  $\wNc$ components. This gives
\beq
q\,:\, (\bar N_f,1)_{N_c/\wNc,N_c/N_f}\qquad
\bar q\,:\, (1,N_f)_{-N_c/\wNc,N_c/N_f} \ .
\eeq{eq:n7again}
Then the fermionic components of these fields have the quantum numbers
\beq
\psi_q\,:\, (\bar N_f,1)_{N_c/\wNc,-1 + N_c/N_f} \qquad
\psi_{\bar Q}\,:\, (1,\bar N_f)_{-N_c/\wNc,-1 + N_c/N_f} \ .
\eeq{eq:o7again}
In addition, the physical fermions of the dual picture include the 
superpartners of $T$ and the $SU(\wNc)$ gauginos, 
\beq
\psi_T \,:\, (N_f,\bar N_f)_{0,1-2N_c/N_f} \qquad
\lambda \,:\, (1,1)_{0,+1} \ .
\eeq{eq:t7}

From this fermion content, it is straightforward to check the matching of
all of the possible anomaly coefficients:
\beq
\begin{array}{rcc}
 & {elementary} & {composite}  \\[1.5ex]
\left(SU(N_f)\right)^3 \, : &  A N_c & A N_c \\[1ex]
\left(SU_L(N_f)\right)^2 U_B(1)\, :   & \half N_c &  \half N_c
      \\[1ex]
\left(SU_L(N_f)\right)^2 U_R(1)\, :   & -\half N_c^2/N_f &  -\half N_c^2/N_f
      \\[1ex]
\left(U_B(1)\right)^2U_R(1)\,: & -2 N_c^2 &  -2 N_c^2 \\[1ex]
\tr \left[ U_R(1) \right] \,:&  -(N_c^2 + 1) &  -(N_c^2 + 1) \\[1ex]
\left(U_r(1)\right)^3 \,:& N_c^2 -1 -2N_c^4/N_f^2 & 
  N_c^2 -1 -2N_c^4/N_f^2  \\[1ex]
\end{array}
\eeq{eq:i7again}
In  the last two of these relations, the dual gauginos  give  a
contribution which is necessary for the success of the consistency check.
Since the value of this contribution is $(\wNc^2-1)\cdot (-1)$, including 
a sum over the dual gauge color quantum numbers, the matching requires us to
take seriously the realization of the  full $SU(\wNc)$ gauge supermultiplet
as a set of physical asymptotic states.

\subsection{Decoupling relations}

To make further checks of Seiberg's proposal, we should ask whether it 
connects correctly, through holomorphic decoupling,
 to the picture we have derived for supersymmetric QCD
with a smaller number of flavors.  In the process of answering this 
questions, we will find two other decoupling relations which also 
provide nontrivial checks of Seiberg's duality.	

The first of these addresses the question of whether the duality  relation
is in fact a duality.  If we act with the relation twice, do we recover the 
original theory?  Start with a supersymmetric Yang-Mills theory  of $SU(N_c)$
with $N_f$ flavors and no superpotential.  By the duality relation, this 
should be equivalent to a Yang-Mills theory with gauge group $SU(\wNc)$, 
an extra chiral multiplet $T$ which is a singlet of the gauge group,  and
the superpotential given in \leqn{eq:w7}.  Carrying out the duality 
transformation once again, we find a Yang-Mills theory with gauge group
$SU(N_c)$, an extra chiral multiplet $U$ which is a singlet of the gauge group,
and a superpotential of the form \leqn{eq:w7} which couples $U$ to the 
quark fields of $SU(N_c)$.  The superfield $U$ is identified with the 
bilinear $\bar q^i q^j$.  Thus, the final theory contains two singlet
 multiplets $T$ and $U$ and the superpotential
\beq
W = q\, T\, \bar q + QU\bar Q = \tr UT + QU\bar Q \ .
\eeq{eq:z7}
The first term in this expression gives mass to all of the components of
$T$ and $U$.  In addition, the $F$-flatness condition for $F_T$ implies
that $U = 0$.  Thus, when $T$ and $U$ decouple, we are left with an 
$SU(N_c)$ gauge theory with zero superpotential, just the theory that we 
started with.

Next, consider the effect of adding a mass term for the last flavor.
I will assume for the moment that $N_f > N_c +2$, so that this decoupling
should connect two theories which would both be expected to exhibit 
Seiberg's duality.  In the original theory, decoupling the $N_f$th flavor
gives a supersymmetric Yang-Mills theory with $(N_f-1)$ flavors and zero 
superpotential.

In the dual theory, the addition of the mass term gives us the superpotential
\beq
W = q\, T\, \bar q + m\, T_{N_fN_f} \ .
\eeq{eq:a8}
The $F$-flatness conditions for the $T_{N_f N_f}$,
 $q_{N_f}$ and $\bar q_{N_f}$ are
\beq
q^a_{N_f}\bar q^a_{N_f} + m = 0 \qquad (T\cdot \bar q^a)_{N_f}= (q^a\cdot
T)_{N_f}=0 \ .
\eeq{eq:b8}
In this equation, I have explicitly written the $SU(\wNc)$ gauge indices
$a$.  To solve the first of these equations, 
$q^a_{N_f}$ and $\bar q^a_{N_f}$ must obtain vacuum expectation values along
a parallel direction of the gauge group.  These expectation values break
$SU(\wNc)$ to $SU(\wNc-1)$.  Then the second and third equations in 
\leqn{eq:b8} imply that the $N_f$th row and column of $T^{ij}$ vanish.
The final result is an $SU(\wNc-1)$ gauge theory with $(N_f -1)$ flavors, 
a gauge singlet superfield $T$ which is an $(N_f-1)\times (N_f-1)$ matrix,
and the superpotential \leqn{eq:w7} coupling $T$ to the quark superfields.
This is the Seiberg dual of supersymmetric Yang-Mills theory with 
$(N_f-1)$ flavors.

We have now seen that holomorphic decoupling correctly connects different
theories with Seiberg's duality in a way that preserves the duality relation.
However, we still need to check that the theories with Seiberg's duality
are correctly connected to the supersymmetric QCD models with a smaller
number of flavors which we have described in earlier sections using different
physical pictures.  To check this connection, consider decoupling the last
flavor in supersymmetric QCD with $(N_c + 2)$ flavors.  In this case, 
the dual Yang-Mills theory has $SU(2)$ gauge symmetry.  The analysis 
of the previous paragraph still applies to this case, leading to a 
superpotential of the form \leqn{eq:w7} with $q_i$, $\bar q_i$ now 
1-component fields for each value of $i = 1, \ldots, (N_c +1)$.

  At the same
time, the expectation values of $q_{N_f}$ and $\bar q_{N_f}$ break the 
$SU(2)$ gauge symmetry completely.  Thus, as in Section 4.4, this case
provides us with a well-defined instanton calculation which potentially adds
another term to the superpotential.  As in that section, we can analyze the 
instanton effect by counting zero modes.  The instanton creates one each 
of the fermions $\psi_{qi}$, $\psi_{\bar q i}$, and 4 dual gauginos $\lambda$.
We can turn four quarks into squark fields  
using the squark-quark-gaugino coupling and replace the squark fields with
$i=N_f$  by the vacuum expectation values of these fields.  We can 
then use the superpotential coupling to convert pairs  
 $\psi_{qi} \psi_{\bar q j}$ to $T^{ij}$ or pairs $q_i \psi_{\bar q j}$ to 
$\psi_{T}^{ij}$.  This allows us to construct  an effective interaction with
 two fermions which contains
 $(N_f-1)$ powers of $T^{ij}$ or $\psi_{T}^{ij}$, $i,j = 1, \ldots (N_f-1)$,
in  a combination invariant
to the residual $SU(N_f-1)\times SU(N_f-1)$ flavor symmetry.  This 
interaction has the form of a superpotential correction
\beq
       \int d^2\theta \Delta W =  \int d^2\theta \det T \ ,
\eeq{eq:b8plus}
up to an overall constant depending on $\Lambda$ and $\VEV{q_{N_f}}$.  
Putting together the two contributions to the superpotential, we find
\beq
W_\eff = (q\cdot T\cdot \bar q - \det T) \ .
\eeq{eq:h8}
This has exactly the form of the superpotential \leqn{eq:a7} which we wrote
for the theory with $(N_c + 1)$ flavors, with the identification
\beq
q_i \rightarrow B_i \qquad \bar q_i \rightarrow \bar B_i \ . 
\eeq{eq:i8}
Now the whole chain of effective descriptions of supersymmetric QCD, 
from $N_f = 0$ to large values of $N_f$, is linked together by 
holomorphic decoupling.

\subsection{Fixed points and asymptotic states}

In the analysis we have just completed, it seemed that Seiberg's duality
could connect supersymmetric Yang-Mills theories with arbitrarily large
values of $N_f$.  But there is a problem here, because, for sufficiently
large $N_f$, the Yang-Mills theory will lose asymptotic freedom.  In this
case, the theory reverts to a weakly-coupled system of quark and gluon 
supermultiplets, interacting through asymptotically decaying forces.  There
does not seem to be a role here for the dual quark and gaugino fields which
I insisted in the previous section should be thought of as physical particles.

To understand how the regime with non-Abelian duality fits together with this 
infrared-free regime, Seiberg proposed an additional interesting hypothesis:
At fixed $N_c$, for some intermediate region of $N_f$, supersymmetric QCD 
is described by a scale-invariant theory which would be a new 
infrared fixed point of the renormalization group.  In fact, this fixed point
is already known in a certain region of $N_f$.  In nonsupersymmetric 
gauge theories, at the point in $N_f$
where the first $\beta$
 function coefficient vanishes, the second $\beta$ function
coefficient has already turned positive.  Thus, for values of $N_f$ just 
below the critical value $N_f^c$ where asymptotic freedom is lost, there
 is an
infrared fixed point at a weak coupling
 $g^2/4\pi \sim (N_f^c - N_f)$.\cite{BZ}  A similar 
result holds in the supersymmetric case.  Seiberg conjectured that this 
fixed point extends downward in $N_f$ through a significant region.

In a supersymmetric field theory, a scale-invariant point necessarily has 
{\em superconformal} invariance, and this extension of the global symmetry
group adds interesting structure to the theory.  Recall that, in 
supersymmetric theories, the energy-momentum tensor $T^{\mu\nu}$ belongs to a 
supermultiplet which also contains the supersymmetry current $S^\mu_\alpha$
and a $U(1)$ current $J^\mu$.  In a classical scale-invariant supersymmetric
theory,  $J^\mu$ is the current of the canonical $R$ symmetry.  Ordinary
supersymmetry implies that $T^{\mu\nu}$ and  $S^\mu_\alpha$ are conserved.
Superconformal invariance implies, in addition,
\beq 
T^\mu_\mu = 0 \qquad 
\gamma^\mu_{\alpha\beta}S_{\mu\beta} = 0 \qquad \partial_\mu
J^\mu = 0 \ . 
\eeq{eq:k8}
At a fixed point of  supersymmetric QCD, then, $J^\mu$ must be the 
conserved current of the anomaly-free $R$ symmetry. The superconformal 
algebra gives restrictions on the eigenvalues of these operators.  In 
particular, the scaling dimension of a field is bounded by its $R$ 
charge,
\beq
d \geq \frac{3}{2} |R| \ ; 
\eeq{eq:l8}
the inequality is saturated for chiral and antichiral superfields.\cite{SCf}
In addition, as is true in the nonsupersymmetric case, the scaling 
dimension of a scalar field must satisfy
\beq
d \geq 1 \ , 
\eeq{eq:m8}
with $d=1$ possible only for a free field.\cite{JFed}  I should note that 
both of these inequalities apply strictly only to gauge-invariant operators.

 Consider the implications of these statements if supersymmetric QCD is 
scale-invariant in a region where it exhibits Seiberg's duality. Since 
the basic objects of our description are chiral superfields, we can work out
their scaling dimensions from their $R$ charges.  In particular,  for the 
gauge-invariant combinations,
\beqa
 Q\cdot \bar Q = T \quad &\mbox{\rm has}& \quad  d= 
3 \left({N_f-N_c\over N_f}\right) \CR
 q\cdot \bar q = U \quad &\mbox{\rm has}& \quad  d= 
3 \left({N_c\over N_f}\right)
\eeqa{eq:n8}
As a check on these relations, the superpotential \leqn{eq:w7} has 
$R=2$, as needed to preserve the $R$ symmetry. By \leqn{eq:l8}, this
superpotential would  also have $d=3$, which is the correct value for this
to be a marginal perturbation. 

In supersymmetric QCD, the $\beta$ function coefficient $b_0$ is given 
by \leqn{eq:g1} and vanishes at $N_f = 3 N_c$. At this point, the 
bilinear $U$ in \leqn{eq:n8} comes down to $d=1$ and becomes a free 
field.  For larger values of $N_f$, the dual theory can no longer be
consistently described as a superconformal fixed point, but this is 
just as well, because the original QCD is known not to be scale-invariant
in this regime.  Rather, it is a theory with weak gauge
interactions whose strength decreases logarithmically at large distances.

In a similar way, the dimension of the bilinear $T$ reaches 1 at 
$N_f = 3N_c/2$.  This value has another significance; since the 
beta function coefficient of the dual theory is
\beq
   b_0 = 3 \wNc - N_f =  2N_f - 3 N_c \ , 
\eeq{eq:o8}
this is the value of $N_f$ below which the dual theory becomes infrared-free.

\begin{figure}
\begin{center}
\leavevmode
\epsfbox{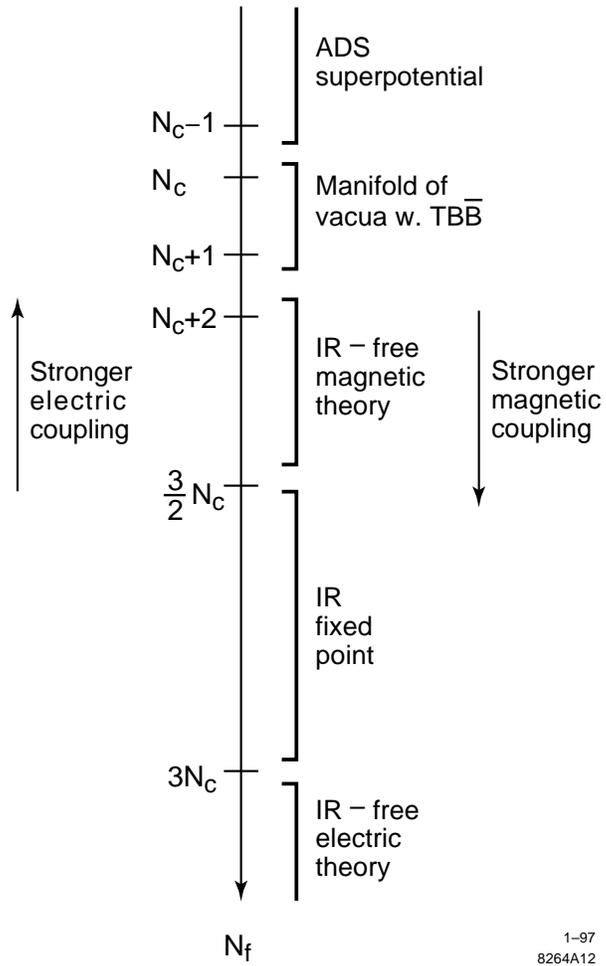}
\end{center}
 \caption{Seiberg's plan of the behaviour of supersymmetric Yang-Mills 
theory as a function of the number of flavors $N_f$.}
\label{fig:twelve}
\end{figure}

From this information, we can put together the following picture of the
behavior of supersymmetric QCD for values of $N_f$ greater than $N_c$.
For $N_f = (N_c + 1)$, the asymptotic particles are the mesons $T$ and 
baryons $B$ and $\bar B$ and their superpartners.  For the next few values 
of $N_f$, the asymptotic particles are the mesons $T$ and the dual quarks
$q$ and $\bar q$, interacting through an infrared-free supersymmetric
 Yang-Mills theory of $SU(\wNc)$.  Above $N_f = 3N_c/2$, however, the 
theory goes to a nontrivial infrared fixed point which is an attractor for
both the original and the dual Lagrangian.  As $N_f$ increases, this fixed
point theory looks less and less like the dual Yang-Mills theory and more and
more like a weakly-coupled version of the original Yang-Mills theory. 
Finally, at $N_f = 3N_c$, the fixed point comes to zero coupling in the 
original supersymmetric Yang-Mills theory.  For still higher values of $N_f$,
the asymptotic particles are the original quarks, interacting 
through an infrared-free supersymmetric  Yang-Mills theory of $SU(N_c)$.
The whole picture of the evolution of supersymmetric Yang-Mills theory 
with $N_f$ is displayed in Figure \ref{fig:twelve}.

An interesting aspect of the plan shown in this figure is that, as $N_f$
decreases, the qualitative behavior of the theory contains increasingly
more strong-coupling, nonperturbative dynamics for the original quarks
and gluons.  We proceed from a free region, to a fixed-point region, to 
a region of confinement, to the extreme region of the Affleck-Dine-Seiberg
superpotential.  On the other hand, along this same axis, the dual theory
changes from a strongly-coupled, confining theory to a free theory.  Where
one coupling is weak, the dual coupling is strong.  This behavior strongly
motivates Seiberg's idea that the relation of the original and dual pictures
is a non-Abelian generalization of electric-magnetic duality.

  Although our analysis in this section has been given for supersymmetric 
QCD, it is highly suggestive that a similar behavior could appear in 
ordinary nonsupersymmetric QCD.   For a sufficiently small number of flavors,
we have color confinement and chiral symmetry breaking due to the expectation
value of the quark bilinear \leqn{eq:d}.  However, for larger values of 
$N_f$, the theory could go to an infrared fixed point which corresponds to an
asymptotic non-Abelian Coulomb phase with no chiral symmetry breaking.
Some time ago, Banks and Zaks argued that such a phase always appears for 
$N_f$ sufficiently close to the critical value at which the theory loses 
asymptotic freedom.\cite{BZ} And there is some evidence from 
numerical lattice simulations that QCD with the gauge group $SU(3)$ 
no longer exhibits confinement  and chiral symmetry breaking
for $N_f > 7$.\cite{Iwasaki}  It will be very interesting to learn whether
the complete picture that Seiberg has assembled for supersymmetric QCD
has a direct analogue in nonsupersymmetric QCD.

To conclude this section, I would like to note two interesting checks of 
Seiberg's duality.  Argyres, Plesser, and Seiberg\cite{APSd}
 have studied
the duality starting from $N=2$ supersymmetric QCD, by introducing
explicit breaking to $N=1$.  They have exhibited a point in the  
Coulomb phase of the $N=2$ 
theory such that the reduction to $N=1$ gives the Seiberg dual theory,
and they have shown that  this point can be continuously connected to
the standard picture of supersymmetric QCD at weak coupling through a path
in the $N=2$ Coulomb phase.  Bershadsky, Johansen, Pantev, Sadov, 
and Vafa\cite{BJPSV} have recently identified Seiberg's duality in a 
stringy context, as a $T$-duality of certain Type IIB compactifications.

\section{Generalizations of non-Abelian Duality}
\label{sec:f}

Seiberg's work described in the previous section gives a unified picture
of the behavior of $N=1$ supersymmetric $SU(N_c)$ Yang-Mills theories
with $N_f$ flavors for the whole range of possible values of $N_f$.
We might draw from this analysis the insight that it is interesting to 
consider the systematics of other families of supersymmetric Yang-Mills
theories with varying numbers of flavors.  In this section, I will briefly
discuss a few interesting cases.  In the past year, many examples of 
strong-coupling behavior in $N=1$ supersymmetric Yang-Mills theories have 
been explored.  There is no space here for a complete review of this 
subject, but I hope that these examples will give an idea of the richness
of the phenomena that have been uncovered.

\subsection{$SO(N_c)$ and $Sp(2n_c)$}

The simplest generalizations of Seiberg's duality occur in vectorlike
$SO(N_c)$ and $Sp(2n_c)$ gauge theories with $N_f$ flavors of quarks and  
squarks in the fundmental representation.  I will now explain how the 
systematics of $SU(N_c)$ gauge theories presented in Section 7 extends to 
these theories. 

Consider first $SO(N_c)$ gauge theories with $N_f$ flavors of quarks
and squarks $Q^i$ in the representation of dimension $N_f$. The global
flavor symmetry of this theory is $SU(N_f) \times U_R(1)$.
  The $\beta$ function of the theory is
given by
\beq
        b_0 = 3(N_c -2) - N_f \ .
\eeq{eq:a9}
The fundamental and adjoint representations of $SO(N_c)$ have anomaly 
coefficients $n$, as in \leqn{eq:h1plus}, equal to 2 and $2(N_c-2)$, 
respectively.  Thus, for this theory we can make a table similar to 
\leqn{eq:uaa}.  Let $A$ represent the anomalous $U(1)$ flavor symmetry
of the $Q^i$.  Let $R$ and $R_{AF}$ represent the canonical and 
non-anomalous $R$ symmetries; $R_{AF}$ is given by 
\beq
R_{AF} = R + \frac{N_f-N_c+2}{N_f} A \ .
\eeq{eq:b9}
Let $T^{ij}$ be the gauge-invariant chiral superfield $Q^i\cdot Q^j$; 
this is a symmetric tensor of the flavor $SU(N_f)$.
Then we have
\beq
\begin{array}{c@{\hspace{.2in}\extracolsep{.1in}}cccl}
                  & A & R    & R_{AF} & \\[2ex]
Q^i            & +1  & 0           & (N_f+2 -N_c)/N_f & \\[1ex]
\lambda  &  0      &  +1           &  +1 & \\[1ex]
\Lambda^{b_0}  &  2N_f  & - 2(N_f + 2 - N_c) & 0  & \\[1ex]
\det T  &  2N_f  & 0 &  2(N_f + 2 - N_c)   & \\[1ex]
\end{array}
\eeq{eq:c9}

A  nonperturbative superpotential 
for this theory must be invariant under $A$ and must have $R$ charge
2.  From the data in the table, the only possibility is 
\beq
W_\eff = c \cdot \left(\frac{\Lambda^{b_0}}{\det T}\right)
^{1/(N_c-2 - N_f)} \ .
\eeq{eq:d9}
Thus, we expect that, for $N_f < (N_c -2)$, a superpotential is generated
in the matter described by Affleck, Dine, and Seiberg, while for 
$N_f \geq N_c$, there is an electric-magnetic duality.

The duality of the theory for large $N_f$ has been worked out by 
Intriligator and Seiberg.\cite{SON}  The dual theory is an $SO(N_f-N_c + 4)$
gauge theory with dual quark superfields $q_i$ in the $\bar{N_f}$ 
representation of the $SU(N_f)$ flavor group, the gauge singlet superfield
$T^{ij}$, and the superpotential
\beq
W =  T^{ij} q_i \cdot q_j\ .
\eeq{eq:e9}
This theory satisfies the 't Hooft anomaly conditions at the origin of 
moduli space in a manner similar to that of the $SU(N_c)$ duality.

For intermediate values of $N_f$, there are some interesting special 
cases.\cite{SON}
  For $N_f = N_c -4$, the theory is described at weak coupling by 
expectation values $\vev{Q^i}$ which generically break $SO(N_c)$ to an 
$SO(4)$ pure gauge theory.  Since $SO(4) = SU(2)\times SU(2)$, this 
theory has two $SU(2)$ gaugino condensates which are equal in magnitude.
For each of these condensates, the $Z_2$ symmetry of the theory
gives two choices $(\pm 1)$ for its phase. 
 If the two condensates are chosen parallel,
we obtain the superpotential \leqn{eq:d9}.  If the two condensates are 
chosen antiparallel, we obtain a second branch of the theory with zero 
superpotential and a nontrivial moduli space.   This second branch is also
found in the case $N_f = N_c -3$, reflecting the possibility of a 
cancellation between the contributions to the superpotential from 
the $SO(3)$ gaugino condensate and from explicit instanton effects.
In this latter case, a new chiral field $q_i$ in the $\bar{N_f}$ representation
of $SU(N_f)$ is needed to satisfy the 't Hooft anomaly condition.
For $N_f = N_c -2$, no superpotential can be generated.  The weak-coupling
description of the theory has $SO(N_c)$ broken to $SO(2) = U(1)$, so 
the theory has a Coulomb phase.  The new fields $q_i$ from the previous
case are generated by decoupling from  magnetic monopoles
in this theory.  For $N_f = N_c -1$, the theory is described by a dual
$SO(3)$ gauge theory with the superpotential
\beq
W =  T^{ij} q_i \cdot q_j  - \det\, T \ .
\eeq{eq:f9}
Beginning with $N_f = N_c$, we find the generic situation for large $N_f$
described in the previous paragraph.  There are additional complications
for the special cases of $N_c = 3,4$.

For $Sp(2n_c)$ gauge theories, the situation is rather more 
straightforward.\cite{SPN}  For these theories, the number of flavors
must be even to avoid discrete gauge anomalies.  Thus, we introduce
an even number $N_f = 2n_f$ of supermultiplets $Q^i$ in the fundamental
$2n_c$-dimensional representation. The global
flavor symmetry of this theory is $SU(2n_f) \times U_R(1)$.
  The $\beta$ function of the theory is
given by
\beq
        b_0 = 3(2n_c + 2) - 2n_f \ .
\eeq{eq:aa9}
The fundamental and adjoint representations of have anomaly 
coefficients $n$ equal to  2 and $4(n_c+1)$, 
respectively.   Let $A$ again represent the anomalous $U(1)$ flavor symmetry
of the $Q^i$.  Let $R$ and $R_{AF}$ represent the canonical and 
non-anomalous $R$ symmetries; $R_{AF}$ is given by 
\beq
R_{AF} = R + \frac{n_f-n_c-1}{n_f} A \ .
\eeq{eq:g9}
Let $T^{ij}$ be the gauge-invariant chiral superfield $Q^i\cdot Q^j$; 
this is an antisymmetric  tensor of the flavor $SU(2n_f)$.
Then the table of quantum numbers reads
\beq 
\begin{array}{c@{\hspace{.2in}\extracolsep{.1in}}cccl}
                  & A & R    & R_{AF} & \\[2ex]
Q^i              & +1  & 0           & (n_f- 1 -n_c)/n_f & \\[1ex]
\lambda       &  0           &  +1 &  +1 &\\[1ex]
\Lambda^{b_0}  &  4n_f  & - 4(n_f - 1 - n_c) & 0  & \\[1ex]
\det T  &  4n_f  & 0 &  4(n_f - 1 - n_c)   & \\[1ex]
\end{array}
\eeq{eq:h9}
Since $T$ is an antisymmetric matrix, its determinant factorizes as the 
square of simpler object, the Pfaffian $\hbox{Pf}\ T$.

A  nonperturbative superpotential 
for this theory must be invariant under $A$ and must have $R$ charge
2. The unique possibility is
\beq
W_\eff = c \cdot \left(\frac{\Lambda^{b_0/2}}{\hbox{Pf}\ T}\right)
^{1/(n_c+1 - n_f)} \ .
\eeq{eq:i9}
This superpotential is generated for all cases $n_f < n_c + 1$. 
 In the 
case $n_f = (n_c +1)$, the theory has a moduli space of vacua with a 
nonperturbatively modified constraint
\beq
{\hbox{Pf}}\ T = \Lambda^{2(n_c +1)} \ .
\eeq{eq:j9}
For $n_f = (n_c + 2)$, the theory has a moduli space of vacua with the 
superpotential
\beq
     W = {\hbox{Pf}}\ T \ .
\eeq{eq:k9}
For $n_f \geq (n_c + 3)$,  the theory is dual to an $Sp(2(n_f-n_c-2))$
gauge theory with quark superfields $q_i$ in the $\bar{2n}_f$ representation
of $SU(2n_f)$ and the superpotential
\beq
W =  T^{ij} q_i \cdot q_j\ .
\eeq{eq:l9}
Thus, the vectorlike supersymmetric Yang-Mills theories based on the 
classical groups $SU(N_c)$, $SO(N_c)$, and $Sp(2n_c)$ all show similar
patterns in their qualitative behavior.

\subsection{Examples with chiral matter content}

The systematics of $N=1$ supersymmetric Yang-Mills theories becomes strang\-er 
when we consider models with more general representations.  An interesting
example to consider next is the $SU(N_c)$ model with a symmetric tensor
multiplet $S$ and $N_f$ multiplets $Q^i$ in the $\bar N_c$ representation.
This is a chiral guage theory, and the cancellation of gauge anomalies
requires $N_f = N_c + 4$.   This theory has a large number of possible 
gauge-invariant chiral fields, of which the two simplest are
\beq
         U = \det S \ , \qquad   M^{ij} = Q^i\cdot S \cdot Q^j \ ,
\eeq{eq:m9}
a singlet and a symmetric tensor of the flavor group $SU(N_f)$.

 Pouliot and Strassler have found that the properties of this theory are
matched by a dual gauge theory with the gauge group $SO(8)$.\cite{spineight}
The dual theory contains $N_f$ multiplets $q_i$ in 
vector representations, one multiplet $p$ in the spinor 
representation, and gauge singlet fields $U$ and $M^{ij}$.  The dual
theory has a nontrivial superpotential
\beq
         W =   M^{ij} q_i\cdot q_j +  U p\cdot p \ .
\eeq{eq:n9}
Reciprocally, the $SO(8)$ theory with the same charged matter content
$q_i$, $p$ and zero superpotential
is dual to an $SU(N_c)$ gauge theory with a symmetric tensor 
multiplet $S$, quarks $Q^i$, and the additional
gauge singlet fields
\beq
         T = p\cdot p \ , \qquad   N^{ij} = q_i\cdot q_j
\eeq{eq:o9}
and the superpotential
\beq
         W =   N_{ij} Q^i \cdot S \cdot Q^j + T \det S\ .
\eeq{eq:n9again}

This is a bizarre transformation.  In the forward direction, we began from a 
chiral gauge theory, but the dual was a vectorlike theory.  In the
reciprocal relation, we began from a vectorlike theory and found a chiral
theory as the dual.  This turns out to be a common phenomenon in the 
more complex examples of non-Abelian duality. The first example was found
by Pouliot in an $SO(7)$ model.\cite{spinseven}

Similar examples can be found in models with antisymmetric tensor 
representations.  Consider, for example, $SU(N_c)$ Yang-Mills theory
with an antisymmetric tensor representation $A^{ij}$, $M$ multiplets $Q^i$
in the $N_c$  representation, and $N$ multiplets $\bar Q_i$
in the $\bar N_c$ reprsentation, and zero  superpotential. Anomaly 
cancellation requires $N = N_c - 4 + M$. 
As $M$ is increased, this theory exhibits a progression of behaviors, 
with a nonperturbative superpotential generated for $M \leq 2$, 
a constrained moduli space with a nonperturbative correction for $M=3$,
and a moduli space of vacua with a superpotential for $M=4$.\cite{PT}
 Pouliot has 
shown that, for $M \geq 5$,
 this theory has as a  dual which is an  $SU(M-3)\times Sp(2(M-4))$ 
gauge theory.  The matter content is rather large.  Let me denote the 
fundamental representation by $f$ and the antisymmetric tensor representation
by $a$, and write for each multiplet the content under the gauge group
and the non-Abelian part of the flavor group.  Then each multiplet belongs
to a representation of  $SU(M-3)\times Sp(2(M-4))\times SU(M)\times SU(N)$.
In this notation, the 
dual theory contains the multiplets
\beqa
   x\ :\ (f,f;1,1)\ ,  &\quad&  p\ :\ (f,1;1,1)\ ,  \CR
  \bar a\ :\ (\bar a,1;1,1)\ ,  &\quad&  \bar q\ :\ (\bar f,1;\bar f,1)\ ,\CR
   \ell \ :\ (1,f;1,\bar f)\ , &\quad&   M\ :\ (1,1;f,f)\ , \CR 
   H \ :\ (1,1;1,a)\ , &\quad&   B\ :\ (1,1;f,1)\  
\eeqa{eq:p9}
interacting through the superpotential
\beq
   W =  M \bar q \ell x + H \ell \ell + B p \bar q + \bar a x^2 \ .
\eeq{eq:q9}

This theory brings us into territory that is interesting for another 
reason.  $SU(N_c)$ gauge theories with antisymmetric tensor representations
provide the simplest examples of supersymmetric Yang-Mills theories with 
spontaneously broken supersymmetry.  Some time ago, Affleck, Dine, and 
Seiberg pointed out that the $SU(5)$ gauge theory with one $10$ and one
$\bar 5$ matter superfield spontaneously breaks supersymmetry.\cite{ADSUfive}
The intuitive reason for this is easy to understand
from the considerations of Section 4:  The origin of field space where
the $10$ and $\bar 5$ have zero vacuum expectation values is destabilized 
by nonperturbative dynamics, as we found there.
  But, since it is not possible to 
build a gauge-invariant chiral field from these ingredients, there are no
$D$-flat directions along which the vacuum can escape to infinity.
Recently, Murayama\cite{Murayama} has made this argument quite concrete
by studying the $SU(5)$ gauge theory with a $10$, two $\bar 5$s, and a $5$
with a mass term that decouples one $5 + \bar 5$ pair.  The argument can be
repeated for every larger odd value of $N_c$.  In those theories, there 
is a $D$-flat direction along which the theory can escape to infinity, but
at the end of this trajectory the theory is broken only to $SU(5)$.  Thus,
there is no possible vacuum state that preserves supersymmetry.

The example just discussed shows the possibility of exploring dynamical 
supersymmetry breaking using duality.  Indeed, Pouliot showed that, when 
one decouples $M$ flavors in the dual picture, the resulting theory has a 
superpotential which does not allow an $F$-flat vacuum 
configuration.\cite{SUNA}

By combining the various ingredients that I have discussed in these 
lectures, working with  non-simple gauge groups and including explicit
as well as dynamical superpotentials,
 it is possible to construct a wide variety of models of 
dynamical supersymmetry breaking.  Intriligator and Thomas have presented a
catalogue of supersymmetry-breaking mechanisms that appear in these 
models,\cite{InT} and many examples are now being generated.

On the other hand, the broad picture of non-Abelian duality in  $N=1$
supersymmetric Yang-Mills theory remains far from clear.  Many examples
of duality have been generated in the past year, many more than I have 
space to review, but as yet there is no broad picture of the systematics
of this phenomenon.  The recent papers \cite{BS,CSSk} are two recent
attempts to bring order to the $N=1$ gauge theories, neither completely
successful.  Most likely, there are many strange things still to 
be learned about these models.

In this atmosphere of promise and confusion, I end these lectures.  I wish
you, the reader, good luck in finding the connections among supersymmetric
Yang-Mills theories that are still hidden.  I hope that we will also be
able to find a place for the 
wealth of phenomenon these theories provide in realistic models of Nature.

\section*{Acknowledgments}

I am grateful to Brian Greene and  K. T. Mahantappa for the opportunity to 
participate in this stimulating school, and to Alex C.-L. Chou, 
Michael Dine, Joseph Minahan, Ann Nelson,
Nati Seiberg,  Michael Shifman,
Scott Thomas, Shimon Yankielowicz, and many other people who have helped
me to understand the topics reviewed here.  This work was supported by the 
Department of Energy under contract DE--AC03--76SF00515.

\end{document}